\documentclass{pinchcr}   

\usepackage{amsmath,amssymb,amsfonts}
\usepackage{graphicx}
\usepackage{array}

\def\texthc{\text{h.c.}}

\def\kF{k_{\text{F}}}\def\vkF{\vec{k}_{\text{F}}}
\def\kB{k_{\text{B}}}

\def\dos{\cN}       

\def\cN{\mathcal{N}}

\def\vk{\vec{k}}

\def\vq{\vec{q}}\def\vr{\vec{r}}

\def\openone{\leavevmode\hbox{\small1\kern-3.3pt\normalsize1}}
\def\down{\downarrow}\def\up{\uparrow}

\newcommand{\step}{\operatorname{\theta}}

\newcommand{\cref}[1]{Chapter~\ref{#1}}
\newcommand{\sref}[1]{Section~\ref{#1}}

\newcommand{\fref}[1]{Fig.~\ref{#1}}

\newcommand{\acomm}[2]{[#1,#2]_+}

\newcommand{\ket}[1]{\left\vert#1\right\rangle}

\newcommand{\braket}[2]{\langle#1\vert#2\rangle}

\newcommand{\hc}[1]{#1^{\dagger}}

\def\eps{\varepsilon}

\def\hn{\hat{n}}
\def\beq{\begin{equation}}
\def\beqno{\begin{equation}\nonumber}
\def\eeq{\end{equation}}

\def\t{t}

\def\li6{$^6$\rm{Li}}

\newcommand{\br}{\mathbf{r}}

\newcommand{\vrp}{\vec{r^\prime}}

\newcommand{\gamk}{\gamma_{\vec{k}}}
\newcommand{\xivk}{\xi_{\vec{k}}}
\newcommand{\xivok}{\xi^0_{\vec{k}}}

\newcommand{\omu}{\mu^*}

\title{Strongly correlated bosons and fermions in optical lattices}

\author{Antoine Georges}
\affiliation{Centre de Physique Th\'eorique, \'Ecole Polytechnique, CNRS, 91128 Palaiseau Cedex, France;
\\ Coll\`ege de France, 11 place Marcelin Berthelot, 75005 Paris, France;
\\ DPMC-MaNEP, University of Geneva, 24 Quai Ernest Ansermet, 1211 Geneva 4, Switzerland}
\author{Thierry Giamarchi}
\affiliation{DPMC-MaNEP, University of Geneva, 24 Quai Ernest Ansermet, 1211 Geneva 4, Switzerland}
\authors{2}

\begin{document}

\maketitle

%
%
%
%
%
\tableofcontents

\maintext





\section{Introduction}
\label{sec:intro}

The effect of interactions on many-particle quantum systems particles has proven to be
of the most fascinating problems in physics.
From the fundamental
physics point of view, this is a formidable challenge that combines the difficulties of quantum mechanics and statistical physics.
Indeed even in very small clumps of matter there are more particles than stars in the universe and when these particles interact one is
thus totally unable to solve by brute force the coupled equations. This is even more so
when one deals with quantum particles, which behaves as interfering waves,
and must in addition obey the principles of symmetrization and antisymmetrization.
As a consequence of this complexity, beautiful new physics emerge from the collective
behavior of these particles, something that could not even be guessed at by simply looking
at the solutions of small numbers of coupled particles. However,
finding the proper tools to even tackle such a type of problems
is an herculean task. Fortunately some important concepts allow us to understand the main
physical properties of many of these systems. However many systems defy our understanding and we need to build new tools to
tackle them.
This forces us to accomplish progress in our way to understand these systems theoretically,
either analytically or numerically.

The pressure to solve these problems goes way beyond the academic realm.
Understanding how electrons behave in solids led to technological revolutions such as
silicon-based electronics and the transistor, the control of spin in magnetic storage and electronics,
or the fascinating applications of
superconductivity. Hence, this endeavor is intimately connected with our ability to engineer and
control solids, and make devices of use in our everyday's life.
 %

Recently a new type of physical systems, cold atoms in optical lattices, has provided a marvelous
laboratory to tackle the effects of strong correlations in quantum systems. These systems
made of light and neutral atoms constitute a welcome alternative to the standard realization
in solid state physics.
Because in these systems interactions are short-ranged and controllable,
because optical lattice can be engineered in a flexible ways and phonon modes are
absent, these systems can be viewed as model realizations.
In addition, they have opened the path to novel physics which uses the control
and flexibility of these systems (mixtures of bosons-fermions, possibility to change rapidly
the potentials, isolated quantum systems etc..) In particular they have allowed to realize quantum systems in
reduced dimensionality in which quite remarkable novel physics can occur.

In these lectures we give an introduction to the physics of interacting quantum
systems, both bosonic and fermionic. We review the main concepts and tools which are cornerstones of our understanding
of such systems and point out the challenges that interactions pose. These lecture cannot of course make any claim of completeness
given the broad scope of the problem, and we encourage the reader to search the literature for more.

The plan of the lectures is as follows. In \sref{sec:lattices} we will give an introduction to the physics of
quantum particles in periodic lattices. The presentation is essentially targeted to the case of cold atomic systems. We will examine how the interactions should be taken into account and define the basic models, such as the Hubbard model,
that can be used to describe such interacting systems. In \sref{sec:mottboson} we examine for the case of bosons,
how the combined effects of lattice and interaction can turn the system into an insulator, the so-called Mott insulator, and discuss the corresponding physics. \sref{sec:boso} discusses what happens when the system is one-dimensional.
In that case the fact that two particles cannot cross without feeling their interaction lead to novel physics effects.
This section discusses this new physics and the methods needed to access it. We then move in \sref{sec:fermiliquid} to the case of fermions.
We discuss first Fermi-liquid theory and the concept of {\it quasiparticles}, a remarkable description
of the low-energy excitations of interacting fermion systems, due to Landau. In a nutshell, this approach
implies that the effect of interactions does not qualitatively change the nature of low-energy excitations as compared
to a non-interacting system, except of course if the interactions are strong enough to lead to a instability of the
system and for example destroy metallic behavior. Fermi liquid theory and the concepts behind it have been the cornerstone
of our understanding of the properties of most solids.
We will then see in \sref{sec:mottfermion} how, in a similar way than for bosons, the combination of a lattice and strong interactions can turn
a Fermi liquid into a Mott insulator. We look in \sref{sec:mottfermion1D} at the properties of one-dimensional fermions, show how Fermi liquid theory fails because low-energy excitations are now collective modes instead of quasiparticles,
and examine the corresponding physics both for the conducting and insulating phases. Finally we draw some conclusions
and give some perspectives in \sref{sec:conclusion}.


\section{Optical lattices}
\label{sec:lattices}


Before dealing with the effects of interactions let us first have a look at the properties
of individual quantum particles. One essential ingredient, both in solids and in cold atom
systems is the presence of a periodic potential. In solids, such a potential occurs naturally for the electrons
because of the presence of the regular array of positively charged nucleus. In cold atomic systems
it can be imposed by the presence of an optical lattice. Such a potential takes usually the form (in the direction of the lattice) \cite{bloch_cold_atoms_optical_lattices_review}
\begin{equation} \label{eq:optical_potential}
 V(x) = V_0 \sin^2(k x)
\end{equation}
where $k = \pi/a$, with $a$ the lattice spacing.
The presence of such a periodic potential changes considerably the properties compared to the one of free particles.

\subsection{Zero kinetic energy (``Atomic'' limit)} \label{sec:atomiclimit}

\begin{figure}
\begin{center}
 \includegraphics[width=\columnwidth]{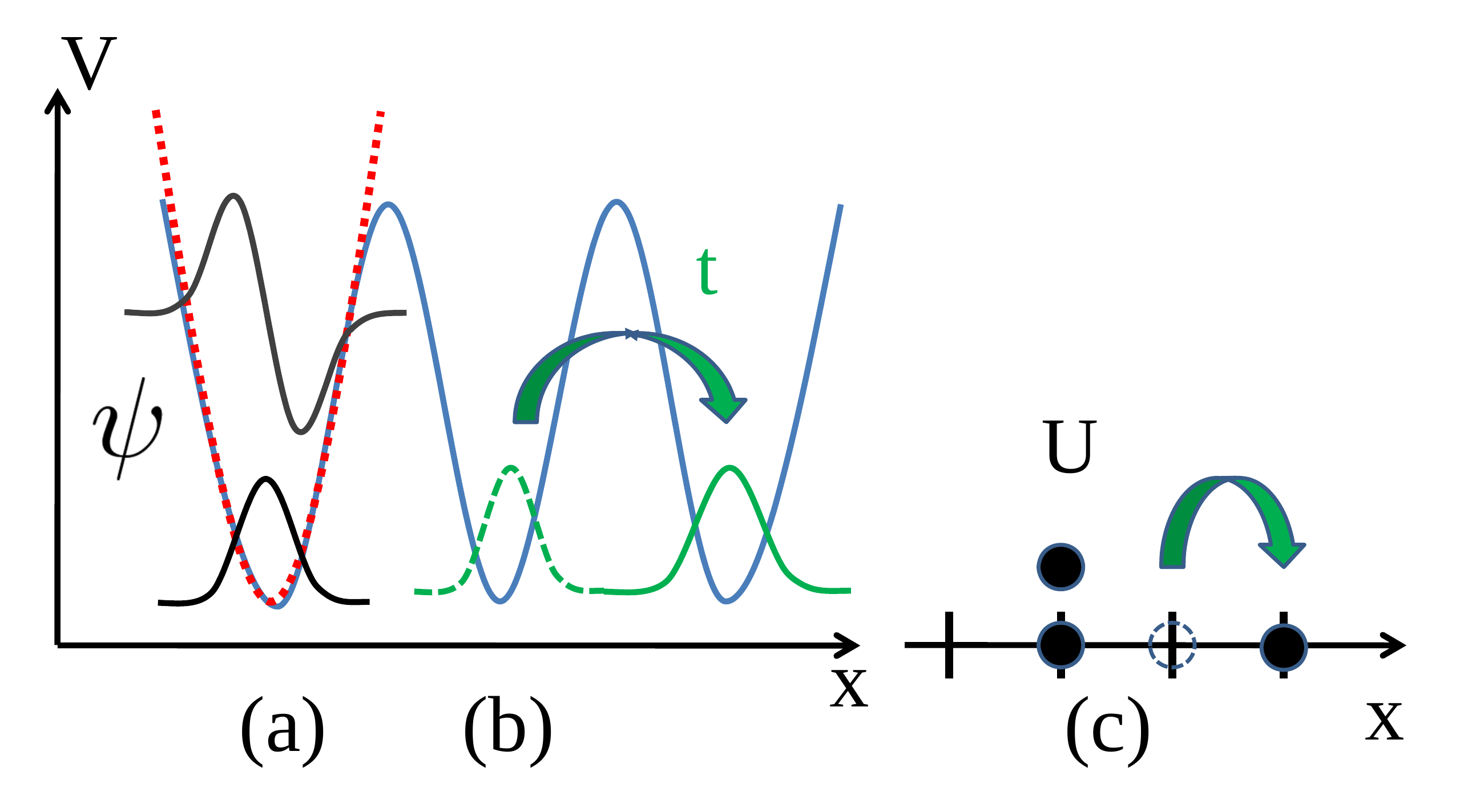}
\end{center}
\caption{\label{fig:atomic} a) If the periodic potential is extremely high compared to the kinetic energy, it
is a good approximation to consider that the particles are essentially localized around the minima
of the potential. In that case one has approximately the solutions of an harmonic oscillator. b) Since the wavefunctions in different
wells have a small overlap there is a finite tunneling amplitude $t$ to go from one well to the next. c) One can thus describe such a system
by particles forced to be on a lattice, with a certain hopping amplitude $t$ which will delocalize them. In addition if two particles are on the same
site they will feel the repulsion and pay an energy $U$.}
\end{figure}

Let us first analyze the effects of the periodic potential by considering a limit in which the
periodic potential would be extremely large, and in particular much larger than the kinetic energy of the particles. In that case, as shown in \fref{fig:atomic} it is a good approximation to consider that the particles remain mostly localized around one of the minima of the potential. Because in the condensed matter context this means that they stay essentially localized around each atom, this limit is called ``atomic limit'', a somewhat confusing term in the cold atom context.
Let us examine the case of the optical lattice potential (\ref{eq:optical_potential}). If the particles stay
around the minima we can expand the periodic potential. The Hamiltonian to solve becomes then (for one minimum)
\begin{equation}
 H = \frac{P^2}{2m} + V_0 (k x)^2
\end{equation}
and is thus the Hamiltonian of an harmonic oscillator. As shown in \fref{fig:atomic} around each minima, there is thus a full set of eigenstates
\begin{equation} \label{eq:eigen_harm}
 \psi_{j,n}(r) = \psi_n(r-R_j)
\end{equation}
which is centered around the $j$th site $R_j = a j$, and is the $n$th excited state with the energy
\begin{equation}
 E_n = \omega_0 (n + \frac12)
\end{equation}
with
\begin{equation} \label{eq:harmfreq}
 \omega_0^2 = \frac{2 V_0 k^2}{m}
\end{equation}
Note that this frequency is associated with each well of the optical lattice in the limit of a deep lattice, and should not be
confused with the frequency associated with the shallow parabolic trap usually present in those systems.
If the barriers are extremely high then the states centers around different sites $j$ are essentially orthogonal and can thus serve as a complete basis of all the states of the system. A convenient way to represent the system, is to use the second quantization representation \cite{mahan2000}. We introduce creation (and destruction) operators $b^\dagger_{j,n}$ which will create (destroy) a particle in the state (\ref{eq:eigen_harm}). Note that this does not mean that the particle is created at the position $R_j$ but with the wavefunction (\ref{eq:eigen_harm}).
The Hamiltonian of the system is then
\begin{equation} \label{eq:ham_atom}
 H = \sum_{j,n} \hbar \omega_0 [\frac12 + n] b^\dagger_{j,n} b_{j,n}
\end{equation}

Although extremely primitive, this limit allows us already to deduce a certain number of parameters. For example one can have an estimate of the interactions among the particles. Optical lattices play, in that respect a central role. To understand that point, let us assume that the microscopic interaction between the atoms can be described by the standard contact interaction \cite{pitaevskii_becbook}:
\begin{equation} \label{eq:intercont}
 U(r) = \frac{4 \pi \hbar^2 a_s}{m} \delta(r) = g \delta(r)
\end{equation}
Starting from such interaction one can define in the continuum a dimensionless ratio which is the typical
kinetic energy relative to the interaction energy. This ratio reads
\begin{equation}
 \gamma = \frac{E_{\rm int}}{E_{\rm kin}} = \frac{g n m}{\hbar^2 n^{2/3}} \simeq 4 \pi n^{1/3} a_s
\end{equation}
in three dimensions, using that the density of particles $n^{-1/3} = a$ the mean interparticle distance.
Typical numbers for the parameter $\gamma$ would be $\gamma= 0.02$. In other words, the interaction is
normally quite weak. In order to see strong interaction effect one thus needs to reinforce it. This can be reached by either increasing the interaction itself, for example by using a Feshbach resonance \cite{bloch_cold_atoms_optical_lattices_review}, or confinement \cite{olshanii_cir}, or by acting on the kinetic energy via an optical lattice as we discuss below.

In the optical lattice, as we saw, wavefunctions on different sites have essentially zero overlap, which means
that the interaction between particles located on different sites is essentially zero. Indeed if we take for example the ground state wavefunction of the harmonic oscillator:
\begin{equation} \label{eq:groundharm}
 \psi_0(x) = \left(\frac{m \omega_0}{\hbar \pi}\right)^{1/4} e^{-\frac{m\omega_0}{2\hbar}x^2}
\end{equation}
The above is for one dimension. In three dimensions the wave function is the product of the above wavefunction for each of the coordinates.
Given the fact that $\omega_0 \sim \sqrt{V_0}$ (see (\ref{eq:harmfreq})) the spatial extension of the wavefunction decreases as $1/V_0^{1/4}$ and can thus be much smaller than the ``intersite'' distance $a$ (see \fref{fig:atomic}) for large
potentials $V_0$. It means that, if we use the second quantization representation of (\ref{eq:ham_atom}), interaction can only involve operators on a given site.

If two particles are present on one given site, one can estimate the energy cost coming from the interactions. Let us assume that both particles are present in the lowest energy state of the harmonic oscillator (see \fref{fig:atomic}). Then the energy cost is
\begin{equation}
 U = \frac12 \int dr_1 dr_2 U(r_1-r_2) |\psi_0(r_1)|^2 |\psi_0(r_2)|^2
\end{equation}
Using the expression for the interaction (\ref{eq:intercont}) and the wavefunction (\ref{eq:groundharm}) extended to the three dimensional case, one obtains
\begin{equation}
 U = \frac{g}{2\sqrt2} \left(\sqrt{m \omega_0}{\hbar \pi}\right)^{3/2}
\end{equation}
Using the second quantization representation, and general expressions for the two-body operators, this leads to an interaction term in the Hamiltonian
of the form
\begin{equation}
 H_{\rm int} = \frac{U}2 \sum_{j} \hn_{j,0} (\hn_{j,0} - 1)
\end{equation}
where
\begin{equation}
 \hn_{j,0} = b^\dagger_{j,0} b_{j,0}
\end{equation}
is the operator counting the number of particles in the state $0$ on site $j$. One thus sees that the higher the barriers the larger is the energy cost of having two particles and more on the same site. This is because the wavefunctions are tighter and tighter confined and thus feel a local repulsion more strongly. Of course this expression, involving only one orbital is only valid if the population of the higher levels is zero. This implies in particular that one should be in a limit where the temperature is small compared to the interlevel separation $T \ll \omega_0$ but also that the interaction parameter $U$ is smaller than the interlevel separation $ U \ll \omega_0$. Otherwise it is
more favorable to promote one of the particle to a higher orbital state, which might reduce in part the overlap of the wavefunctions.
This is energetically more favorable than paying the full repulsion price. If these conditions are not met one needs to involve several orbitals to build the model.

\subsection{Tight binding approximation} \label{sec:tight-binding}

The approximation of the previous chapter essentially remove the kinetic energy of the particles, that remains localized around one site. This is clearly an oversimplification. Given the fact that there is some level of overlap of the wavefunctions on different sites there is a finite probability of tunneling between two sites. We can thus build a theory including this tunneling starting from the basis of wavefunctions localized around each site, defined in the previous section. This method is known as a the tight-binding approximation \cite{ashcroft_mermin_book,ziman_solid_book}. It is specially transparent and contains all the main features of exact solutions in periodic potential that we will detail in the next section. We will thus examine it in details.

Let us for simplicity restrict ourselves to the lowest orbital $\ket{0}$ on each site. Generalizing to several orbital per site poses no problem. Let us take a system with $N$ sites. We can write the full wavefunction of the problem as a linear combination of all the wavefunction on each site since we consider that they are essentially orthogonal
\begin{equation}
 \psi(r) = \frac1{\sqrt{N}} \sum_{j=0}^{N-1} \alpha_j \psi_0(r - R_j)
\end{equation}
where the $\alpha_j$ are coefficients to determine.
Since we want the problem to be invariant by a translation of $a$, the wavefunction can only be multiplied by a phase if we translate $x$ by $a$
\begin{equation} \label{eq:bloch}
\psi(x+a) = e^{i k a} \psi(x)
\end{equation}
which defines the parameter $k$. This parameter, which of course depends on the wavefunction $\psi$ is known as the pseudomomentum of the system. Note that this constraint is in fact an exact statement, known as the Bloch theorem. In order to satisfy the constraint (\ref{eq:bloch}) it is easy to see that we can take
\begin{equation}
 \psi_k(r) = \frac1{\sqrt{N}} \sum_{j=0}^{N-1} e^{i k R_j} \psi_0(r - R_j)
\end{equation}
which ensures also the proper normalization of the wavefunction. In order to have \emph{independent} wavefunctions we should not take values of $k$ leading to the same coefficients. Since $R_j = a j$ values of $k$ differing by $2\pi/a$ would lead to the same coefficients. We must thus restrict the values of $k$ to an interval of size $2 \pi/a$, called the first Brillouin zone. Typically one takes $k \in [-\pi/a,\pi/a]$. All physical quantities are thus periodic over this interval. In addition not all values of $k$ are allowed. Because of the system is of size $N$, $k$ must be quantized. The precise quantization depends on the boundary conditions. For example for periodic boundary conditions $\psi(x + L a) = \psi(x)$ imposes that $k$ is a multiple of an integer:
\begin{equation}
 k = \frac{2 \pi p}{N} \quad,\quad p \in \mathbb{Z}
\end{equation}
There are thus in the first Brillouin zone exactly $N$ values of $k$ and thus $N$ independent functions $\psi_k(x)$.

\subsection{More general relations}

Many of the relations or properties that we have obtained within the tight binding approximation
are in fact general and exact. Let us briefly review them here.

The first one is the Bloch theorem which states that in a periodic potential there exist a quantum number $k$ labeling the eigenfunctions such that
\begin{equation} \label{eq:bloch_full}
 \psi_k(r) = e^{i k r} u_k(r)
\end{equation}
where $u_k(r)$ is a periodic function
\begin{equation}
 u_k(r+a) = u_k(r)
\end{equation}
The constraints on the pseudomentum $k$ that we have established in the previous section hold.

In the same way the tight binding wavefunction has the right structure. One can represent the eigenfunction
under a form known as Wannier functions \cite{ashcroft_mermin_book,ziman_solid_book}:
\begin{equation}
 \psi_k(r) = \frac1{\sqrt{N}} \sum_j \phi(r - R_j)
\end{equation}
The Wannier function is given by
\begin{equation}
 \phi(r - R_j) = \phi_{R_j}(r) = \frac1{\sqrt{N}} \sum_k e^{-i k R_j} \psi_k(r)
\end{equation}
Two Wannier functions centered on two different sites are \emph{exactly} orthogonal
\begin{equation}
 \braket{\phi_{R_i}}{\phi_{R_j}} = \delta_{i,j}
\end{equation}
and the wavefunction $\phi_{R_j}(r)$ is essentially localized around the site $R_j$. We see that in the limit of high barriers, the local functions around one of the minimum of the potential provide an approximate form for the Wannier function.

Let us for example look at a Wannier function that would correspond to (\ref{eq:bloch_full}) with a $u_k(r)$ independent of $k$. In that case the Wannier function would be (in one dimension)
\begin{equation}
 \phi_{R_j}(x) = u(r) \frac{\sqrt{N}}{\pi} \frac{\sin(\pi (x-R_j)/a)}{x-R_j}
\end{equation}
showing the localization around the site $R_j$.

\subsection{Hubbard and related models}

Optical lattices thus provide a natural realization for certain models of interacting quantum systems with local interactions. In condensed matter these models are approximation of the realistic situations. Indeed in a solid the basic interaction is normally the Coulomb interaction between the electrons. However in a metal this interaction is screened, with a quite short screening length, of the order of the lattice spacing in a good metal \cite{ashcroft_mermin_book,ziman_solid_book}. It is thus tempting to replace the interaction with a local one. It is however in principle a caricature of the reality since the screening length can vary, hence the need to take into account interactions with a range longer than a single site etc. When comparing a certain solution of these models with reality, it is thus difficult to known if discrepancies are due to the approximations made in the solution or in the approximations made in the model. Optical lattices at least provide a reasonably clean realization of such models that can be compared directly with theoretical predictions. Let us examine some of these models

\subsubsection{Bosonic Hubbard model:}

We already obtained this model in \sref{sec:tight-binding}. It is
\begin{equation} \label{eq:bose-hub}
 H = - t \sum_{\langle i,j\rangle} (b^\dagger_i b_j + \texthc) + \frac{U}2 \sum_j \hn_j (\hn_j - 1)
\end{equation}
where $\langle\rangle$ denotes nearest neighbors. $t$ is the hopping amplitude from one site to the next, and $U$ the energy cost of putting two particles on the same site. This model describes quantum particles (typically bosons) hopping on a lattice and paying the interaction price $U$. This is essentially the simplest model that contains all the important elements of the competition between kinetic energy and interactions in a solid: i) the kinetic energy; ii) the notion of filling of a band (which would not be present in a continuum); iii) the interaction. This model known as the Hubbard model was introduced in 1963 \cite{hubbard63_model} for fermions (see below). The model (\ref{eq:bose-hub}) which applies to bosons is sometimes referred to as the Bose-Hubbard model to distinguish it from its venerable ancestor.
One can of course add several perturbations to the above model. The most common ones are the confining potential or any local potential, such as disorder. This would lead to
\begin{equation}
 H_\mu = \sum_j \mu_j \hn_j
\end{equation}
For the confining potential the chemical potential term is of the form $\mu_j \propto j^2$ and takes of course
any suitable form depending on the perturbation.
Optical lattices allow an easy control of the hopping amplitude $t$ while Feshbach resonance changes $U$ \cite{bloch_cold_atoms_optical_lattices_review}.
These two methods allows for a large variation of the ratio $U/t$ which controls the strength of the interaction effects.

As mentioned already (\ref{eq:bose-hub}) is a faithful description of the system in the optical lattice provided the
temperature $T$ and interaction $U$ are smaller than the distance between the lowest orbital and the first excited one, an energy of order $\hbar \omega_0$. Otherwise one must generalize the above model to a multiorbital one.
Note that if the optical lattice is not deep enough, or if the scattering length is too large, additional terms will
appear in the hamiltonian and the simple one-band Hubbard model is no longer valid: for a discussion,
see e.g. \cite{werner_cooling_2005}.

\subsubsection{$t-V$ model:}
For spinless fermions (\ref{eq:bose-hub}) would not contain any interaction since the Pauli principle forbids double occupancy of a given site. For spinless fermions one can thus consider an interaction of the form:
\begin{equation} \label{eq:int_tV}
 H_V = V \sum_{\langle i,j\rangle} \hn_j \hn_j
\end{equation}
In condensed matter this is merely taking into account the long range nature of the interactions. In cold atoms it is rather difficult to realize but could be relevant for systems with longer range interactions such as dipolar molecules. The model with kinetic energy on the lattice and the interaction (\ref{eq:int_tV}) is known as $t-V$ model and is also related to models for spins as we will see below.

\subsubsection{Hubbard model:}
Since electrons in solids have a spin $1/2$, i.e. an internal degree of freedom, it is important to consider the generalization of this class of models to the case of two species of particles. This is the canonical Hubbard model for fermions \cite{hubbard63_model}. Hopping conserves the internal degree of freedom (that we will call ``spin'' for simplicity), while a local interaction can only exist
between two opposite spins since the Pauli principle prevents two fermions of the same spin to be on the same site. The model is thus
\begin{equation} \label{eq:hubdef}
 H = - t \sum_{\langle ij \rangle, \sigma = \up\down} (c^\dagger_{i\sigma} c_{j\sigma} + \texthc) +
 U \sum_i \hn_{i,\up} \hn_{i,\down}
\end{equation}
where $\up,\down$ denote the two eigenstates of opposite spin (for example the two eigenstate of the spin along the $z$ direction). In the cold atom context the ``spin'' degree of freedom can denote any two possible internal state. This model contains the essential ingredients of the physics of strongly correlated quantum systems. Even if it is extremely simple to write it is extremely challenging to solve.

\subsubsection{Generalizations:}
Of course this Hamiltonian can be complicated in several ways, by putting for example state dependent hopping amplitudes $t_\up$ $t_\down$
\cite{cazalilla_coupled_fermions}, by adding longer range interactions to the system or by considering a larger number of internal degrees of freedom.
All these models can be (or have been already) potentially realized in cold atomic systems.

In addition to the fermionic Hubbard model, cold atomic systems have also allowed to realize bosonic systems with internal degrees of freedom. This has lead to several interesting models, in particular the one of a two component Bose-Hubbard model. Contrarily to the case of fermions for which the Pauli principle prevent the occupation of a site by two particles of the same species Bosons can have such terms. The interaction term for the two component Bose-Hubbard model thus involves three different interactions
\begin{equation}
 H = \frac{U_{\up\up}}2 \sum_i \hn_{i\up}(\hn_{i\up}-1) + \frac{U_{\down\down}}2 \sum_i \hn_{i\down}(\hn_{i\down}-1) + U_{\up\down} \sum_i \hn_{i\up}\hn_{i\down}
\end{equation}
As we will see in the next section, the combination of these three interactions can lead to a wide range of physical behaviors.

With respect to these models, cold atoms, given the local nature of the interactions and the degree of control on the lattice, interactions and the nature of the particles are a fantastic laboratory to realize and test these models. There are however several limitations or points to keep in mind. We have already mentioned some of them. Let us summarize them here:
\begin{enumerate}

\item If one wants to be able to use a single band model the separation between level in one of the optical lattice well must be larger than the interaction.
This is not a major problem when the lattice is deep, and when the interaction is reasonably small, but it can become a serious limitation if the interaction is
increased by a Feshbach resonance.

\item If one want to use the optical lattice to reduce the kinetic energy in order to change the ratio of the kinetic energy/interactions then one has to worry about
the temperature. Indeed if the kinetic energy becomes small compared to the temperature one has a essentially a classical system.

\item Finally the confining potential which corresponds to a locally varying chemical potential is both an advantage and a serious limitation. Indeed as we will discuss the physics of such systems depends strongly on the filling. So controlling the chemical potential and/or the number of particles per site is of course crucial.
    Having a confining potential has the advantage that in the system there are many different values of the chemical potential and thus one does not need (it would be in practice extremely difficult) to control exactly the number of particles compared to the number of sites. On the other hand the system is inhomogeneous which mean
    that most measurements will give an average response over many different phases, obscuring deeply the physics. Clearly this question is related to the ability or not to probe the system locally.
\end{enumerate}

\subsection{Superexchange}

The models of the previous sections describe the behavior of itinerant quantum particles on a lattice. Particularly interesting behavior occurs when these particles
can have internal degrees of freedom such as in the Hubbard model. In that case it is possible as we will discuss in the following sections that due to the interactions
the charge of the particles gets localized for special filling of the lattice, for example one particle per site (Mott transition). In such a case as we will discuss later
the repulsion between the charges ($U$ in the Hubbard model (\ref{eq:hubdef}) can lead to an insulating phase in the case of one particle per site (Mott transition).
For the Bose-Hubbard model with one component such a ground state would be featureless. But for systems with two (or more) components, both fermionic and bosonic,
the ground state is a priori quite complex since on each site one has to choose the state of the internal degree of freedom (which we will call spin in all this section).

\begin{figure}
\begin{center}
 \includegraphics[width=\columnwidth]{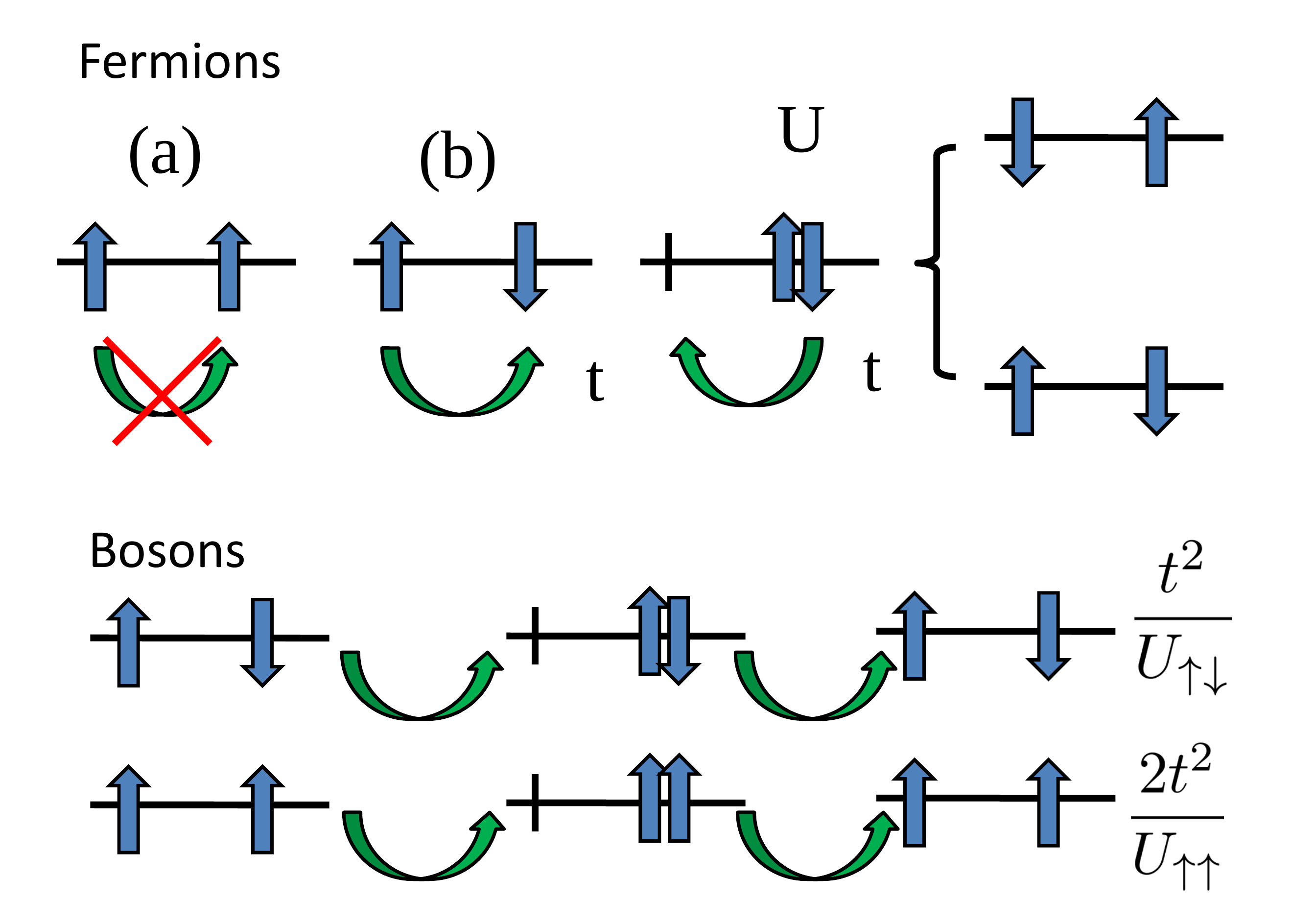}
\end{center}
\caption{\label{fig:superex} For a large repulsion $U$ and one particle per site, charge excitations cost an energy of order $U$, but
virtual processes allow to gain some kinetic energy.
a) For fermions the Pauli principle completely blocks hopping if the spins are parallel. b) For opposite spins virtual hopping is possible.
This leads to a superexchange that is dominantly antiferromagnetic (see text).
For bosons both processes are possible and depend on the intra- and inter-species interactions.
Bosonic factors favor parallel spins. Thus if all interactions
are equal for bosons the superexchange is dominantly ferromagnetic (see text). Changing the interactions between the two type of species allow to go from the ferromagnetic exchange to the antiferromagnetic one.}
\end{figure}

As shown in \fref{fig:superex}
if the repulsion is very large charge excitations which would put two particle per site would cost an energy of order $U$ and are thus essentially forbidden. On the other hand since $U$ is not infinite there could be virtual processes that allow the system to benefit from the kinetic energy, while leaving the system in a sector with exactly one
particle per site. These processes are the so-called superexchange processes. Here we will not give the full derivation of the superexchange term, this can be found in quite details in \cite{giamarchi_singapore_lectures} for example. We simply give here a qualitative argument.

Since the charge is essentially frozen one can stay in the Hilbert space in which each site has exactly one particle per site and only the spin degree of freedom remain. It means that on each site we have need to states to fully describe the Hilbert space. We can thus reduce the complete Hamiltonian (\ref{eq:hubdef}) to an \emph{effective} hamiltonian acting only on the spin degrees of freedom. For fermions it is easy to see that if one has two parallel spins on neighboring sites, no kinetic energy process can take place. On the other hand if the spins are antiparallel second order perturbation theory (see \fref{fig:superex}) can lead back to the initial state or lead to a state in which the two spins have been exchanged. The matrix element involved if of
order $ J = t^2/U$ since each hopping has an amplitude $t$ and the intermediate state is of energy $U$. The first process can be described by the effective Hamiltonian (written only for two spins)
\begin{equation} \label{eq:supsz}
 H_1 = J S_1^z S_2^z - \frac{J}4
\end{equation}
where we have introduced the spin operators $S^\alpha = \frac12 \sigma^\alpha$ where the $\sigma^\alpha$ are the three Pauli matrices. As usual we introduce the two eigenstate of $S^z$ and the hermitian conjugate operators $S^+ = S^x + i S^y$ and $S^- = S^x - i S^y$. These operators verify
\begin{equation}
\begin{split}
 S^z \ket{\up,\down} &= \pm \frac12 \ket{\up,\down} \\
 S^+ \ket{\down} &= \ket{\up} \quad,\quad S^+ \ket{\up} = \ket{\down}
\end{split}
\end{equation}
The equation (\ref{eq:supsz}) shows that the energy of two antiparallel spins is lowered by an energy $-J/2$ while the one of two parallel spins remains zero. The second process leads to an exchange of the two spins and can be written as
\begin{equation}
 H_2 = \frac{J}2 [S^+_1 S^-_2 + S^-_1 S^+_2]
\end{equation}
Putting the two processes together and taking proper care of the numerical factors one obtains for the full effective Hamiltonian (up to a constant energy term)
\begin{equation} \label{eq:heisdef}
 H = \frac{J}2 \sum_{\langle ij \rangle} [S^+_i S^-_j + S^-_i S^+_j] + J \sum_{\langle ij \rangle} S^z_i S^z_j
   = J \sum_{\langle ij \rangle} \vec{S}_i \cdot \vec{S}_j
\end{equation}
where $J \simeq 4 t^2/U$ for large values of $U$. This hamiltonian is known as the Heisenberg hamiltonian. We thus see that the combination of kinetic energy, Pauli principle and interaction leads to a remarkable exchange term between the spins which looks very much like the dipolar one that would exist for the direct magnetic exchange between magnetic moments. However there are also remarkable differences. This exchange, nicknamed superexchange, is responsible for many of the magnetic properties of the strongly interacting quantum systems in solids \cite{auerbach_book_spins}. Some noteworthy points are as follows
\begin{enumerate}
 \item Compared to an exchange between magnetic moments this superexchange is isotropic in the spins variables and will not couple the lattice
 direction with the spin directions. In that sense it is even simpler than a normal dipolar exchange. The spin rotation invariance of (\ref{eq:heisdef}) comes of course from the spin rotation invariance of the original Hubbard hamiltonian (\ref{eq:hubdef}).

 \item Quite importantly the order of magnitude of typical interactions are quite different. Direct magnetic exchange are quite ridiculous in solids.
 If one take spins on typical lattice spacing distance in a solid one obtains direct magnetic exchange of less than $1K$. On the contrary  since kinetic energy is typically $1eV$ and interactions of the order of $ \sim 10eV$ leads for solids to a $J$ of the order $J \sim 1000K$. Superexchange is thus in solids by far the dominant term and is at the root of the magnetic properties that we can observe in nature.
 In cold atoms the ``spin'' is of course merely an internal degree of freedom so the superexchange is the only term that can exist.

 \item For fermions because of the Pauli principle $J > 0$ which means that the Fermionic Hubbard model lead to antiferromagnetic phases.
 The situation is quite different for Bosons as indicated in \fref{fig:superex}. In that case, both species can hop, so the sign of the effective exchange $J$ will depend on the relative values of the intra- and inter-species interactions. If the intra-species $U_{\up\up}$ and $U_{\down\down}$ is the largest, than it is very much like a Pauli principle and one recovers an antiferromagnetic superexchange. On the contrary if the inter-species interaction $U_{\up\down}$ is the largest then one has a \emph{ferromagnetic} (i.e. a negative $J$) superexchange. In the case where all the interactions are equal then the bosonic factors still favor a ferromagnetic exchange \cite{DuanLukin2003}. Multicomponent bosonic systems will thus offer a particularly rich physics \cite{kleine_2velocities_bosons,zvonarev_ferro_cold}.
\end{enumerate}


\section{The Bose-Hubbard model and the superfluid to Mott insulator transition}
\label{sec:mottboson}
%

In this section, we make our first encounter with the Mott phenomenon: strong repulsive
interactions between particles can prevent the formation of an itinerant state and favour
a situation in which particles are localized.
This phenomenon is of key importance to the physics of strongly correlated materials.
Many remarkable physical properties
are found for those materials which are close
to a Mott insulating state.
For example, high-temperature superconductivity is found in copper oxides when a metallic
state is induced by introducing a relatively small amount of charge carriers into a Mott insulator.

\begin{figure}
\begin{center}
 \includegraphics[width=\columnwidth]{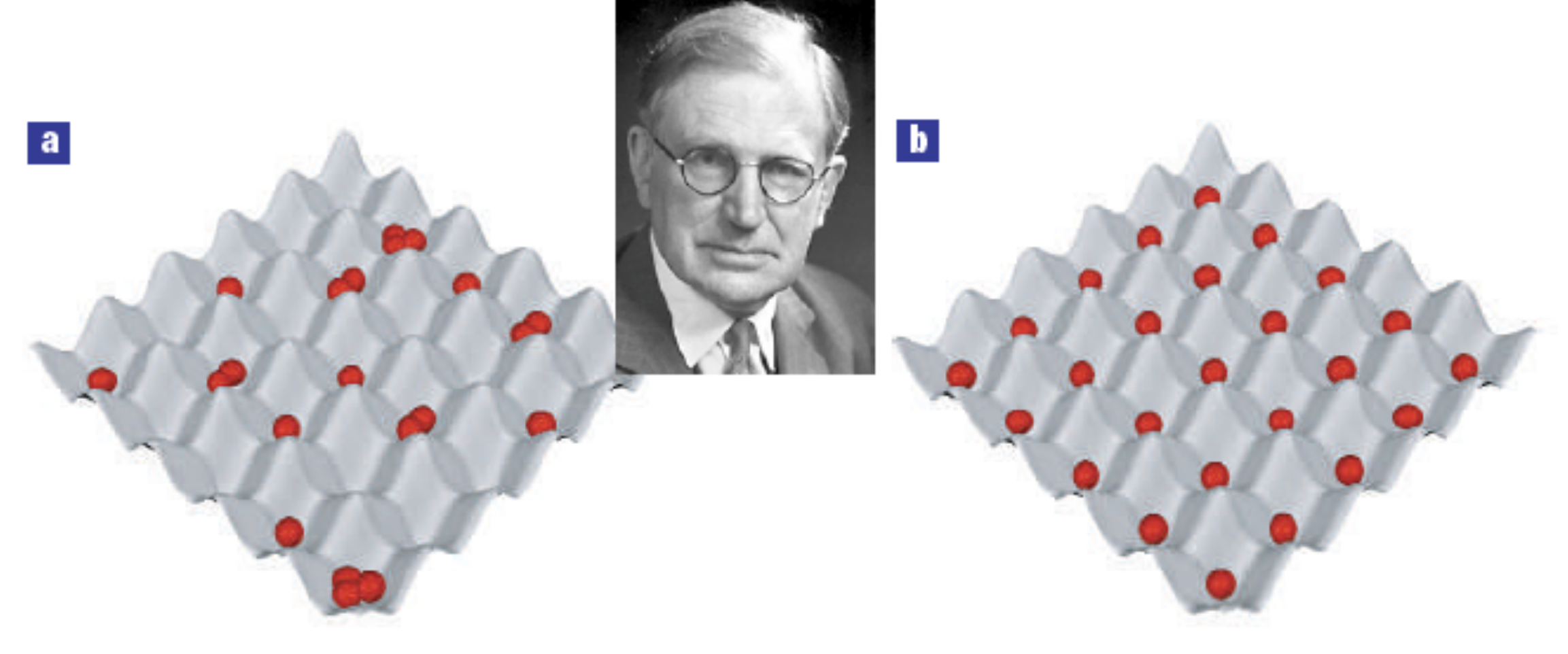}
\end{center}
\caption{\label{fig:mott_cartoon}(a) Typical real-space configuration of particles in an itinerant (metallic or superfluid) state.
(b) Typical real-space configuration in the Mott insulating state, in which double-occupancies are strongly suppressed. (Center:) Sir Nevil Mott. Adapted in part from \protect\cite{bloch_review_natphys_2005}.}
\end{figure}

In such circumstances, particles ``hesitate'' between itinerant and localized
behavior, making quantum coherence more difficult to establish and leading
to a number of possible instabilities.
From a theoretical viewpoint, one of the key difficulties is to describe
consistently an entity which is behaving simultaneously in a wave-like (delocalized) and
particle-like (localized) manner. Viewed from this perspective, strongly correlated quantum
systems raise fundamental questions in quantum physics.
%
Because the Mott phenomenon is so important,
the theoretical proposal~\cite{jaksch98_bose_hubbard}
and experimental observation~\cite{Greiner2002}
of the Mott transition in a gas of ultra-cold bosonic atoms in an
optical lattice have truly been pioneering works establishing a bridge between
modern issues in condensed matter physics and ultra-cold atomic systems.

In this section, we deal with this phenomenon in the simplest possible context: that of
the Hubbard model for bosonic atoms in an optical lattice. The case of fermions will be
considered later, in \sref{sec:mottfermion} and \sref{sec:mottfermion1D}.
The hamiltonian of this model reads (see also (\ref{eq:bose-hub})):
\begin{equation}\label{eq:ham_bose_hubbard}
H=-\sum_{ij} \t_{ij}\, b^\dagger_ib_j +
\frac{U}{2}\sum_i \hn_i(\hn_i-1) +
\sum_i v_{\mathrm{trap}}(i)\, \hn_i
-\mu \sum_i \hn_i
\end{equation}

\subsection{General considerations: lifting a macroscopic degeneracy}
\label{sec:general_bose}

Let us first consider a homogeneous system ($v_{\mathrm{trap}}=0$) in the limit where
there is no hopping $\t_{ij}=0$ (very deep lattice) as discussed in the \sref{sec:atomiclimit}.
The hamiltonian is then diagonal in occupation-number basis and has eigenstates
$|n\rangle$ with energies $E_n^0=\frac{U}{2} n(n-1)-\mu n$.
These energy levels cross at specific values of the chemical potential $\mu_n^0=nU$ at which
$E_{n}^0=E_{n+1}^0$. Hence, the nature of the ground-state depends crucially on the value of
the chemical potential:
\begin{itemize}
\item If $\mu\in \left] (n-1)U , nU\right[$, the ground-state is {\it non-degenerate}, with exactly $n$
bosons on each lattice site.
\item If $\mu=nU$, having $n$ or $n-1$ bosons on each lattice site is equally probable. Hence,
the ground-state has a {\it macroscopic degeneracy} $2^{N_s}$ (with $N_s$ the number of sites in the
lattice).
\end{itemize}
The number of particles per site in the ground-state as a function of chemical potential
has the form of a ``staircase'' made of plateaus of width $U$ in which $\langle \hn \rangle$
remains constant, separated by steps at $\mu_n^0=\mu n$ at which it jumps by one unit
(\fref{fig:phasediag_bose}).
In the context of mesoscopic solid-state devices, this is called the ``Coulomb staircase'': in order to
increase the charge by one unit, a Coulomb charging energy must be paid due to the electrostatic
repulsive interactions between electrons.

Within a given plateau $\mu\in \left] (n-1)U , nU\right[$, the first excited state (at constant total particle number)
consists in moving one boson from one site to another one, leaving a site
with occupancy $n-1$ and another one with occupancy $n+1$. The energy of this excitation is:
\begin{equation}
\Delta_g^0 = E^0_{n+1}+E^0_{n-1}-2E^0_n \,=\,U
\end{equation}
Hence, the ground-state is separated from the first excited state by a {\it finite energy gap}.
(In passing, we note that this gap can be written as
$\Delta_g^0 = (E^0_{n+1}-E^0_n)+(E^0_{n-1}-E^0_n)$ which in chemist's terminology
corresponds to ionization energy minus affinity). Adding or removing an electron also requires
a finite amount of energy: hence the system is {\it incompressible}. Indeed, each
plateau has a vanishing compressibility:
\begin{equation}
\kappa = \left(\frac{\partial^2 E}{\partial n^2}\right)^{-1}\,=\,\frac{\partial n}{\partial\mu}
\end{equation}

Having understood the zero-hopping limit, we can ask what happens when a small hopping amplitude
is turned on. Obviously, a non-degenerate incompressible ground-state separated by a gap from
all excitations is a quite protected state. Hence, we expect that the system will remain incompressible
and localized when turning on a small hopping, for $\mu$ well within a given charge plateau.
In contrast, the hopping amplitude is likely to be a singular perturbation when starting
from the macroscopically degenerate ground-state at each of the degeneracy points $\mu^0_n=nU$.
One natural way for the perturbation to lift the degeneracy is to select a unique ground-state which
is a superposition of the different degenerate configurations, with different number of particles
on each site. If the mixing between the different charge states corresponds to a state with small
phase fluctuations (the phase is the conjugate variables to the local charge), the resulting state
will be a superfluid. Hence, we expect that a superfluid state with Bose condensation will occur
already for infinitesimal hopping at the degeneracy points $\mu=nU$. These expectations are entirely confirmed
by the mean-field theory presented in the next section.
We note in passing that interesting phenomena often happen in condensed-matter physics when a
perturbation lifts a large degeneracy of the ground-state (the fractional quantum Hall effect is another
example).

\subsection{Mean-field theory of the bosonic Hubbard model}
\label{sec:meanfield_bose}

As usually the case in statistical mechanics, a mean-field theory can be constructed
by replacing the original hamiltonian on the lattice by an effective single-site
problem subject to a self-consistency condition. Here, this is naturally achieved by
factorizing the hopping term~\cite{Fisher1989,sheshadri_bosehubbard_epl_1993}:
$b^\dagger_ib_j \rightarrow \mathrm{const.} +
\langle b^\dagger_i\rangle b_j + b^\dagger_i \langle b_j\rangle + \cdots$ in which
``$\cdots$'' denote fluctuations which are neglected.
Another essentially equivalent formulation is based on the
Gutzwiller wavefunction~\cite{rokhsar_bosehubbard_prb_1991,krauth_bosehubbard_prb_1992}.
The effective 1-site hamiltonian for site $i$ reads:
\begin{equation}
h_{\rm{eff}}^{(i)}=
-\lambda_i b^\dagger -\lambda_i b +
\frac{U}{2} \hn(\hn-1)
-\mu \hn
\label{eq:singlesite_bose}
\end{equation}
In this expression, $\lambda_i$ is a ``Weiss field'' which is determined
self-consistently by the boson amplitude on the other sites of the lattice through the
condition:
\begin{equation}
\lambda_i = \sum_j \t_{ij}\, \langle b_j \rangle
\label{eq:scc_bose}
\end{equation}
For nearest-neighbour hopping on a uniform lattice of connectivity $z$, with all sites
being equivalent, this reads:
\begin{equation}
\lambda = z\,\t\,\langle b \rangle
\label{eq:scc_bose_uniform}
\end{equation}
These equations are easily solved numerically, by diagonalizing the effective
single-site hamiltonian (\ref{eq:singlesite_bose}), calculating $\langle b \rangle$ and
iterating the procedure such that (\ref{eq:scc_bose_uniform}) is satisfied.
The boson amplitude $\langle b \rangle$ is an order-parameter associated with
Bose condensation in the $\vk=0$ state: it is
non-zero in the superfluid phase.

For densities corresponding to an integer number $n$ of bosons per site on average, one finds
that $\langle b \rangle$ is non-zero only when $\t/U$ is larger than a critical
ratio $(\t/U)_c$ (which depends on the filling $n$). For $\t/U < (\t/U)_c$, $\langle b \rangle$
(and $\lambda$) vanishes, signalling a non-superfluid phase in which
the bosons are localized on the lattice sites: the Mott insulator.
For non-integer values of the density, the system is a superfluid for
all $\t/U>0$. This fully confirms the expectations deduced on a qualitative basis at the
end of the previous section.

\subsubsection{Perturbative analysis}

It is instructive to analyze these mean-field equations close to the critical
value of the coupling: because $\lambda$ is then small, it can be treated in (\ref{eq:singlesite_bose})
as a perturbation of the zero-hopping hamiltonian .
%
%
Considering a given plateau $\mu\in ](n-1)U,nU[$, the perturbed ground-state reads:
\begin{equation}
|\psi_0\rangle = |n\rangle -
\lambda\,\left[\frac{\sqrt{n}}{U(n-1)-\mu}|n-1\rangle+
\frac{\sqrt{n+1}}{\mu-Un}|n+1\rangle\right]
\end{equation}
so that:
\begin{equation}
\langle\psi_0|b|\psi_0\rangle =
- \lambda\,\left[\frac{n}{U(n-1)-\mu}+
\frac{n+1}{\mu-Un}\right]
\end{equation}
Inserting this in the self-consistency condition yields:
\begin{equation}
\lambda = - z\,\t\,\lambda\,\left[\frac{n}{U(n-1)-\mu}+
\frac{n+1}{\mu-Un}\right]+\cdots
\end{equation}
where ``...'' denotes higher order terms in $\lambda$.
This equation can be viewed as the linear term in the expansion of the equation of state
for $\lambda$.  As usual, the critical value of the
coupling corresponds to the vanishing of the coefficient of this linear term
(corresponding to the quadratic or mass term of the expansion of the Landau free-energy).
Hence the critical boundary for a fixed average (integer) density $n$ is given by:
\begin{equation}
\frac{z\t}{U}\,=\,\frac{(n-\mu/U)(\mu/U-n+1)}{1+\mu/U}
\label{eq:critical_bose}
\end{equation}

\subsubsection{Phase diagram.}
\begin{figure}
\begin{center}
\includegraphics[width=0.9\linewidth]{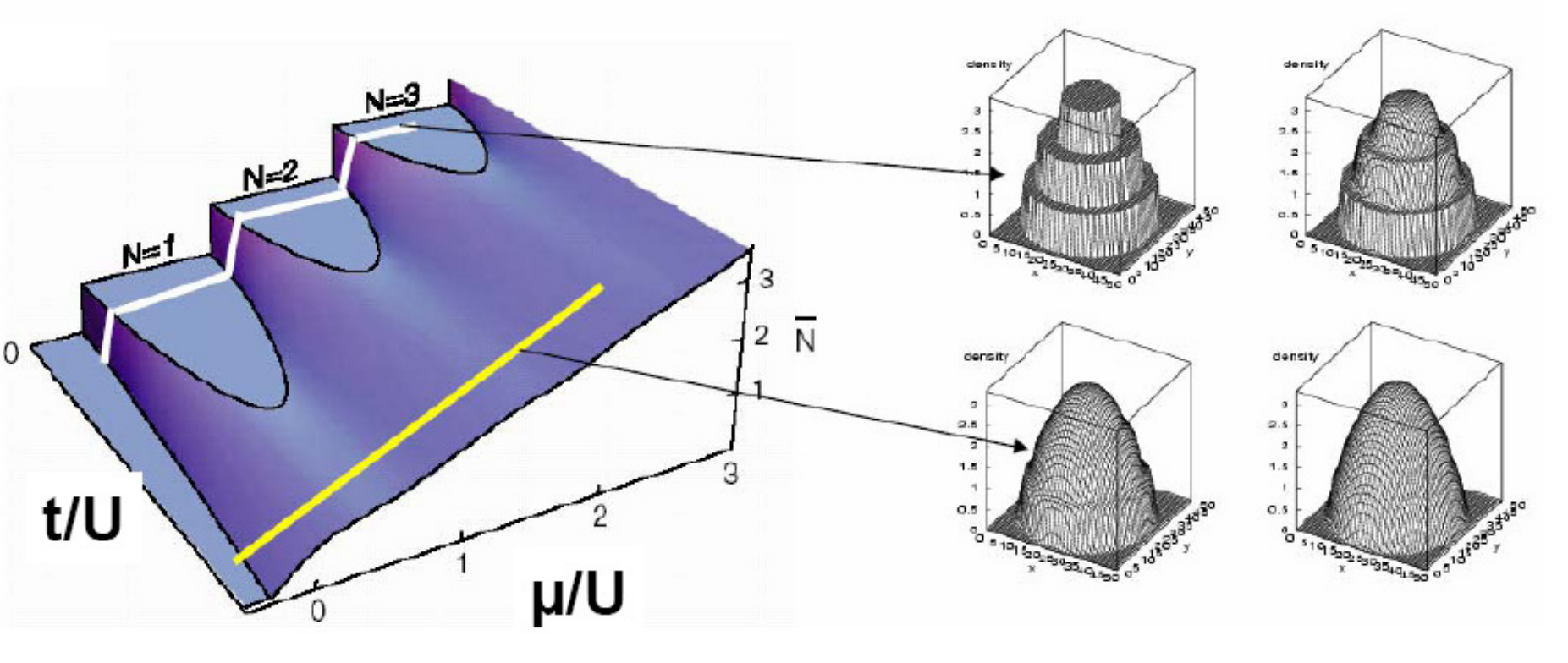}
\caption{Left: phase diagram of the Bose Hubbard model as a function of chemical potential
$\mu/U$ and coupling $\t/U$. An incompressible Mott insulator is found within each lobe of
integer density. Right: density profiles in a harmonic trap. The ``wedding cake'' structure
(see text) is
due to the incompressibility of the Mott insulator (numerical calculations courtesy of
H.Niemeyer and H.Monien, figure courtesy F.Gerbier).
}
\label{fig:phasediag_bose}
\end{center}
\end{figure}

This expression gives the location of the critical boundary as a function of the
chemical potential.
As expected, it vanishes at the degeneracy points $\mu^0_n=nU$ where the system becomes
a superfluid for infinitesimal hopping amplitude.
In the ($\t/U,\mu/U$) plane, the phase diagram
(Fig.~\ref{fig:phasediag_bose})
consists of lobes inside which the density is integer and the system is a Mott insulator.
Outside these lobes, the system is a superfluid. The tip of a given lobe corresponds to the
the maximum value of the hopping at which an insulating state can be found. For
$n$ atoms per site, this is given by:
\begin{equation}
\frac{z\t}{U}|_{c,n}= \mathrm{Max}_{x\in[n-1,n]}
\frac{(n-x)[x-n+1]}{1+x} = \frac{1}{2n+1+2\sqrt{n(n+1)}}
\end{equation}
So that the critical interaction strength is
$(U/z\t)_c\simeq 5.8$ for $n=1$, and increases as $n$ increases ($(U/z\t)_c\sim 4n$ for large $n$).

%
%
\subsubsection{Mott gap.}
The gap in the Mott insulating state is of course reduced by the hopping from its zero-hopping
value $\Delta_g^0=U$. We can obtain its mean-field value from the extension of the density plateau:
\beq
\Delta_g(n) = \mu_{+}(n)-\mu_{-}(n)
\eeq
where $\mu_{\pm}$ are the solutions of the quadratic equation corresponding to
(\ref{eq:critical_bose}), i.e:
\beq
(\mu/U)^2-[2n-1-(z\t/U)] (\mu/U) + n(n-1)+(z\t/U) = 0
\eeq
yielding:
\beq
\Delta_g(n)=U\,\left[(\frac{z\t}{U})^2-2(2n+1)\frac{z\t}{U}+1\right]^{1/2}
\eeq
The Mott gap is $\sim U$ at large $U/\t$ and vanishes at the critical coupling
($\propto\,(U-U_c)^{1/2}$ within mean-field theory).

\subsubsection{Incompressibility and ``wedding-cake'' shape of the density profile in the trap}

The existence of a gap means that the chemical potential can be changed within the
gap without changing the density. As a result, when the system is placed in a trap,
it displays density plateaus corresponding to the Mott state, leading to
a ``wedding cake'' structure of the density profile (Fig.~\ref{fig:phasediag_bose}).
This is easily
understood in the local density approximation, in which the local chemical potential
is given by:
$\mu(r)=\mu-v_{\mathrm{trap}}(r)=\mu-m\omega_0^2 r^2/2$, yielding a maximum extension of the
plateau: $\sim (2\Delta_g/m\omega_0^2)^{1/2}$. Several authors have studied
these density plateaus beyond the LDA by numerical simulation (see e.g
\cite{batrouni_domains_prl_2002}), and they have also been imaged
experimentally, see e.g. \cite{folling_shellstructure_prl_2006}.

\subsection{Mean-field theory: the wave-function viewpoint}
\label{sec:mft_wavefunction}

An alternative, but equivalent, viewpoint on the above mean-field theory is to formulate it as a
variational ansatz for the
ground-state wave-function~\cite{rokhsar_bosehubbard_prb_1991,krauth_bosehubbard_prb_1992}.

In the zero-hopping limit, the ground-state wave-function within a given density
plateau reads:
\begin{equation}
\Psi_0^{\t=0} = \prod_i |n\rangle_i = \prod_i \frac{1}{\sqrt{n!}}(b_i^\dagger)^n |0\rangle
\end{equation}
In the opposite limit of a non-interacting system ($U=0$), the ground-state wave-function is
obtained by placing all bosons in the $\vk=0$ state:
\begin{equation}
\Psi_0^{U=0} = \frac{1}{\sqrt{N!}}(b^\dagger_{\vk=0})^N |0\rangle
= \frac{1}{\sqrt{N!}}\left[\frac{1}{\sqrt{N_s}}\sum_i b_i^\dagger\right]^N|0\rangle
\end{equation}
In the limit of large $N,N_s$, the ground-state wavefunction for the non-interacting case can alternatively
be formulated (by letting $N$ fluctuate) as a product of coherent states on each site:
\begin{equation}
\Psi_0^{U=0} = \prod_i |\alpha\rangle_i\,\,\,,\,\,\,
|\alpha\rangle = e^{-|\alpha|^2/2}\sum_{n=0}^\infty \frac{\alpha^n}{\sqrt{n!}} |n\rangle
\end{equation}
with $|\alpha|^2 = \langle n \rangle = N/N_s$. In this limit, the local density obeys
Poisonnian statistics $p(n)=e^{-|\alpha|^2}|\alpha|^{2n}/n!
=e^{-\langle n \rangle}\langle n\rangle^{n}/n! $.

We note that in both limits, the ground-state wave-function is a product of individual
wave-functions over the different lattice sites. The individual wave-functions have
a very different nature however in each limit: they are number state for $\t=0$ while they
are a phase-coherent superposition of number states in the $U=0$ limit.

A natural variational ansatz is then to assume that the wave-functions remains an uncorrelated
product over sites for arbitrary $U/\t$, namely:
\begin{equation}
\Psi_0^{\mathrm{var}} = \prod_i \left[\sum_n c_n |n\rangle_i\right]
\end{equation}
The variational principle then leads to equation for the coefficients $c_n$ which are identical to the
mean-field equations above. The trial wave-function interpolates between the Poissonian statistics
$c_n=\alpha^{n}/\sqrt{n!}$ for $U=0$ and the zero-fluctuation limit $c_n=\delta_{n,n_0}$ as the
insulator is reached. The fact that $n$ has no fluctuations throughout the Mott phase
is of course an artefact of the mean-field.

%
%

The derivation of the above results rest heavily on the fact that one can build a mean-field theory, and in particular
that a well defined superfluid phase, with perfect order of the phase exists. It is thus interesting to see how the above physics and competition
between the superfluid and Mott insulating phase would be modified in situation where phase fluctuations are very strong and the mean-field
theory is invalid. This is clearly the case if the dimension of the system is getting smaller, since in low enough dimensions is it impossible
to break a continuous symmetry (the so call Mermin-Wagner theorem \cite{mermin_theorem}), and thus no true superfluid phase -- which would correspond to a breaking of the phase symmetry of the wavefunction -- can exist. Since cold atoms systems allow an excellent control on the dimensionality of the problem by changing the strength of the optical lattice, they allow in particular to tackle these questions in the one dimensional situation for which one can expect novel effects to occur. We will thus examine in \sref{sec:boso} the case of one dimensional quantum systems.

\subsection{Probing Mott insulators: shaking of the optical lattice} \label{sec:shaking}

In order to probe the above physics it is important to have good probes. The time of flight measurements, which give access to the single particle
correlations is of course one of them and we will examine several others in these notes. In this section we want to discuss a relatively simple probe,
but which gives extremely useful information for such systems and which consists in shaking of the optical lattices.

\begin{figure}
\begin{center}
  \includegraphics[width=0.8\linewidth]{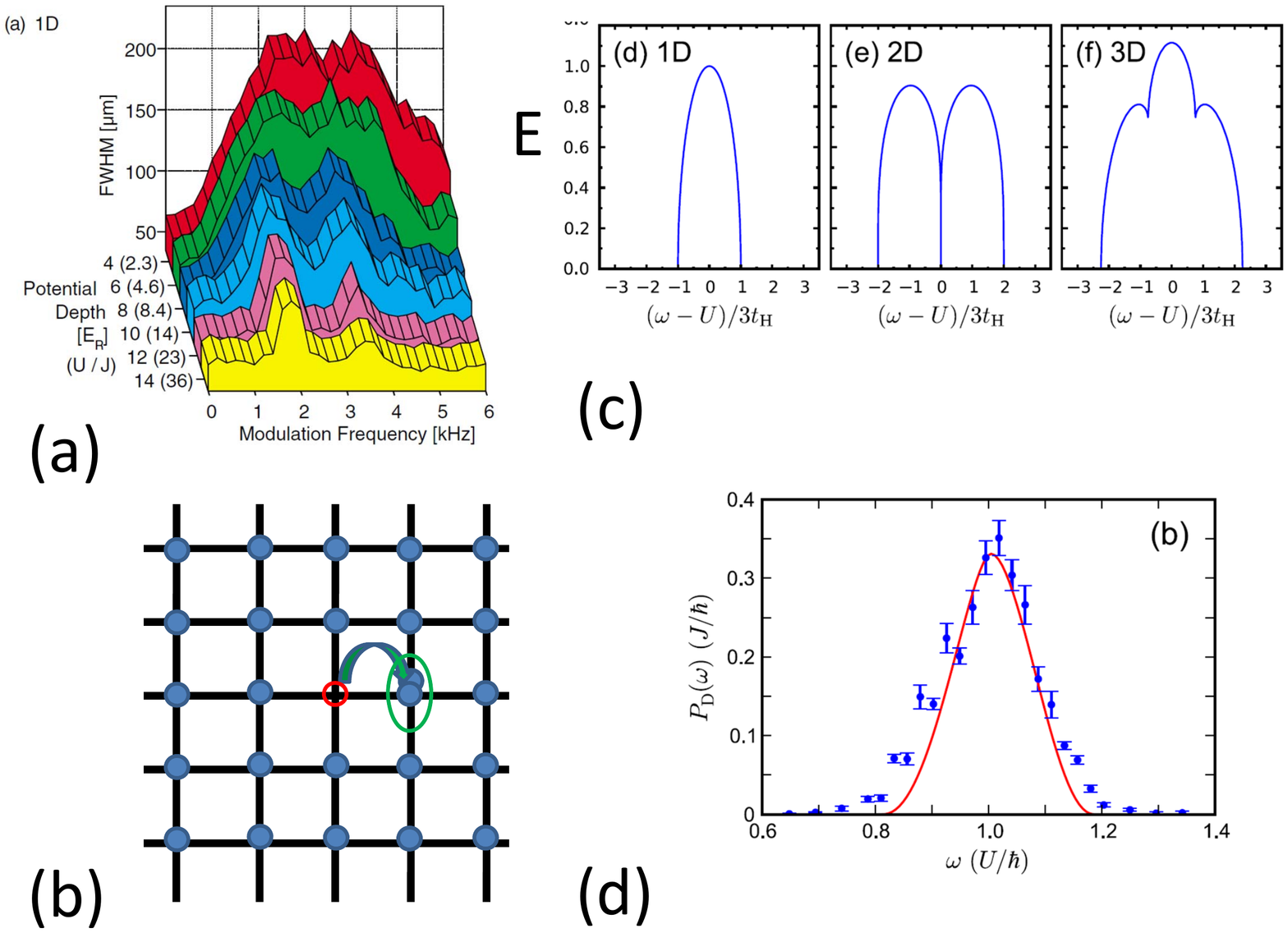}
\end{center}
 \caption{\label{fig:shaking}
 a) shaking of the optical lattice for a system of bosons. One sees marked differences depending on the depth of the optical lattice. In the Mott insulating phase a peak structure is observed. [After {\protect\cite{stoferle_tonks_optical}}]. b) Deep in the Mott phase the structure can be explained by considering the creation of a doublon (doubly occupied site) and a holon (empty site) due to the modulation of the kinetic energy by the shaking. c) structure of the peak depending on the dimension. This structure is located at an energy around the Mott gap, and the width is reflecting the kinetic energy of the doublon and holon. [After {\protect\cite{tokuno_shaking_phase}}]. d) for fermions similar results can be obtained by considering the creating of doubly occupied states which makes it a very sensitive probe. Fitting to a slave boson theory gives excellent agreement with the data and can give some access to the temperature of the system. [After {\protect\cite{tokuno_shaking_slave_fermions}}]}
\end{figure}

The idea is to modulate in a time dependent way the amplitude of the optical lattice \cite{stoferle_tonks_optical} for a given amount of time and then to measure the energy deposited in the system by such a process, as a function of the modulation frequency. This corresponds to adding a term in the Hamiltonian of the form
\begin{equation}
 H_L = \int dx [V_L + \delta V_L \cos(\omega_0 t)] \cos(Q x) \rho(x)
\end{equation}
The results for such an operation are shown in \fref{fig:shaking}.
One sees marked differences depending on the strength of the interactions. In particular in the Mott insulator one can recognize a peak structure.

Interpreting such data is of course not easy since one deals with a full time dependent Hamiltonian, making it difficult to deal with analytically and numerically. Analytically it is possible to use linear response to study the effects of the shaking \cite{iucci_absorption}. The results crucially
depend on whether the lattice is weak or strong. We will concentrate here on the case of the strong lattice and refer the reader to the literature for the other limit. In that case the main effect of modulating the lattice is to change, in the resulting effective Bose-Hubbard model (\ref{eq:bose-hub})
the hopping $t$ and the interaction $U$. Indeed as we saw in the \sref{sec:lattices} these terms are directly determined by the shape of the wavefunctions and thus by the depth of the lattice. One can even realize that the main effect will occur on the tunneling term \cite{reischl05_shaking_temperature_mott} which depends exponentially on the lattice depth. In the case of the strong lattice the main consequence is thus a modulation of the kinetic energy in the Hubbard model
\begin{equation}
 H_K = H_K^0 + \delta H_K(t) = [t_0 + \delta t \cos(\omega_0 t)] \sum_{\langle i,j\rangle} (b^\dagger_i b_j + \texthc)
\end{equation}
It is thus possible to study the effects of the shaking by considering the linear response in this term \cite{iucci_absorption,kollath_bosons_shaking_dmrg,huber_shaking_bosons,tokuno_shaking_phase}.
We will enter in the detail of the analysis but give again the main ideas.
In linear response the energy absorbed is directly related to the imaginary part of the Fourier transform of the equilibrium correlation function
\begin{equation}
 \chi(t) = - i \langle [\delta H_K(t),\delta H_K(0)] \rangle
\end{equation}
We thus see that the shaking of the lattice measures the kinetic energy-kinetic energy correlations. In other words we have to consider the processes that are shown in \fref{fig:shaking}. We transfer at time zero a particle from one site to the neighboring one, then this excitation propagates and at a later time we undo it by applying the kinetic energy operator again. Deep in the Mott phase we start with one particle per site. The application of the kinetic energy term thus creates a doubly occupied site and an empty site. The energy of this excitation is of the order of the Mott gap $\Delta_M \sim U$. We can thus expect that the system absorbs energy when the frequency of the modulation matches the Mott gap $\hbar \omega_0 = \Delta_M$. The shaking of the lattice thus allows to directly measure the Mott gap of the system. In addition the doublon and holon can propagate and thus have their own kinetic energy of the order of $t_0$. This will broaden the peak in a way that reflects this propagation. Such a propagation can be computed by properly taking into account the fact that the holon and doublon cannot be at the same site without recombining and give the remarkable peak structure of \fref{fig:shaking}. Not taking such a constraint into account leads to incorrect results. Such a structure reflects the van Hove singularities in the density of state. We refer the reader to the literature for more details and references on the subject.

A variant of the shaking of the optical lattice, namely a modulation of the phase of the lattice rather than its amplitude can be treated by similar methods. Quite remarkably modulating the phase leads to the current-current correlation function instead of the kinetic energy-kinetic energy one. It is thus giving a direct access to the frequency dependent conductivity of the system \cite{tokuno_shaking_phase}, something that allows to make a direct connection with comparable experiments done in the condensed matter context. It will be interesting to practically implement such a probe.

The shaking is thus an extremely useful probe for Mott insulating physics. One drawback for the bosons, is that measuring the energy absorbed is difficult. As a result, one needs to modulate with a relatively large intensity, which takes the system out of the linear response regime. In order to describe the absorption in this limit it is thus necessary to perform a numerical analysis of the system, something not trivial given the out of equilibrium nature of the problem. In one dimension DMRG studies using the possibility to tackle fully time dependent hamiltonian have been performed \cite{kollath_bosons_shaking_dmrg} and have allowed to elucidate the nature of the higher peaks in the experimental data shown in \fref{fig:shaking}. In order to circumvent the difficulty caused by the measure of the energy (and in particular as we will see in \sref{sec:fershaking} for the fermions this is extremely difficult) it has been suggested \cite{kollath_shake_fermions_DMRG} to measure instead the production of the doubly occupied states as a function of time. This allows for much more precise measurements. We will come back to this point in \sref{sec:fershaking}.


\section{One-dimensional bosons and bosonization}
\label{sec:boso}


Let us now turn to one dimensional systems, for which very special
effects arise. Indeed as we discussed already for the case of bosons,
and will see in \sref{sec:fermiliquid} for fermions, the effects of interactions are crucial.
For bosons, interactions lead both to the superfluid state and to the Mott insulating one.
As one can naively expect in one dimension the effects of interactions will be maximum since the
particles cannot avoid each other, while in three dimensions one can naively expect that particles will see each other much less.
In addition, as we already mentioned it is impossible to
break a continuous symmetry, so even at $T=0$ a true ordered superfluid ground state cannot exist.
On the other hand a bosonic system will still retain strong superfluid tendencies. One can thus expect
that quantum system in one dimension exhibit a radically different physics than for their higher
dimensional counterparts. Cold atomic systems have been remarkable in showing such a physics given the remarkable control
over dimension and interactions.

We will examine some of the aspects of this novel physics in this section. Of course there is much too much to be examined
in these few pages. This sections will thus simply be a general presentation, and will not pretend to be exhaustive.
The interested reader can find much more details in a whole book on the subject of one dimensional systems
\cite{giamarchi_book_1d} where a complete description of the various one dimensional systems and physical effects and methods is given.
In addition, for the specific case of bosons in cold atoms several lecture notes also contain complementary material
\cite{giamarchi_bosons_salerno,giamarchi_singapore_lectures}. Finally these notes will not make attempts in giving a comprehensive list of references
since an extensive review on the subject of one-dimensional bosons exists \cite{cazalilla_review_bosons_1D}.

\subsection{Peculiarities of one dimension}

Before we embark on the one dimensional world, let us briefly recall some of the points of the typical solution
for a bosonic system in higher dimension. As discussed in \sref{sec:mottboson} for a high (meaning $d \geq 1$) dimensional
system one can expect that there is a well defined superfluid order. As a result the wavefunction can be written as
\begin{equation} \label{eq:meansingle}
 \psi(x) = \sqrt{\rho(x)} e^{i \theta(x)}
\end{equation}
where $\rho(x)$ is the density of particles at point $x$ and $\theta(x)$ the phase of the wavefunction at the same point.
The present of superfluid order implies that we can use $\rho(x) \to \rho_0$ and $\theta(x)$ acquires a finite expectation value
$\theta(x) \to \theta_0$ so that the wavefunction has a coherent phase through the whole sample. Fluctuations above this ground state
can be described by the Bogoliubov theory \cite{pitaevskii_becbook}. We will not recall the theory here but just give the results.
The Bogoliubov spectrum is linear at small $k$ with a velocity $u$ of the excitations which represent the Goldstone mode of the superfluid.
The velocity $u$ depends on the interactions among the particles. This linear modes is the hallmark of the superfluidity in the system.
At larger $k$ the dispersion gives back the $k^2$ dispersion of free particles.
\begin{equation}
 E(k) = \sqrt{ u^2 k^2 + (k^2/(2 m))^2}
\end{equation}
An important point is that the mode is a well defined dispersive mode, which characterize excitations that have a well defined relation between
their momentum and energy.

Given the superfluid order the single particle correlation function tends to a constant
\begin{equation}
 g_1(r) = \lim_{r\to 0} \langle \psi(r) \psi^\dagger(0) \rangle \to {\rm Cste}
\end{equation}
and as a result the occupation factor $n(k)$ which is the fourier transform of the above correlation function
\begin{equation}
 n(k) = \int dr g_1(r)
\end{equation}
has a $\delta$-function divergence at $k=0$.

\begin{figure}
\begin{center}
  \includegraphics[width=0.8\linewidth]{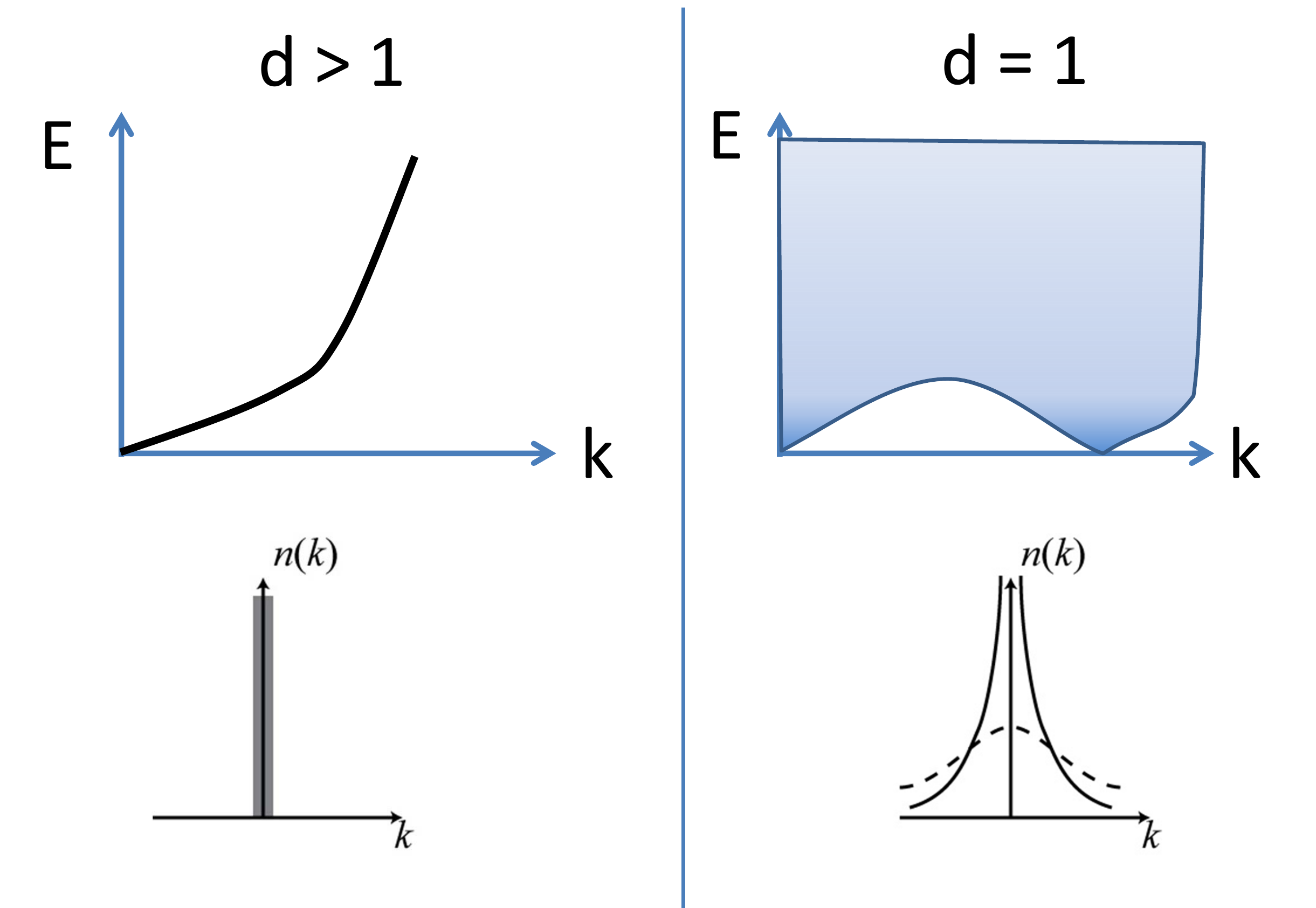}
\end{center}
 \caption{\label{fig:1dvs3d}
 (left) behavior in high dimension ($d \geq 1$). One expects an ordered superfluid
 state for which the phase of the wavefunction is well defined (see text). The excitation spectrum is made of Bogoliubov
 excitations with a linear dispersion at small $k$. The single particle correlation $g_1(k)$ (see text) has a divergent $\delta$-peak
 at $k=0$. (right) in $d=1$ no state will a fully ordered phase can exist and correlation functions are usually decaying as powerlaws at
 $T=0$ and exponentially at finite $T$. The spectrum has a continuum of excitations and low energy modes at
 $k = 2\pi \rho_0$ where $\rho_0$ is the average density. The single particle correlation has (at $T=0$)
 a powerlaw divergence that characterize the quasi-long range order of the superfluid. At finite temperatures this turns into an exponential decay and thus a lorentzian like behavior for $n(k)$}
\end{figure}

These results are summarized in the \fref{fig:1dvs3d}. We will contrast them with the results in one dimension in the subsequent sections.

\subsection{Realization of one dimensional systems}

\begin{figure}
\begin{center}
  \includegraphics[width=0.8\linewidth]{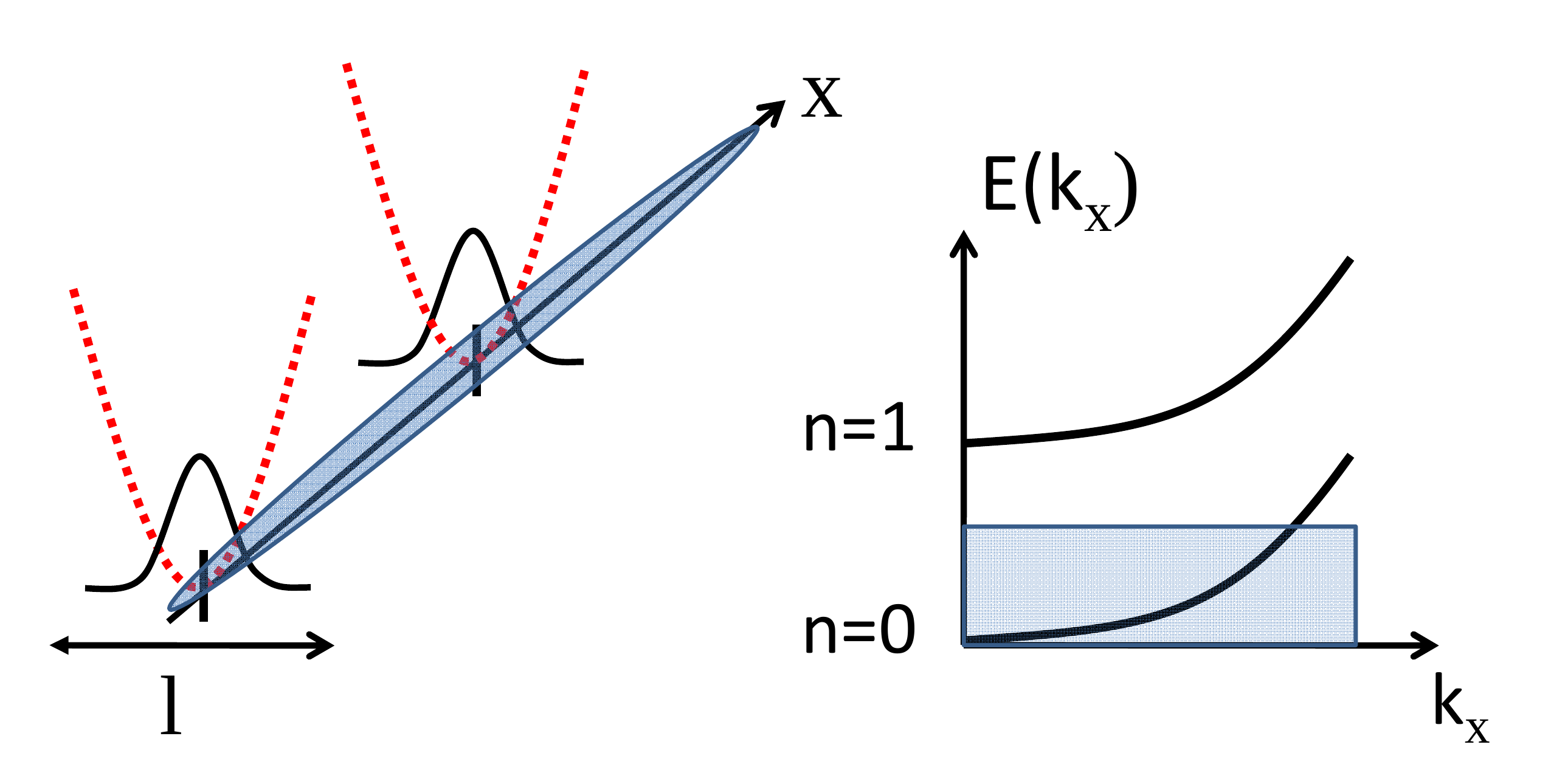}
\end{center}
 \caption{\label{fig:confwire} (left) Confinement of the electron gas in a one-dimensional tube of transverse size $l$.
 $x$ is the direction of the tube. Only one transverse direction of
 confinement has been shown for clarity. Due to the transverse confining
 potential the transverse degrees of freedom are strongly quantized.
 (right) Dispersion relation $E(k)$. Only half of the dispersion relation is shown for clarity.
 $k$ is the momentum parallel to the tube direction.
 The degrees of freedom transverse to the tube direction lead to the formation of minibands,
 labeled by a quantum number $n$. If only one miniband is populated,
 as represented by the gray box, the system is equivalent
 to a one dimensional system where only longitudinal degrees of
 freedom can vary.}
\end{figure}

The possibility to obtain ``one dimensional'' systems is deeply rooted in the quantum nature of the problem.
Indeed the objects themselves are much smaller than the possibility to confine them, so one could naively think
that it is always possible for them to avoid each other. The answer comes from the quantization of wavefunction.
In the presence of an optical lattice one has the wavefunction (\ref{eq:groundharm}) with one frequency
$\omega_0$ in the longitudinal direction and $\omega_\perp$ in the two other directions (see \fref{fig:confwire}).
If the confinement along the longitudinal direction is very weak, one can consider that the wavefunction is essentially
a plane wave in the longitudinal direction, leading to a wavefunction of the form
\begin{equation}
 \psi(x,r_\perp) = e^{i k x} \phi(r_\perp)
\end{equation}
where $\phi$ depends on the precise form of the confining
potential For an infinite well, as show in \fref{fig:confwire},
$\phi$ is $\phi(y) = \sin((2n_y+1)\pi y/l)$, whereas it would
be a gaussian function (\ref{eq:groundharm}) for an harmonic
confinement. The energy is of the form
\begin{equation}
 E = \frac{\hbar^2 k^2}{2 m} + \hbar \omega_\perp (n_\perp + \frac12)
\end{equation}
Due to the narrowness of the transverse channel $l$, the transverse quantization energy is sizeable
while the energy along the longitudinal direction is nearly continuous.
This leads to minibands as shown in \fref{fig:confwire}. If the
distance between the minibands is larger than the temperature
or interactions energy one is in a situation where only one
miniband can be excited. The transverse degrees of freedom are
thus frozen and only $k_x$ matters. The system is a
one-dimensional quantum system.

This is quite similar to the conditions established at the end of \sref{sec:lattices} for the use of a single band model.
In addition to the cold atom situation similar conditions have been met in condensed matter systems in a variety of problems
such as spin chains and ladders, organic superconductors, nanotubes, edge states in the quantum hall effect, quantum wires in semiconducting
structures, Josephson junction arrays, Helium in nanopores. For more details on these systems we refer the reader to \cite{giamarchi_book_1d}.
Quantum systems thus allow to realize situations where, although of course the physical system is three dimensional, all the important properties can be described purely from a one-dimensional description. Solving one dimensional problems is thus not jut a theorist game but has deep consequences for a large number of physical systems. Let us note that in addition to realizing purely one dimensional systems one can have by including a larger and larger number of minibands an intermediate between the one dimensional world the two dimensional one.

\subsection{1D techniques} \label{sec:1dtech}

Treating interacting particle in one dimension is a quite
difficult task, since we loose most of the techniques (mean field theory
for example) that we used to have to handle higher dimensional systems.
Fortunately there are some techniques that have proved very efficient, and which
combined together allowed to make significant progress in our understanding of such systems.
We will of course not detail these techniques and refer the reader to \cite{giamarchi_book_1d,cazalilla_review_bosons_1D}
for details and references. Here is a brief summary
\begin{enumerate}
\item {\bf Exact solutions:} Some models in one dimension are exactly solvable by a technique known as Bethe-Ansatz (BA).
This technique is limited to special models. For example the fermionic Hubbard model or the $t-V$ model are BA solvable, but the bosonic
Hubbard model is not. These exact solutions allow to extract relatively easily the spectrum of excitations, and thus with some effort the
thermodynamics properties of the system. It is an herculean task to go beyond this and in particular to compute the correlation functions.
Fortunately significant progress could be accomplished in this domain and some correlation functions have been obtained by BA in the recent years.
This technique can be potentially extended to out of equilibrium situations as well.

\item {\bf Numerical techniques:} Numerical techniques to deal with quantum interacting particles suffer from notorious convergence problems
(specially for fermions) or have a hard time to deal with real-time dynamics. Fortunately in one dimension, a special technique, the Density Matrix Renormalization Group technique, introduced by S. White in the 90's, allows extremely precise results without suffering from essential convergence problems. Recently this technique has been extended to deal with dynamical correlation functions as well. It is thus a method of choice to tackle one dimensional systems. As with any numerical techniques, it is well adapted to give short and intermediate range physics, but has the advantage to be able to deal with additional complications such as the trap or other modifications of the model without too much problems.

\item {\bf Low energy techniques:} As for high dimensional materials (see in particular the next section \sref{sec:fermiliquid} for the Fermi liquid theory) there is in one dimension a way to extract a universal description of the physical properties of the problem at low energy. This technique, resting on something called bosonization, is thus complementary of the two above mentioned techniques. It allows from the start to obtain the asymptotic properties of the system, as a function of space, time at zero or finite temperature. It also provided a nice framework to understand the new physical properties of one dimensional systems. It depends on parameters that can be efficiently determined by the two above techniques or extracted directly from experiments.
\end{enumerate}

These three lines of approach are directly complementary. In this section we will mostly discuss the bosonization technique since it is the one that gives the most direct physical representation of the physics of the problem.

The idea behind the bosonization technique is to reexpress the
excitations of the system in a basis of collective excitations.
Indeed in one dimension it is easy to realize that single
particle excitations cannot really exit. One particle when
moving will push its neighbors and so on, which means that any
individual motion is converted into a collective one.
Collective excitations should thus be a good basis to represent
a one dimensional system.

\begin{figure}
\begin{center}
  \includegraphics[width=0.8\linewidth]{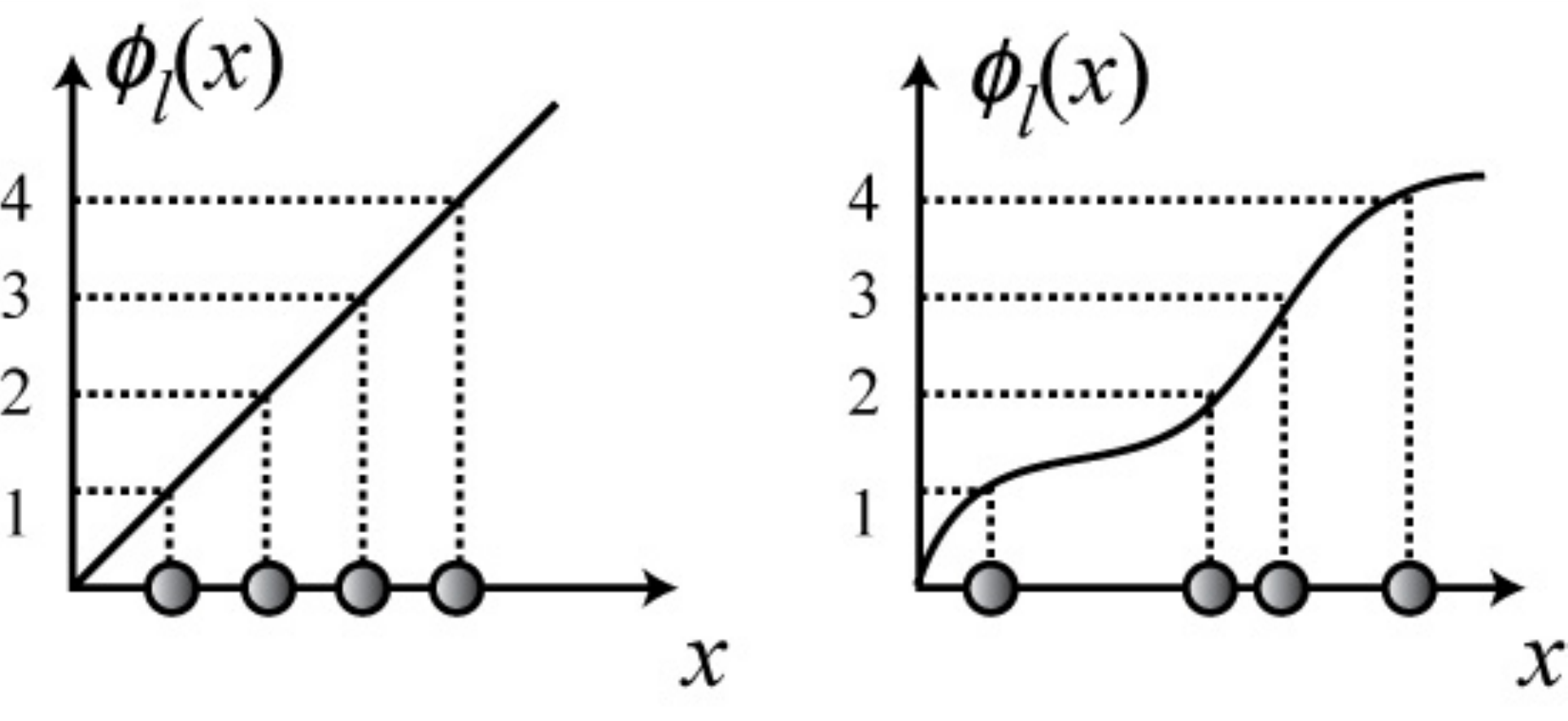}
\end{center}
 \caption{Some examples of the labeling field $\phi_l(x)$. If the
 particles form a perfect lattice of lattice spacing $d$, then
 $\phi_l^0(x) = 2\pi x/d$, and is just a straight line.
 Different functions
 $\phi_l(x)$ allow to put the particles at any position in space. Note that $\phi(x)$ is always
 an increasing function regardless of the position of the particles. [From \protect\cite{giamarchi_book_1d}]}
 \label{fig:labelfield}
\end{figure}

To exploit this idea, let us start with the density operator
\begin{equation}\label{eq:densmoche}
 \rho(x) = \sum_i \delta(x-x_i)
\end{equation}
where $x_i$ is the position operator of the $i$th particle. We
label the position of the $i$th particle by an `equilibrium'
position $R_i^0$ that the particle would occupy if the
particles were forming a perfect crystalline lattice, and the
displacement $u_i$ relative to this equilibrium position. Thus,
\begin{equation}
 x_i = R_i^0 + u_i
\end{equation}
If $\rho_0$ is the average density of particles,
$d=\rho_0^{-1}$ is the distance between the particles. Then,
the equilibrium position of the $i$th particle is
\begin{equation}
 R_i^0 = d i
\end{equation}
Note that at that stage it is not important whether we are
dealing with fermions or bosons. The density operator written
as (\ref{eq:densmoche}) is not very convenient. To rewrite it
in a more pleasant form we introduce a labeling field
$\phi_l(x)$ \cite{haldane_bosons}. This field, which is a
continuous function of the position, takes the value
$\phi_l(x_i) = 2\pi i$ at the position of the $i$th particle.
It can thus be viewed as a way to number the particles. Since
in one dimension, contrary to higher dimensions, one can always
number the particles in an unique way (e.g. starting at
$x=-\infty$ and processing from left to right), this field is
always well-defined. Some examples are shown in
\fref{fig:labelfield}.
Using this labeling field one can rewrite the density as
\begin{eqnarray}
 \rho(x) &=& \sum_i \delta(x-x_i) \nonumber \\
  &=& \sum_n |\nabla \phi_l(x)| \delta(\phi_l(x) - 2\pi n)
  \label{eq:denslab}
\end{eqnarray}
It is easy to see from \fref{fig:labelfield} that $\phi_l(x)$
can always be taken as an increasing function of $x$, which
allows to drop the absolute value in (\ref{eq:denslab}). Using
the Poisson summation formula this can be rewritten
\begin{equation}
 \rho(x) = \frac{\nabla \phi_l(x)}{2\pi} \sum_p e^{i p \phi_l(x)}
\end{equation}
where $p$ is an integer. It is convenient to define a field
$\phi$ relative to the perfect crystalline solution and to
introduce
\begin{equation}
 \phi_l(x) = 2 \pi \rho_0 x - 2\phi(x)
\end{equation}
The density becomes
\begin{equation} \label{eq:locintbos}
 \rho(x) = \left[\rho_0 -\frac{1}{\pi} \nabla \phi(x)\right] \sum_p e^{i 2 p (\pi \rho_0 x - \phi(x))}
\end{equation}
Since the density operators at two different sites commute it
is normal to expect that the field $\phi(x)$ commutes with
itself. Note that if one averages the density over distances
large compared to the interparticle distance $d$ all
oscillating terms in (\ref{eq:locintbos}) vanish. Thus, only
$p=0$ remains and this smeared density is
\begin{equation} \label{eq:smeardens}
 \rho_{q\sim 0}(x) \simeq \rho_0 - \frac1\pi \nabla\phi(x)
\end{equation}
The formula (\ref{eq:locintbos}) has the following semiclassical interpretation:
the field $\phi(x)$ is essentially the displacements of the particles compared to a perfect
crystalline order with a distance $a= \rho_0^{-1}$. The $p=0$ term is essentially the standard elastic
representation of the density of particles. In addition the density is composed of density waves with wavevectors
$2 \pi \rho_0 p$ (the lowest one is simply the one corresponding to a maximum on each particle). The field $\phi(x)$ give the phase
of these density waves.

We can now write the single-particle creation operator
$\psi^\dagger(x)$. Such an operator can always be written (note the similarity with (\ref{eq:meansingle}) as
\begin{equation} \label{eq:singlephen}
 \psi^\dagger(x) = [\rho(x)]^{1/2} e^{-i \theta(x)}
\end{equation}
where $\theta(x)$ is some operator. In the case where one would
have Bose condensation, $\theta$ would just be the superfluid
phase of the system. The commutation relations between the
$\psi$ impose some commutation relations between the density
operators and the $\theta(x)$. For bosons, the condition is
\begin{equation} \label{eq:comphen}
 [\psi^{\phantom\dagger}_B(x),\psi_B^\dagger(x')] = \delta(x-x')
\end{equation}
If we assume quite reasonably that the field $\theta$ commutes
with itself ($[\theta(x),\theta(x')] = 0$) a sufficient condition to satisfy (\ref{eq:comphen}) is thus
\begin{equation}\label{eq:commutbas}
 [\rho(x),e^{-i\theta(x')}] = \delta(x-x')e^{-i\theta(x')}
\end{equation}
It is easy to check that if the density were only the smeared
density (\ref{eq:smeardens}) then (\ref{eq:commutbas}) is
obviously satisfied if
\begin{equation} \label{eq:conjphi}
 [\frac1\pi \nabla\phi(x),\theta(x')] = -i \delta(x-x')
\end{equation}
One can show that this is indeed the correct condition to use
\cite{giamarchi_book_1d}. Equation (\ref{eq:conjphi}) proves
that $\theta$ and $\frac1\pi \nabla\phi$ are canonically
conjugate. Note that for the moment this results from totally
general considerations and does not rest on a given microscopic
model. Such commutation relations are also physically very
reasonable since they encode the well known duality relation
between the superfluid phase and the total number of particles.
Integrating by part (\ref{eq:conjphi}) shows that
\begin{equation}
 \pi\Pi(x) = \hbar\nabla\theta(x)
\end{equation}
where $\Pi(x)$ is the canonically conjugate momentum to
$\phi(x)$. To obtain the single-particle operator one can substitute
(\ref{eq:locintbos}) into (\ref{eq:singlephen}). Since the
square root of a delta function is also a delta function up to
a normalization factor the square root of $\rho$ is identical
to $\rho$ up to a normalization factor that depends on the
ultraviolet structure of the theory. Thus,
\begin{equation} \label{eq:singlebos}
 \psi^\dagger_B(x) = [\rho_0 - \frac1\pi\nabla \phi(x)]^{1/2}
 \sum_{p} e^{i 2 p (\pi \rho_0 x - \phi(x))}e^{-i \theta(x)}
\end{equation}
where the index $B$ emphasizes that this is the representation
of a \emph{bosonic} creation operator.

The fact that all operators are now expressed in terms of
variables describing \emph{collective} excitations is at the
heart of the use of such representation, since as already
pointed out, in one dimension excitations are necessarily
collective as soon as interactions are present. In addition the
fields $\phi$ and $\theta$ have a very simple physical
interpretation. If one forgets their canonical commutation
relations, order in $\theta$ indicates that the system has a
coherent phase as indicated by (\ref{eq:singlebos}), which is
the signature of superfluidity. On the other hand order in
$\phi$ means that the density is a perfectly periodic pattern
as can be seen from (\ref{eq:locintbos}). This means that the
system has ``crystallized''. The representation (\ref{eq:locintbos}) and (\ref{eq:singlebos})
and the commutation relation (\ref{eq:conjphi}) is thus a dictionary allowing to reexpress every
term in any interacting one-dimensional bosonic problem in terms of the new collective variables
$\theta$ and $\phi$. Although this does not solve the problem, but simply reexpress it,
because these are the good excitations of the system we can expect the theory to be much simpler in these variables.
We will see that this is indeed the case, and that we can extract some universal physical behavior.

\subsection{Universal physics: Luttinger liquids}

To determine the Hamiltonian in the bosonization representation
we use (\ref{eq:singlebos}) in the kinetic energy of bosons. It
becomes
\begin{equation} \label{eq:expkin}
\begin{split}
 H_K &\simeq \int dx \frac{\hbar^2\rho_0}{2m}(\nabla e^{i\theta})(\nabla e^{-i\theta}) + \frac{\hbar^2 (\nabla \rho(x))^2)}{2 m \rho(x)} \\
     &= \int dx \frac{\hbar^2\rho_0}{2m} (\nabla\theta)^2 + \frac{\hbar^2}{2 m \pi^2 \rho_0} (\nabla^2 \phi(x))^2
\end{split}
\end{equation}
the first part is the part coming from the single-particle operator
containing less powers of $\nabla\phi$ and thus the most
relevant. We have also kept here the second (less relevant term) which allows to make
the connection with Bogoliubov's theory.
Using (\ref{eq:locintbos}) the interaction term becomes
\begin{equation} \label{eq:intbosbos}
 H_{\rm int} = \int dx V_0 \frac{1}{2\pi^2} (\nabla\phi)^2
\end{equation}
plus higher order operators. Keeping only the above lowest
order shows that the Hamiltonian of the interacting bosonic
system can be rewritten as
\begin{equation} \label{eq:luthamphen}
 H = \frac{\hbar}{2\pi}\int dx [\frac{u K}{\hbar^2} (\pi \Pi(x))^2 + \frac{u}{K}
 (\nabla\phi(x))^2]
\end{equation}
where we have put back the $\hbar$ for completeness. This leads
to the action
\begin{equation} \label{eq:lutacphen}
 S/\hbar = \frac1{2\pi K} \int dx \;d\tau [\frac1u (\partial_\tau\phi)^2 +
 u (\partial_x\phi(x))^2]
\end{equation}
This hamiltonian is a standard sound wave one. The fluctuation
of the phase $\phi$ represent the ``phonon'' modes of the
density wave as given by (\ref{eq:locintbos}). One immediately
sees that this action leads to a dispersion relation, $\omega^2
= u^2 k^2$, i.e. to a linear spectrum. $u$ is the velocity of
the excitations. Note that keeping the second term in (\ref{eq:expkin})
gives the dispersion
\begin{equation}
 \omega^2 = u^2 k^2 + A k^4
\end{equation}
which is exactly similar to the Bogoliubov dispersion relation.
Note however that the theory is quite different from the Bogoliubov one
given the highly non-linear representation of the operators in terms of the fields
$\theta$ and $\phi$.

$K$ is a dimensionless parameter whose role
will be apparent below.  The parameters $u$ and $K$ are used to
parameterize the two coefficients in front of the two
operators. In the above expressions they are given by
\begin{equation} \label{eq:ukpert}
\begin{split}
 u K & = \frac{\pi  \hbar \rho_0}m \\
 \frac{u}{K} &= \frac{V_0}{\hbar \pi }
\end{split}
\end{equation}
This shows that for weak interactions $u \propto (\rho_0
V_0)^{1/2}$ while $K \propto (\rho_0 / V_0)^{1/2}$. In
establishing the above expressions we have thrown away the
higher order operators, that are less relevant. The important
point is that these higher order terms will not change the form
of the Hamiltonian (like making cross terms between $\phi$ and
$\theta$ appears etc.) but {\it only} renormalize the
coefficients $u$ and $K$ (for more details see
\cite{giamarchi_book_1d}).

The low-energy properties of interacting quantum fluids are
thus described by an Hamiltonian of the form
(\ref{eq:luthamphen}) \emph{provided} the proper $u$ and $K$
are used. These two coefficients \emph{totally} characterize
the low-energy properties of massless one-dimensional systems.
The bosonic representation and Hamiltonian
(\ref{eq:luthamphen}) play the same role for one-dimensional
systems than the Fermi liquid theory that will be discussed in \sref{sec:fermiliquid}
plays for higher-dimensional systems. It is an effective low-energy
theory that is the fixed point of any massless phase,
regardless of the precise form of the microscopic Hamiltonian.
This theory, which is known as Luttinger liquid theory
\cite{haldane_bosonisation,haldane_bosons}, depends only on the
two parameters $u$ and $K$. Provided that the correct value of
these parameters are used, \emph{all} asymptotic properties of
the correlation functions of the system then can be obtained
\emph{exactly} using (\ref{eq:locintbos}) and
(\ref{eq:singlebos}) or (\ref{eq:singlephenfer}).

Computing the Luttinger liquid coefficient can be done very
efficiently. For small interaction, perturbation theory such as
(\ref{eq:ukpert}) can be used. More generally one just needs
two relations involving these coefficients to obtain them.
These could be for example two thermodynamic quantities, which
makes it easy to extract from either Bethe-ansatz solutions if
the model is integrable or numerical solutions. The Luttinger
liquid theory thus provides, coupled with the numerics, an
incredibly accurate way to compute correlations and physical
properties of a system (see e.g.
\cite{klanjsek_nmr_ladder_luttinger} for a remarkable example).
For more details on the various procedures and models see
\cite{giamarchi_book_1d,cazalilla_review_bosons_1D}. But, of course, the most important
use of Luttinger liquid theory is to justify the use of the
boson Hamiltonian and fermion--boson relations as starting
points for any microscopic model. The Luttinger parameters then
become some effective parameters. They can be taken as input,
based on general rules (e.g. for bosons $K=\infty$ for non
interacting bosons and $K$ decreases as the repulsion
increases, for other general rules see
\cite{giamarchi_book_1d}), without any reference to a
particular microscopic model. This removes part of the
caricatural aspects of any modelization of a true experimental
system. The Luttinger liquid theory is thus
an invaluable tool to tackle the effects of perturbations on an
interacting one-dimensional electron gas (such as the effect of
lattice, impurities, coupling between chains, etc.). We refer
the reader to \cite{giamarchi_book_1d} for more on those
points.

Let us now examine in details the physical properties of such a
Luttinger liquid. For this we need the correlation functions. We just give the results here.
More detailed calculations and functional integral methods are given in \cite{giamarchi_book_1d}.

The density-density and the single particle correlations are given by
\begin{equation} \label{eq:singleboscor}
\begin{split}
  \langle T_\tau \psi(r) \psi^\dagger(0)\rangle &= A_1
  \left(\frac{\alpha}{r}\right)^{\frac1{2K}} + \cdots \\
  \langle T_\tau \rho(r)\rho(0)\rangle &= \rho_0^2 + \frac{K}{2\pi^2}
  \frac{y_\alpha^2 - x^2}{(y_\alpha^2 + x^2)^2} +
  A_3 \cos(2\pi\rho_0 x) \left(\frac{1}{r}\right)^{2K} + \cdots
\end{split}
\end{equation}
where $r = \sqrt{x^2 + y^2}$ and $y = u \tau$ and $\tau$ the standard imaginary time.
Here, the lowest distance in the theory is $\alpha \sim \rho_0^{-1}$. The
amplitudes $A_i$ are non-universal objects. They depend on the
precise microscopic model, and even on the parameters of the
model. These amplitudes can be computed either by BA or by the DMRG calculations.
Contrary to the amplitudes $A_n$, which depend on the
precise microscopic model, the power-law decay of the various
terms are \emph{universal}. They \emph{all} depend on the
unique Luttinger coefficient $K$. The fluctuations of long wavelength decay with a universal
power law. These fluctuations correspond to the hydrodynamic
modes of the interacting quantum fluid. The fact that their
fluctuation decay very slowly is the signature that there are
massless modes present. This corresponds to the sound waves of
density described by (\ref{eq:luthamphen}). However the
density of particles has also higher fourier harmonics. The
corresponding fluctuations also decay very slowly but this time
with an interaction dependent exponent that is controlled by the LL
parameter $K$. This is also the signature of the presence of a
\emph{continuum} of gapless modes, that exists for Fourier components
around $Q = 2n\pi \rho_0$. For bosons $K$ goes to infinity when the interaction goes to zero
which means that the correlations in the density decays
increasingly faster with smaller interactions. This is
consistent with the idea that the system becoming more and more
superfluid smears more and more its density fluctuations. This is shown in \fref{fig:1dvs3d}.

The single particle correlation function decays with distance. This reflects that no true superfluid order
exists. For the non-interacting system $K=\infty$ and we recover that the
system possesses off-diagonal long-range order since the
single-particle Green's function does not decay with distance.
The system has condensed in the $k=0$ state. As the repulsion
increases ($K$ decreases), the correlation function decays
faster and the system has less and less tendency towards
superconductivity. The occupation factor $n(k)$ has thus no
delta function divergence but a power law one, as shown in
\fref{fig:1dvs3d}. Note that the presence of the condensate or not is not directly
linked to the question of superfluidity. The fact that the
system is a Luttinger liquid with a finite velocity $u$,
implies that in one dimension an interacting boson system has
always a linear spectrum $\omega= u k$, contrary to a free
boson system where $\omega \propto k^2$. Such a system is thus
a \emph{true} superfluid at $T=0$ since superfluidity is the
consequence of the linear spectrum \cite{mikeska_supra_1d}. Of
course when the interaction tends to zero $u\to 0$ as it should
to give back the quadratic dispersion of free bosons.

Correlation functions can be computed as easily at finite temperatures using either standard methods
or a conformal mapping. We refer the reader to \cite{giamarchi_book_1d} for these calculations. Essentially the correlation
functions now decrease exponentially as $e^{- C \beta x}$ where $\beta$ is the inverse temperature and $C$ some constant related to the
velocity $u$ and the LL parameter $K$. This will transform the occupation factor into a Lorentzian one as shown in \fref{fig:1dvs3d}.

Note that one finds very often an approximation called ``quasi-condensates'' used. This approximation
consists in assuming that the density is essentially $\rho(x) = \rho_0$ but that the phase can fluctuate.
As is obvious from the representations of \sref{sec:1dtech} this is an approximation compared to the true LL representation.
It is a very accurate one in the limit where $K$ is large (small interactions) since in that case the density-density correlation decay extremely
fast with distance. However for larger interactions the fluctuations of density affect the decay of the correlation functions as described by
(\ref{eq:singleboscor}) and the full theory must be retained.

\begin{figure}
\begin{center}
  \includegraphics[width=0.8\linewidth]{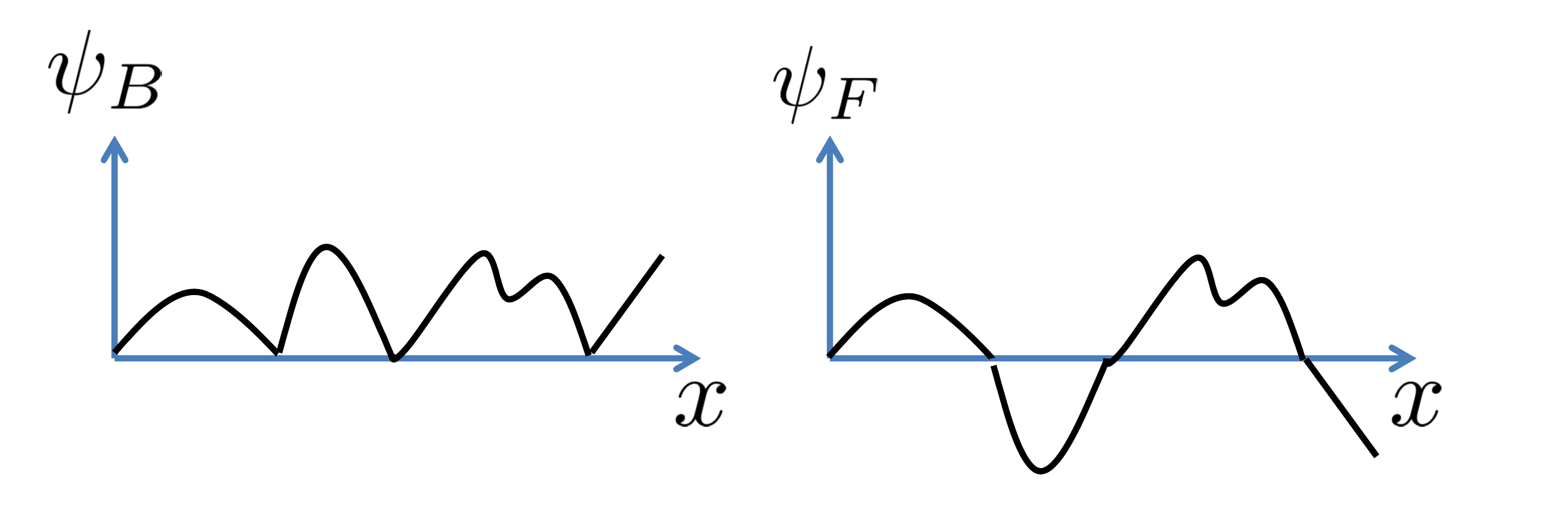}
\end{center}
 \caption{\label{fig:tonkswave}
 If the repulsion between bosons is infinite, one can replace this problem by a free fermion problem, since the Pauli principle will impose a node
 at the position of each particle. However the two problems differ by the sign of the wavefunction across each particle. The properties depending on the square of the wavefunctions are thus identical between the two problems, while single particle properties are quite different.}
\end{figure}

One specially interesting limit to investigate is the so called Tonks-Girardeau limit \cite{girardeau_bosons1d,lieb_bosons_1D}
for which the repulsion between bosons is going to infinity. In that case the repulsion of between the bosons acts as constraint
forbidding two fermions to be at the same point. The wavefunction of one particle has thus a node at the position of each other.
We can thus imagine to replace the repulsion by the Pauli principle of a fake spinless fermion problem and thus map the infinitely repulsive
boson problem into a free fermion one. The price to pay, as show in \fref{fig:tonkswave} is that the wavefunction of the first problem is totally symmetric while the one of the second is totally antisymmetric. Thus the sign of the wavefunction differs between each particles. It means that properties which only depends on the square of the wavefunction -- such as the thermodynamics, and the density correlations -- are the same,
while the ones directly depending on the wavefunction (such as the single particle correlations) will be of course more complicated to compute.
There are direct methods to exploit this limit. Let us see here how the LL theory allows simply to have the correlations. We see that choosing $K=1$ ensures that the density-density correlations decay as $1/r^2$ and have oscillations at $2 \pi \rho_0$. This is exactly what one expects for a free fermion system \cite{ashcroft_mermin_book,ziman_solid_book}. The mapping on the free fermion problem allows here to unambiguously fix the LL parameter $K$ to $K=1$. Of course a direct determination as a function of the interaction also shows that this is the good limit for this parameter when the interaction becomes infinite. The single particle correlation function is not easy to obtain even in the Tonks-Girardeau limit given the change of signs of the wavefunction and one must need to use rather sophisticated techniques (see e.g. \cite{cazalilla_review_bosons_1D} for more details). However the LL directly gives that the single particle correlation decays as $1/\sqrt{r}$. We see on this particularly clear example the universal features that one can extract for the physics of one dimensional interacting systems.

The Luttinger liquid theory has been checked in various context both in condensed matter and in the cold atom systems. In condensed matter, the first
evidence of a LL powerlaw was obtained for organic superconductors \cite{schwartz_electrodynamics}, followed by experiments on nanotubes \cite{yao_nanotube_kink}. Many additional tests have been made in other systems, see \cite{giamarchi_book_1d,cazalilla_review_bosons_1D} for more systems and references. Recently spin-ladder systems have provided remarkable systems in which a quantitative test of the exponents could be performed \cite{klanjsek_nmr_ladder_luttinger}. In cold atomic systems beautiful experiments could probe of one-dimensional interacting bosonic systems.
Coupled one dimensional tubes could be obtained \cite{stoeferle_coldatoms1d} where the role of the superfluid-Mott transition was investigated and the single particle correlation function measured. The existence of the Tonks-Girardeau limit could be checked by investigating the thermodynamics of the system on a single tube \cite{kinoshita_1D_tonks_gas_observation}. In such a system the interaction was raised by using the transverse confinement \cite{olshanii_cir}. The Tonks-Girardeau limit was also observed in systems with optical lattices \cite{paredes04_tonks_gas}. In such systems the ratio between kinetic energy and interactions was controlled by the optical lattice. The single particle correlation functions were measured and the data is roughly compatible with the $n(k) \sim 1/\sqrt{k}$ that one would expect. However the inhomogeneities of density, both in a single tube and between the tubes makes the comparison more complicated. For the particular case of $U=\infty$ the mapping to free fermions allows this averaging to be done allowing a reasonable fit to the experiment. It would however be very interesting to have local measurements of single tube ones, to also check the intermediate interactions regimes for which no comparison with the LL theory has yet been done.
Finally a remarkable system to test such predictions is provided by atom chips. Indeed in such systems the homogeneity is very good, and one can do measurements on a single tube. Interference experiments \cite{gritsev_interferences_chips,hofferberth_full_counting_chip} on condensates have shown excellent agreements with the LL theory, both for the correlation functions and even the full counting statistics. Unfortunately the interactions are small
so that $K$ remains very large and it is difficult to make the difference between LL and simple quasi-condensates. Clearly further experiments will be interesting in this remarkable experimental systems.

\subsection{Mott insulators in one dimension}

Let us finally examine how the Mott transition can take place in one dimension, and compare with the results of \sref{sec:mottboson}. Although we have already shown that the superfluid phase is quite different in one and in higher dimensions, we can certainly expect the basic arguments in favor of the Mott transition of \sref{sec:mottboson} to still be valid. We can thus expect the existence of a Mott transition in one dimension as well. One would then go from a quasi-long range order of the phase (powerlaw decay of the superfluid correlations) to a system with one (or an integer number of) bosons per site which would be an insulator. The LL formalism provides a remarkable way to study such transition. As in the previous section we only sketch the solution and refer the reader to \cite{giamarchi_book_1d,cazalilla_review_bosons_1D} for the gist of the calculations and for references.

\begin{figure}
\begin{center}
  \includegraphics[width=0.8\linewidth]{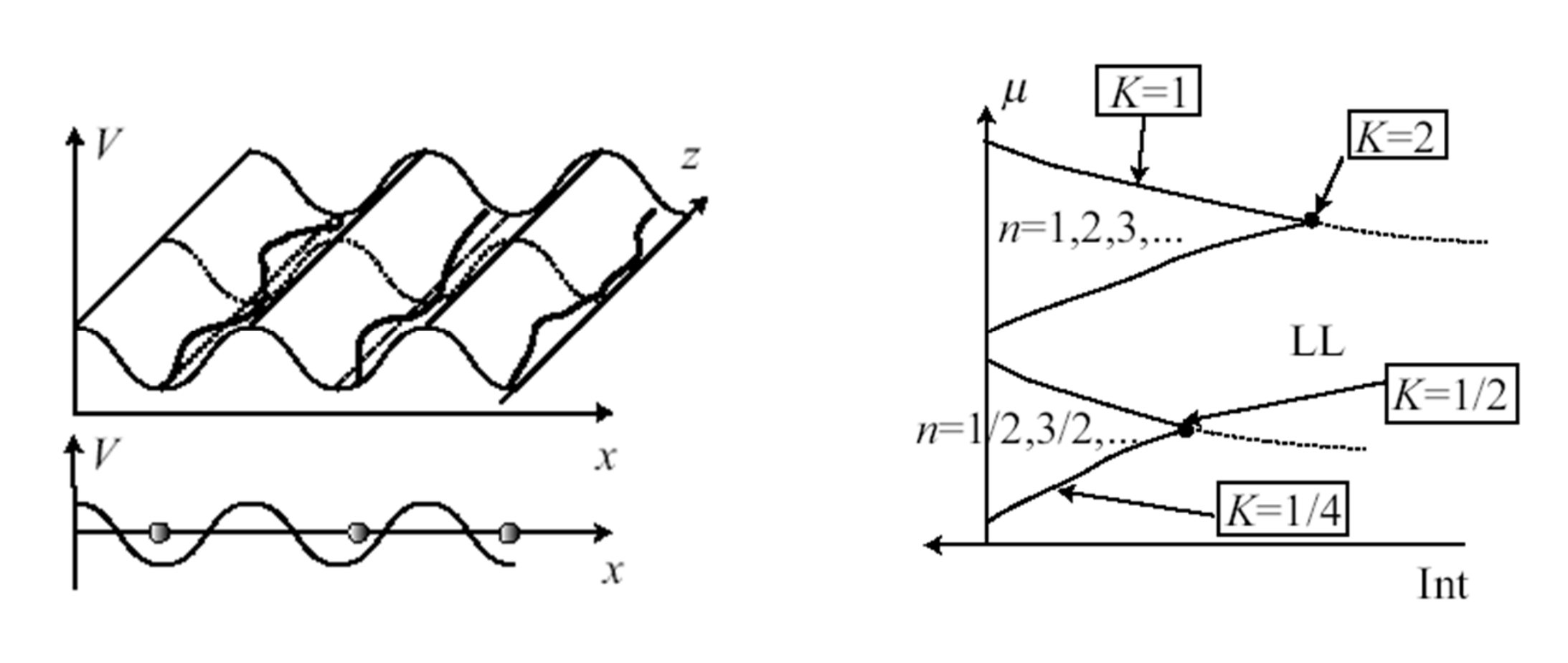}
\end{center}
 \caption{\label{fig:mottper}
 (left) If the periodic potential $V_0 \cos(Q x)$ is commensurate with the particle density $Q=2\pi\rho_0$ then a Mott phase can appear. This problem is equivalent to the localization of elastic lines in a periodic potential or to the Berezinskii-Kosterlitz-Thouless transition in a two dimensional XY model. It occurs for sufficiently repulsive interactions $K \leq 2$. (right) the phase diagram showing the universal value taken by the LL parameter at the transition. Note the presence of two transitions: the Mott-U transition at constant density upon variation of the interactions; the Mott-$\delta$ transition where the interactions are fixed and the system is doped. These corresponds to different universality classes. [After \protect\cite{giamarchi_book_1d}]}
\end{figure}

In the absence of a lattice, the interacting one dimensional system is described by the quadratic action (\ref{eq:lutacphen}). In order to determine the effect of a lattice we just have to add to this action the interaction with a lattice. If we represent the lattice by the potential $V(x) = V_0 \cos(Q x)$, then such a term is
\begin{equation} \label{eq:perio}
 H_V = - V_0 \int dx \cos(Q x) \rho(x)
\end{equation}
We can then use the representation of the density (\ref{eq:locintbos}) to see that terms of the form
\begin{equation} \label{eq:oscbase}
 V_0 \int dx e^{i (Q  - 2 p \pi \rho_0)x} e^{-i2p\phi(x)}
\end{equation}
appear from (\ref{eq:perio}). We thus see that there are two possibilities. A first possibility the wavevector $Q$ of the periodic potential is not commensurate with the density of particles $Q \neq 2 \pi \rho_0$. In that case one does not have exactly one particle per site. In that case the terms in the integral (\ref{eq:oscbase}) oscillate fast and essentially kill the extra term in the action. In that case, the lattice potential is irrelevant and one recovers a LL (superfluid) phase with renormalized parameters $u$ and $K$. This is exactly similar to the case described in the \sref{sec:mottboson}, where the Mott phase could only occur for one particle per site. The Mott phase can potentially appear if $Q$ is commensurate with the particle density. If $Q = 2 \pi \rho_0$, this means, as shown in \fref{fig:mottper}, that there is exactly one particle per site.
In that case the oscillations go away and (\ref{eq:perio}) becomes
\begin{equation} \label{eq:mottcos}
 H_V = - V_0 \int dx \cos(2 \phi(x))
\end{equation}
There are potentially terms with higher $p$ which correspond to higher commensurabilities (one particle every two sites etc.). We will not deal with them here and refer the reader to \cite{giamarchi_book_1d,cazalilla_review_bosons_1D} for these cases.

The effect of the term (\ref{eq:mottcos}) is quite remarkable. There is one one hand the quadratic action (\ref{eq:lutacphen}) which allows the field $\phi$ to fluctuate. These fluctuations are responsible for the decay of the density correlations. On the other hand (\ref{eq:mottcos}) wants to pin the field $\phi$ in one of the minima of the cosine. If the field $\phi$ is pinned it means: a) that the density does not fluctuate any more. We have thus a phase with one particle per site, this is the Mott phase; b) that the field $\theta$ which is conjugate will fluctuate wildly and thus that the superfluid correlations are killed exponentially fast. We thus see that the combination of (\ref{eq:lutacphen}) and (\ref{eq:mottcos}), known as the sine-Gordon model is the model giving the description of the Mott transition in one dimension. This model has connections with several other models \cite{giamarchi_book_1d,cazalilla_review_bosons_1D}. As shown in \fref{fig:mottper}, it is connected to the fluctuations of classical lines, in a tin-roof potential. It is also connected, in a less obvious way, to the classical Beresinskii-Kosterlitz-Thouless transition in the XY modelm the operator $\cos(2\phi)$ being the vortex creation operator in such a model. We will not detail the connection between these models and refer the reader to \cite{giamarchi_book_1d,cazalilla_review_bosons_1D}. The transition can occur if the strength of the potential $V_0$ increases beyond a certain value or if the interaction becomes large enough. In particular one can show that if the fluctuations small enough, i.e. if $K \leq 2$ even an \emph{infinitesimal} $V_0$ is able to pin the field $\phi$ and one goes in the Mott phase. This is a quite remarkable feature since it shows that large enough repulsion between the particle can lead to a Mott phase even if the lattice is very weak. This can be viewed as the pinning of the charge-density wave of the bosons by the periodic potential of the lattice and is a true quantum effect. Of course if the lattice is deep, we also recover our usual intuition of the Mott transition.

\begin{figure}
\begin{center}
  \includegraphics[width=0.8\linewidth]{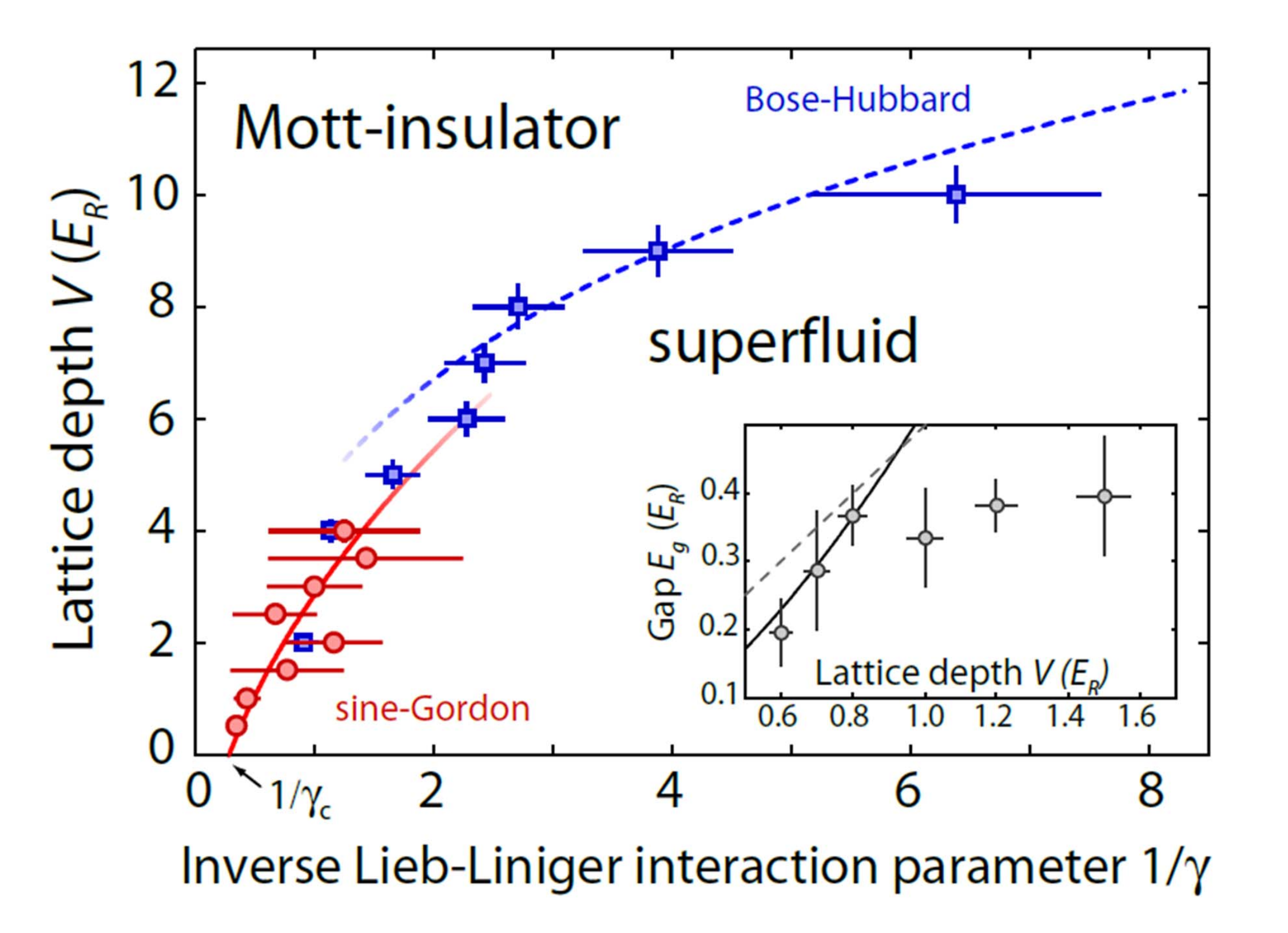}
\end{center}
 \caption{\label{fig:mottexp}
 Phase diagram of a one dimensional system gap as a function of the interaction parameter $\gamma$ (relative to the kinetic energy) and the strength of the optical lattice $V$, as probed by a shaking of the optical lattice. The finite gap indicates the existence of a Mott phase. One sees that regardless of the strength of the lattice, a Mott transition can occur provided that the repulsion is large enough, in agreement with the LL predictions. After \protect\cite{haller_mott_1d}.}
\end{figure}

We thus see that the Mott transition is one dimension is quite similar to its higher dimensional counterpart. One important difference it that it can also occur for weak lattices provided that the repulsion is large enough. At the transition, as indicated in \fref{fig:mottper}, $K$ takes the \emph{universal} value $K=2$. At the transition the transition is in the universality class of the two dimensional XY model. This feature persists even to higher dimensions \cite{Fisher1989}. The fact that in one dimension we have the bosonized representation of the Hamiltonian allows to compute all the correlation functions, both in the superfluid and in the Mott phase. We refer the reader to \cite{giamarchi_book_1d,cazalilla_review_bosons_1D} for further informations on that point. An important point to note is that there are in fact two types of Mott transitions \cite{giamarchi_mott_shortrev} (see also \fref{fig:phasediag_bose})
\begin{enumerate}
\item One can stay at commensurate values for the density, and vary the interactions. This Mott-U transition is in the universality class of the $d$-dimensional XY model. In one dimension it is described by the sine-Gordon theory and lead to the universal values shown in \fref{fig:mottper}.

\item One can have interactions corresponding to being inside the Mott phase, but dope the system, i.e. vary the density. This Mott-$\delta$ transition is in a different universality class. In one dimension it corresponds to a universality class known as the commensurate-incommensurate phase transition \cite{giamarchi_book_1d,cazalilla_review_bosons_1D} and leads to different critical exponents indicated in \fref{fig:mottper}
\end{enumerate}
In one dimension these two universality classes have been confirmed by DMRG calculations where the phase diagram and the LL exponents have been obtained \cite{KuehnerMonien2000}. In the cold atom context, the existence of the Mott transition in one dimension for arbitrarily small lattice but repulsive enough interactions has been checked in a remarkable experiment \cite{haller_mott_1d}. The gap of the Mott phase, probed by the shaking method described in \sref{sec:shaking}, is shown in \fref{fig:mottexp}.

The term (\ref{eq:mottcos}) has another remarkable consequence. The object that is ordered is not simply the density but the field $\phi$ itself.
Given the relation (\ref{eq:smeardens}) in a way the field $\phi$ is the integral of the density, and the changes in density ($\delta$-function peaks at the particle positions) correspond to kinks in the field $\phi$. The fact that $\phi$ itself is ordered means that any function of the form $e^{i \alpha \phi}$ is tending to a constant. This is a much stronger statement than just imposing the density fixed on each site. In fact the order of $\phi$ and be traced to the existence of non-local string order parameters \cite{berg_haldane_cold_bosons}. Such string order, being non local are of course notoriously difficult to measure. However the recent possibilities of local addressability in cold atomic systems has allowed to directly probe such string orders, and a recent experiment has shown for the Mott transition the existence of such an order parameter \cite{endres_string_mott_cold}.


\section{From free fermions to Fermi liquids}
\label{sec:fermiliquid}


This section is based on graduate courses given in Geneva (together with C.
Berthod, A. Iucci, P. Chudzinski) and in Paris (together with O.Parcollet).
\ For more details, see the course notes:\\ \verb+http://http://dpmc.unige.ch/gr_giamarchi/Solides/solides.html+ \\ and \\
  \verb+http://www.cpht.polytechnique.fr/cpht/correl/teaching/teaching.htm+

\subsection{Non-interacting fermions}
\label{sec:freefermions}

Let us start by recalling some well-known but important facts about non-interacting fermion systems.
We shall state these facts without detailing the calculations, since they can be found in every textbook
on solid-state physics~\cite{ashcroft_mermin_book,ziman_solid_book}.

We consider independent electrons described by the Hamiltonian
\begin{equation} \label{eq:kinetic3}
 H_{\rm kin} = \sum_{\vk\nu\sigma} \eps_{\vk\nu} \hc{c}_{\vk\nu\sigma}c_{\vk\nu\sigma}
\end{equation}
When considering fermions in a lattice, the sum over $\vk$ runs over the first Brillouin zone,
and $\nu$ is a  band index (that we shall sometimes omit when focusing on a single band).
It is important to keep in mind that the creation/destruction operators in this expression refer to
{\it single-particle wave-functions}. In a lattice, those wave-functions are of the form
(Bloch's theorem):
$\phi_{\vk\nu}(\vr)=u_{\vk\nu}(\vr)e^{i\vk\cdot\vr}$ with $u_{\vk\nu}$ a Bloch function having the periodicity of
the lattice, while in the continuum $\phi_{\vk}=e^{i\vk\cdot\vr}/\sqrt{\Omega}$.
The fermion-creation field operator at point $\vr$ is expanded onto these wave-functions as
$\psi_\sigma^\dagger(\vr)=\sum_{\vk\nu}\phi^*_{\vk\nu}(\vr)\hc{c}_{\vk\nu\sigma}$.

The eigenstates of (\ref{eq:kinetic3}) are Slater determinants of single-particle wave functions, of the form:
$\mathrm{det}\{\phi_{\vk_i\nu_i}(\vr_j)\}$, which can conveniently be represented in occupation number
basis (Fock representation) as $|\{n_{\vk\nu\sigma}\}\rangle$ with $n_{\vk\nu\sigma}=0,1$ when the single-particle
state is empty or occupied, respectively.
The ground-state for $N$ fermions corresponds to filling all single-particle states with those fermions,
starting from the lowest possible single-particle energy and placing two fermions with opposite spin
per state. Hence, the ground-state is the `Fermi-sea':
\begin{equation} \label{eq:fermi_sea}
|\mathrm{FS}\rangle = \prod_{\vk\nu,\eps_{\vk\nu}<\eps_F} \hc{c}_{\vk\nu\up} \hc{c}_{\vk\nu\down}
|\emptyset\rangle
\end{equation}
In this expression, $\eps_F$ is the {\it Fermi energy}: the largest single-particle energy corresponding to
an occupied state. It is also the zero-temperature limit of the chemical potential (assuming a metallic, or liquid, state):
$\mu(T\rightarrow 0)=\eps_F$.
One often incorporates the chemical potential in the energy and define $\xi^0_{\vk\nu}= \eps_{\vk\nu} -\mu$.

In momentum-space, the condition $\eps_{\vk\nu}=\eps_F$ ($\xi^0_{\vk\nu} = 0$) defines the
{\it Fermi surface}. It has important physical significance, since it defines the loci in momentum-space of
{\it zero-energy excitations}. Hence, the presence of a Fermi-surface (FS) is a distinctive aspect of a metallic
(or liquid) state, in which excitations with arbitrarily low energy are present, in contrast to an insulating
state, in which the ground-state is separated from excited states by an energy gap.

\begin{figure}
\begin{center}
\includegraphics[width=0.9\columnwidth]{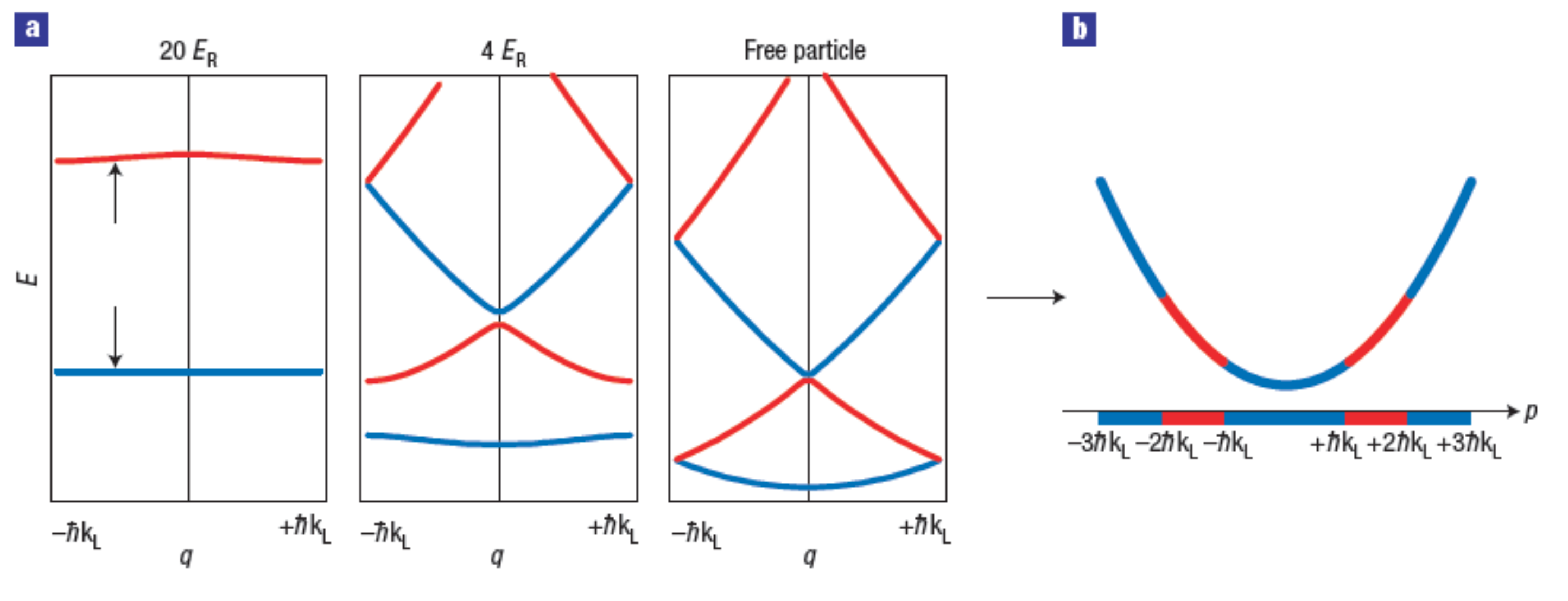}
\caption{``Unfolding'' the bandstructure, from a deep lattice to continuum space, while conserving quasi-momentum.
Fig. adapted from \protect\cite{bloch_review_natphys_2005}}
\label{fig:unfolding_bands}
\includegraphics[width=0.9\columnwidth]{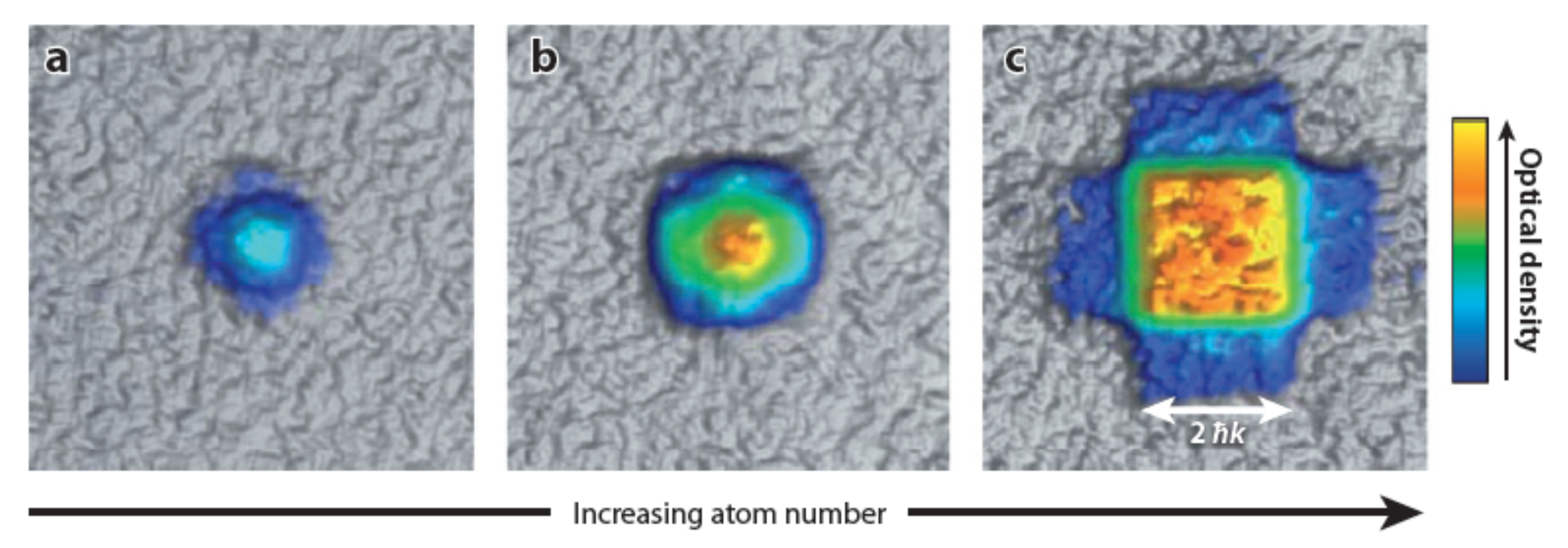}
\caption{Three Fermi-surface images corresponding to low (a), intermediate (b) and higher (c) density of particles per site.
In the latter case, the FS extends beyond the first (square) Brillouin zone.
From \protect\cite{esslinger_annrev_2010} and \protect\cite{kohl_fermisurface_prl_2005}.
}
\label{fig:FS_images}
\end{center}
\end{figure}

In the context of cold fermionic atoms in optical lattices, a direct imaging of the Fermi surface
is possible, as first demonstrated in a remarkable experiment by
M.~K\"{o}hl and coworkers~\cite{kohl_fermisurface_prl_2005}  for a two-dimensional lattice.
The idea is to switch-off the lattice potential adiabatically, so that the quasimomentum of a fermion in a given
quasi-momentum state in the lattice is transferred to the corresponding momentum in the
continuum (i.e quasimomentum is conserved to a good approximation)
(\fref{fig:unfolding_bands}).
A time-of-flight expansion
and absorption imaging are performed in order to obtain the starting quasi-momentum distribution of the
atoms inside the lattice (\fref{fig:FS_images}).
By changing the lattice depth, a gradual increase of the size of the FS was observed, resulting
at some point into a transition from a metal to a band-insulator when the FS coincides exactly with the first
Brillouin zone of the lattice. Note that this effect is a consequence of the presence of a confining potential.
Indeed, in a homogeneous system, the FS is entirely determined by the number of particles present in
the system, and for a given $N$ will remain unchanged if the depth of the lattice (hopping amplitude $t$) is
varied. In contrast, in the presence of a harmonic potential of frequency $\omega_0$, the effective
density is $\rho=N(a/l)^3$ with $l=\sqrt{t/m\omega_0^3}$ (see below), and $\rho$ can be changed
by changing $t$.
%

It is important, at this stage, to have in mind the order of magnitude of key physical
parameters (such as density and mass of particles) relevant to different physical systems of interest, namely:
liquid Helium 3 (for which Landau Fermi-liquid theory was first developed historically), electrons in solids, and
cold atomic gases. Those are summarized in Table~\ref{tab:parameters}. This table illustrates the fact that
the study of degenerate quantum Fermi gases can be undertaken in physical systems with widely different
values of the key energy scale: the Fermi energy. While the Fermi temperature ($=\eps_F/k_B$) is
a few degrees Kelvin in $^3$He, it is several tens of thousands of Kelvin for electrons in solids ($\eps_F$ is a few eV's)
and in the range of a micro- to a nano-Kelvin for cold atomic gases in optical lattices ! The natural unit in the latter
case is the recoil energy of the atoms in the lattice laser beams $\hbar^2k_L^2/2m$, typically of order a $\mu\mathrm{K}$.
By those standards it is considerably more difficult to reach the low-temperature (quantum degenerate) regime
$T\ll T_F$ for ``ultra-cold'' atomic gases than for electrons in solids, for which $T/T_F$ is always of order $10^{-2}$
or less in usual conditions !
This remark can also be turned into an advantage: while the high-temperature crossover between a quantum degenerate and
a classical gas cannot be observed easily in solids, it is easily observed with atomic gases.
The theoretical description of this crossover (and of the `incoherent' regime $T\gtrsim T_F$ requires to handle not
only very low-energy excitations but also higher-energy excited states, which as explained below, are outside the scope
of low-energy effective theories such as Landau Fermi-liquid theory.
\begin{table}
\centering
\begin{tabular}{|c|c|c|c|c|}
\hline  & Mass & Lattice spacing & Density & $T_F=\eps_F/k_B$ \\
\hline Liquid $^3$He & $5.10^{-27}$kg &  & $\sim 2.\,10^{22}$ cm$^{-3}$& A few K \\
\hline Electrons in solids & $9.1\, 10^{-31}$kg & A few$\,\AA\sim 10^{-10}$m & $10^{21}-10^{23}$cm$^{-3}$ & A few$\times 10^4$K \\
\hline Cold atoms  & $\mathrm{m}(^{40}K)\sim 66\, 10^{-27}$kg & $\sim \mu\mathrm{m}$ & $\sim 10^{11}-10^{12}\mathrm{cm}^{-3}$ & $\sim \mathrm{nK}-\mu\mathrm{K}$ \\
\hline
\end{tabular}
\caption{\label{tab:parameters}Typical values of physical parameters for three physical systems. Note that, in the absence of a lattice,
$\eps_F\propto n^{2/3}/m$.}
\end{table}

\begin{figure}
\begin{center}
\includegraphics[width=0.9\linewidth]{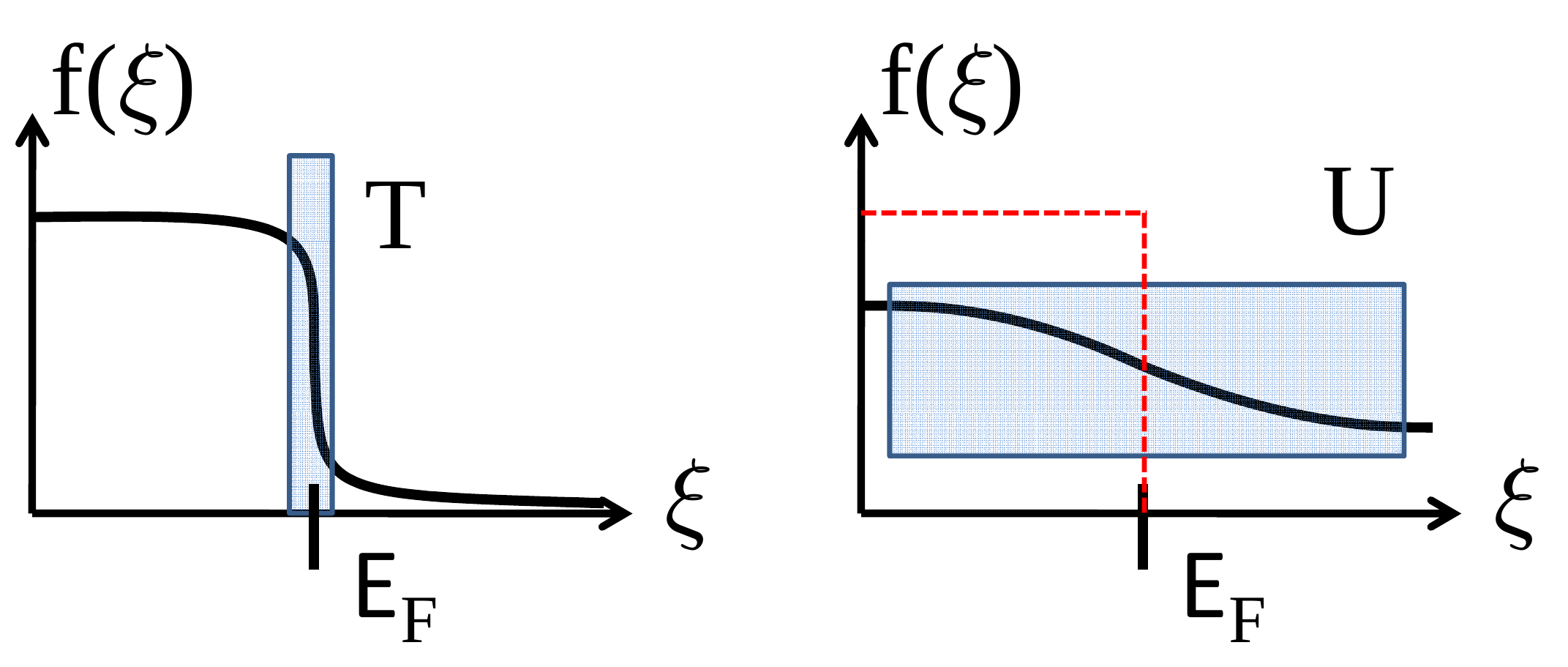}
\end{center}
\caption{\label{fig:broadenings} Broadening of the Fermi distribution due to the temperature $T$.
Left: when $T\ll T_F$, only a tiny fraction of particles close to the Fermi level can be excited.
These low-energy particle-hole excitations control all the physical properties of the system.
Right: in the regime $T\gg T_F$, the Fermi factor is broad and high-energy excitations
become relevant.}
\end{figure}

At finite temperature, single-particle states of a non-interacting Fermi gas are occupied with a probability
given by the Fermi factor (\fref{fig:broadenings}):
\begin{equation}
f(\xi^0_{\vk}) =\langle \mathrm{FS}|\hc{c}_{\vk} c_{\vk}|\mathrm{FS}\rangle = \frac1{e^{\beta\xi^0_{\vk}}+1}
\end{equation}
%
In the quantum degenerate regime $T\ll T_F$, the broadening of the
Fermi distribution is extremely small. The important states are thus the ones in which particles are excited
in a tiny shell close to the Fermi level, as shown in \fref{fig:broadenings}.
The other excitations are completely blocked by the Pauli principle. This strong constraint on available excited states is
of course what confers to fermionic systems their unique properties, and make them so
different from a classical system, or from a bosonic quantum system.
As a consequence, the specific heat $C=dU/dT=TdS/dT$ is linear with temperature (contrarily to the
case of a classical gas for which it would be a constant)
\begin{equation}
 C(T) \propto \kB^2 \dos(\eps_F) T\,\,\,,\,\,\,(T\ll T_F)
\end{equation}
where $\dos(\eps_F)$ is the density of states at the Fermi level.
The compressibility of the fermion gas:
\begin{equation}\label{eq:compress_freefermions}
\kappa \propto \frac{\partial n}{\partial \mu} \propto \dos(\eps_F) \,\,\,,\,\,\,(T\ll T_F)
\end{equation}
 reaches a constant value in the limit $T \to 0$ (in contrast to a band insulator for which $\eps_F$ lies
 within a gap and is hence incompressible $\kappa\propto\dos(\eps_F)=0$).
Similarly, the spin susceptibility which measures the magnetization
$M$ of the electron gas in response to an applied magnetic field $H$,
 reaches  a constant value (Pauli behaviour):
\begin{equation}
 \chi = \left(\frac{\partial M}{\partial H}\right)_T \propto \dos(\eps_F)\,\,\,,\,\,\,(T\ll T_F)
\end{equation}
Note that a system made of independent spins would have instead a
divergent spin susceptibility $\chi\propto 1/T$ when $T\to 0$ (Curie behaviour) instead of a
constant one. In a non-interacting Fermi gas, the slope of the specific heat, the
compressibility and the spin susceptibility are all controlled by the same quantity, namely the density
of states at the Fermi level.

\subsection{Quasiparticles and the $(N\pm 1)$-particle problem}
\label{sec:quasiparticles}

We now consider the effect of interactions between fermions. We focus on a system which is in a compressible
liquid (or metallic) state, with no symmetry breaking of any sort (apart from the translational symmetry breaking
due to the lattice). Possible transitions into a (Mott) insulating state induced by interactions, as well as magnetic
ordering, will be discussed in Sec.~\ref{sec:mottfermion}.

The first important observation is that the ground-state wave-function, which was simple in the
absence of interactions (the Fermi sea), now becomes exceedingly complex. There are very few cases in which this ground-state
wave-function can be found exactly (one such example is the one-dimensional Hubbard model, thanks to Bethe ansatz).
Even numerically, the problem is very difficult. The size of the Hilbert space grows exponentially with the size of the
system (for example, for a single-band Hubbard model, it has dimension $4^{N_s}$ with $N_s$ the number of lattice sites).
Hence, exact diagonalization (e.g. Lanczos) methods can only handle small systems (say, $N_s\lesssim 12$).
As to quantum Monte-Carlo simulations, they are faced with the infamous `minus-sign problem' which severely limit their use, at
least when doing direct simulations without further approximations.

The second important observation is that we may not care so much, after all, about the detailed form of the
ground-state wave-function. What most experiments actually probe are the {\it excitations} above the ground-state\cite{Nozieres_book}.
Furthermore, if the system is at low temperature ($T\ll T_F$) and probed in a gentle-enough manner, only low-energy
excitations matter. So what we really want is a description of these low-energy excitations. This is fortunate, since general wisdom
(backed up by renormalization-group ideas) teaches us that low-energy phenomena (or, equivalently, phenomena involving
long time scales) have a large degree of {\it universality}. Hence, the nature of the low-energy excitations may not depend on all
the microscopic details of the specific problem at hand,
and a universal effective theory of those low-energy excitations may be in sight.
For interacting fermions in more than one dimension, this universal theory is Landau's Fermi-liquid theory\cite{landau_fermiliquid_theory_static,landau_fermiliquid_theory_dynamics}. For one-dimensional systems,
it is Luttinger liquid theory (Secs.~\ref{sec:boso},\ref{sec:mottfermion1D}).

A word of warning, however: being effective theories of low-energy excitations, their applicability is limited to... low-energy. Hence, these
descriptions come with a characteristic scale above which they are no longer valid and a more detailed quantitative description
is required. This scale is associated, as we shall see, with the lifetime of quasiparticles: for energies above a certain coherence scale,
long-lived quasiparticles no longer make sense and Landau Fermi liquid theory does not apply. In strongly correlated systems, this
coherence scale may be quite low, making the range of validity of effective theories too limited to explain all experimental observations.
Also, experiments that perturb the system too strongly (e.g. in a pump-probe experiment)
require tools beyond low-energy effective theories.

\subsubsection{Quasiparticles: qualitative picture} The basic idea behing Landau Fermi liquid theory is that the low-energy excited states can
be constructed by combining together elementary excitations, with combination rules and quantum numbers identical to that of free particles.

For free fermions, this is clearly the case: the simplest low-energy excitation (at constant
particle number $N$) is obtained by considering a single Slater determinant which is obtained from the Fermi sea by exciting an electron
from a state just below the FS to a state just above. Hence, a `particle-hole' excitation has been created, which is a combination of a hole-like excitation
(removing a particle) and a particle-like excitation (adding a particle). In a free system, adding a particle in an empty state yields an eigenstate:
such an excitation does not care about the presence of all the other electrons in the ground state
(otherwise than via the Pauli principle which prevents from
creating it in an already occupied state).

\begin{figure}
\begin{center}
\includegraphics[width=0.3\linewidth]{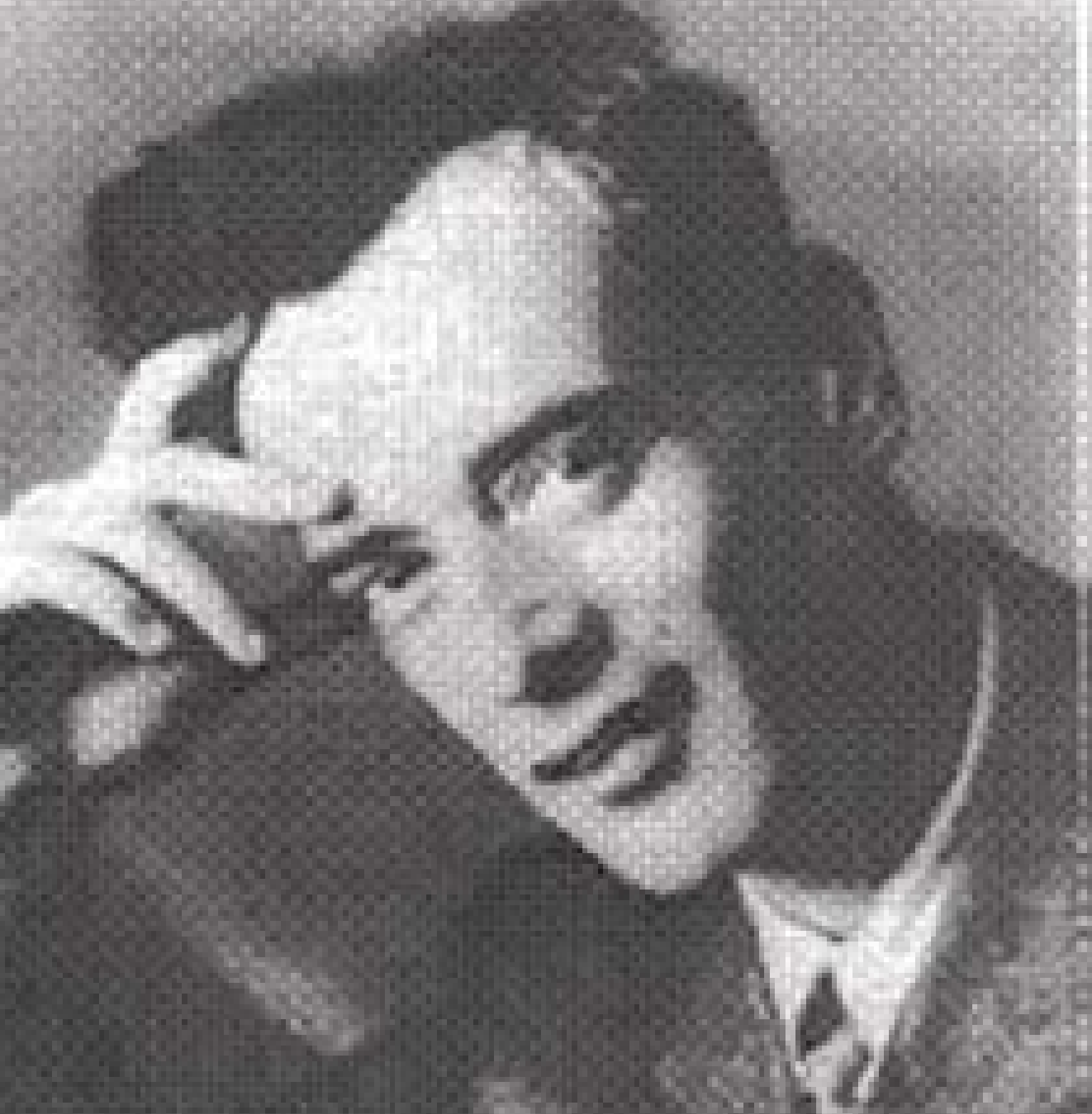}
\includegraphics[width=0.7\linewidth]{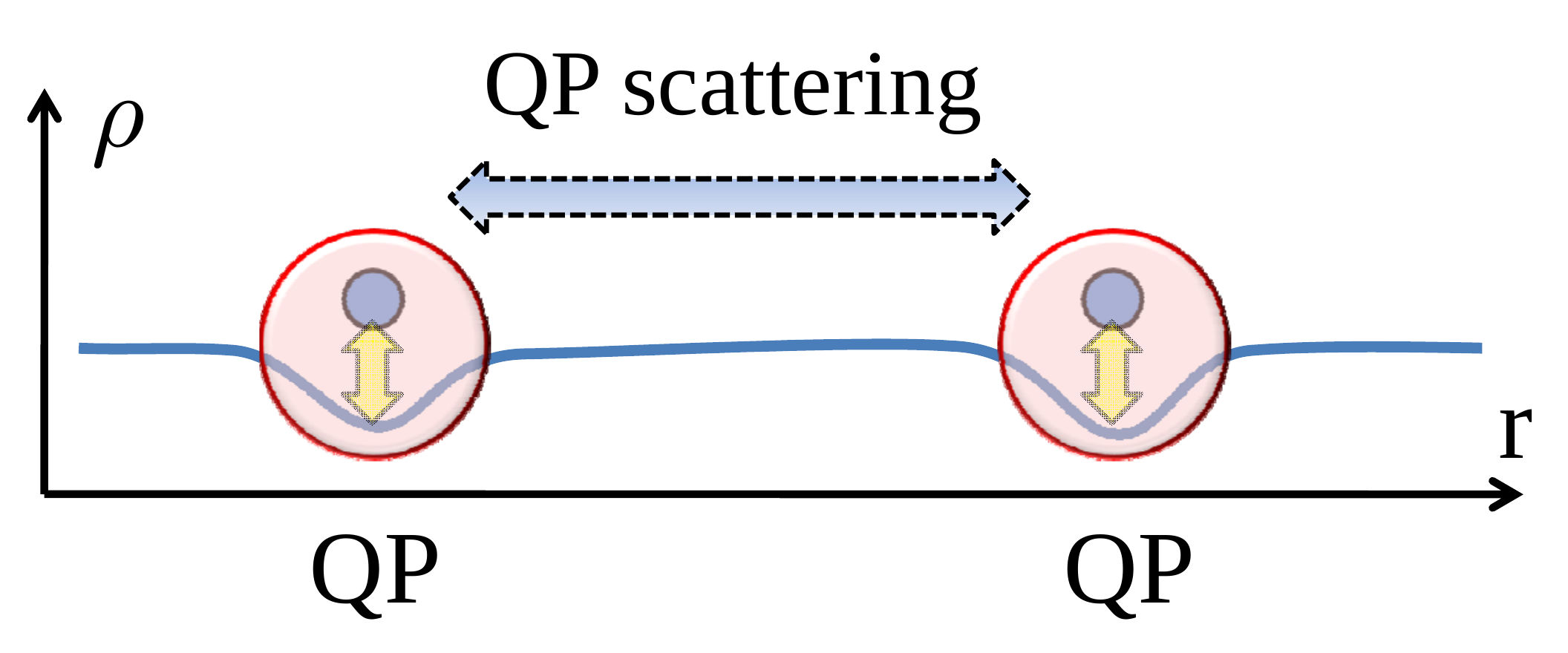}
\end{center}
\caption{\label{fig:fermiliquid} Top: Lev Landau, the man behind the Fermi liquid theory (and many other things).
Bottom: In a Fermi liquid, the few excitations above the ground
state can interact strongly with all the other electrons
present in the ground state. The effect of such interactions is
strong and lead to a strong change of the parameters entering the low-energy effective theory,
compared to free electrons. The combined object behaves as a long-lived
particle, named a quasiparticle: its
characteristics depend strongly on interactions. However
the scattering of the quasiparticles is blocked by the Pauli
principle leaving a very small phase space for scattering. The
lifetime of the quasiparticles is thus extremely large.
Furthermore, a low-energy excited state involves a low-density of excited quasiparticles.
This is the essence of the Fermi liquid theory.}
\end{figure}

In the presence of interactions this will not be the case and the added particle
interacts with the existing particles in the ground state. For
example for repulsive interactions one can expect that this
excitation repels other electrons in its vicinity. This is
schematically represented in \fref{fig:fermiliquid}.
On the other hand if one is at low temperature (compared to the
Fermi energy) there are very few such excitations and one
can thus neglect their mutual interactions (or rather treat them in a mean-field manner).
This defines a new composite
object (fermion or hole surrounded by its own polarization
cloud). This complex object essentially behaves as a particle,
with the same quantum numbers (charge, spin) than the original
fermion, albeit with renormalized parameters, for example its
mass. This image thus strongly suggests that even in the
presence of interactions good elementary excitations looking like free
particles, still exists. These particle resemble free fermions
but with a renormalized excitation energy $\xi_{\vk}$, different from $\xi^0_{\vk}$.

Since our system is gapless, the excitation energy $\xi_{\vk}$ must vanish on a certain surface
in momentum-space. This defines the Fermi surface {\it of the interacting system}. Note that, in
contrast to the free system, we have established no connection between the FS and the ground-state
wave-function: we have been referring only to excitation energies. Under quite general assumptions,
a remarkable property does hold however (even when quasiparticles do not exist in the Landau sense) :
the momentum-space volume encompassed by the FS is identical to that of the free system
({\it Luttinger theorem}), and hence entirely fixed by the number of particles. The {\it shape of the FS}, however, can be changed
by interactions (except in the continuum, where it is a sphere specified only by its radius $k_F$ given by
$2\times \frac{4}{3}\pi k_F^3 /h^3 = \frac{N}{\Omega}$).

The dispersion relation $\xi_{\vk}$ specifying the quasiparticles excitation energy can be expanded around a given
point $\vk_F$ of the FS as:
\begin{equation}
\xi_{\vk} = \nabla_{\vk}\xi|_{\vk_F}\cdot (\vk-\vk_F) + \cdots
\end{equation}
which defines a renormalized Fermi velocity of the quasiparticles at this point:
$\vec{v}^*(\vk_F) = \nabla_{\vk}\xi|_{\vk_F}/\hbar$.
In the continuum, the Fermi velocity is identical on all points of the spherical FS (by isotropy) and it is customary
to define the {\it effective mass} of quasiparticles by analogy to the free-particle dispersion relation
$\xi^0_{\vk} = \hbar^2 k^2/2m-\hbar^2 k_F^2/2m \sim \hbar k_F (k-k_F)/m + \cdots$ :
\begin{equation}
\xi_{\vk} = \frac{\hbar k_F}{m^*} (k-k_F) + \cdots
\end{equation}
The effective mass controls the low-temperature behavior of the specific heat of a Landau Fermi-liquid, which
has the same linear-temperature dependence than a free fermion gas:
\begin{equation}\label{eq:spheat_FL}
\frac{(C/T)}{(C/T)_0} = \frac {\dos _{\mathrm{QP}}} {\dos}|_{\mathrm{FS}}
= \frac{m^*}{m}
\end{equation}
While the low-temperature compressibility and susceptibility are again constant:
\begin{equation}\label{eq:chi_FL}
\frac{\kappa}{\kappa_0}=\frac{m^*/m}{1+F_0^s}\,\,\,,\,\,\,
\frac{\chi}{\chi_0}=\frac{m^*/m}{1+F_0^a}
\end{equation}
but involve distinct renormalizations, which parametrize the effective interaction between quasiparticles in the
low-energy effective theory (Landau parameters $F_0^s$ and $F_0^a$).
Note that expressions (\ref{eq:spheat_FL},\ref{eq:chi_FL}) have been written for the isotropic case
(continuum). In a anisotropic lattice case, the Landau parameters acquire  a more complex angular dependence and
these expressions have less predictive power.

\subsubsection{The $N\pm 1$-particle problem: Green's function, spectral function and self-energy}
\begin{figure}
 \begin{center}
 \includegraphics[width=0.5\linewidth]{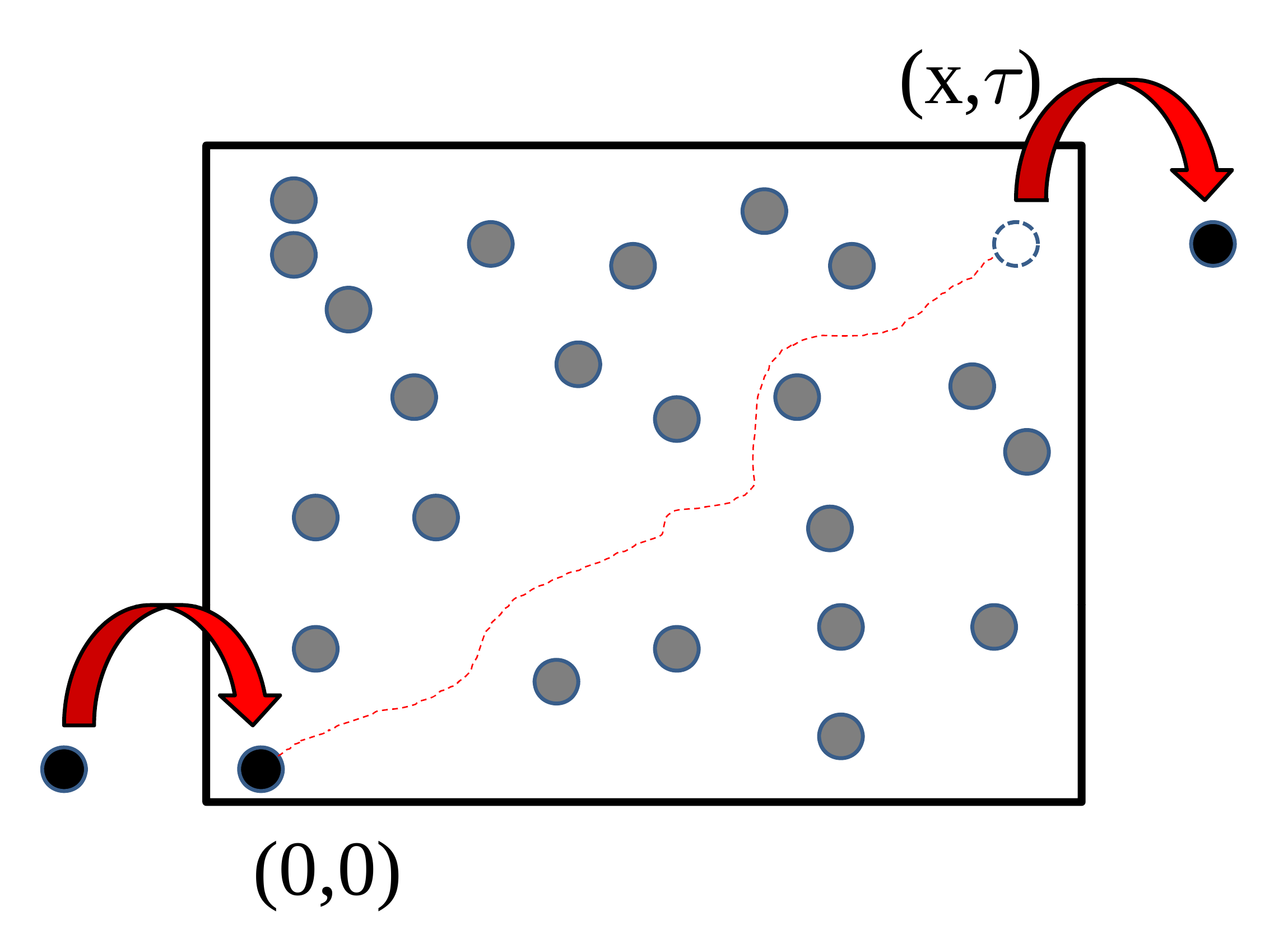}
 \end{center}
 \caption{\label{fig:gedankensingle}
 A {\it gedanken} experiment probing the excitations of an interacting system. A particle is injected from outside
 at point $\vr$ and time $t=0$. It propagates through the system and partially decays by creating excitations.
 The particle is removed at position $\vrp$ and time $t$. The amplitude of this process contains a wealth of
 information about single-particle excitations of the system, encapsulated in the spectral function $A(\vk,\omega)$.
 ``Resemblance'' of the final state with the original bare
 particle signals the existence of long-lived quasiparticle excitations.
 }
\end{figure}

Of course, the above are qualitative ideas. Let us now turn to a more formal treatment. To this aim, we imagine
probing the excitations of the $N$-particle system by removing a fermion from the system (at, say, point $\vr$ and time $0$) and sending it to the
outside vacuum, or by injecting a fermion from the outside (\fref{fig:gedankensingle}).
Starting from the (complicated) ground-state $|\Psi_0\rangle$
of the $N$-particle system, we thus prepare the wave-function (for simplicity, we consider a one-band system and
one spin component, so that we omit both indices $\nu,\sigma$):
\begin{equation}\label{eq:state1}
|\Psi(\vr,0)\rangle \equiv
\psi^\dagger(\vr,t) |\Psi_0\rangle = \sum_{\vk}
\phi^*_{\vk}(\vr)\, c^\dagger_{\vk} |\Psi_0\rangle
\end{equation}
In a non-interacting system, the wave functions on the right-hand side are eigenstates of the system with
$N+1$ particles. But this is not the case in the presence of interactions. Hence, in order to understand the time-evolution of this
wave-function, we expand it onto the {\it exact eigenstates} of the $N+1$-particle system:
\begin{equation}
|\Psi(\vr,0)\rangle = \sum_{\vk} \phi^*_{\vk}(\vr) \sum_A \langle\Psi_A|c^\dagger_{\vk} |\Psi_0\rangle |\Psi_A\rangle
\end{equation}
We then time-evolve this state with the evolution operator $\exp{-\frac{i}{\hbar}(\hat{H}-\mu\hat{N})}$, with $\hat{H}$ the
full (interacting) many-body hamiltonian, yielding the state $|\Psi(\vr,t)\rangle$. We compare this state with the state obtained by
injecting a fermion into the ground-state directly at time $t$, and point $\vrp$, $\psi^\dagger(\vrp)|\Psi_0(t)\rangle$, by forming the
overlap of the two states $\langle\Psi_0(t)\psi(\vrp)|\Psi(\vr,t)\rangle$.
This overlap can also be viewed as the amplitude for injecting a particle at $\vr$ at time $0$ and removing it
at $\vrp$ at time $t$ (\fref{fig:gedankensingle}).
It reads:
\begin{equation}
\langle\Psi_0| \psi(\vrp,t)\psi^\dagger(\vr,0)|\Psi_0\rangle =
\sum_{\vk} \phi_{\vk}(\vrp)\phi^*_{\vk}(\vr)
\sum_A |\langle\Psi_A|c^\dagger_{\vk} |\Psi_0\rangle|^2\,
e^{-\frac{i}{\hbar}[E_A-(E_0+\mu)]t}
\end{equation}
Note that $\mu=\partial E_0/\partial N$ and thus the ground-state energy of the $N+1$-particle
system is (for a large gapless system) $E_0^{N+1}\simeq E_0+\mu$. Hence, the frequencies appearing in the
time-evolution on the r.h.s involve the excitation energies $\hbar\omega=E_A-E^{N+1}_0>0$ of the $N+1$-particle system.
We see that this {\it gedanken} experiment provides information on the excitation of the system, more precisely on the
excited states to which $|\Psi_0\rangle$ couples by injecting a particle.
It is very useful to introduce the {\it one-particle spectral function}, which condenses all this information,
and is defined (at $T=0$) as:
\begin{eqnarray}\nonumber
A(\vk,\omega)&\equiv&\sum_{A(N+1)} |\langle\Psi_A|c^\dagger_{\vk} |\Psi_0\rangle|^2\,
\delta\left[\omega-\frac{1}{\hbar}(E_A-E_0-\mu)\right]\,\,\,,\,\,\,(\omega>0)
\\
&\equiv&\sum_{B(N-1)} |\langle\Psi_B|c_{\vk} |\Psi_0\rangle|^2\,
\delta\left[\omega-\frac{1}{\hbar}(E_0-\mu-E_B)\right]\,\,\,,\,\,\,(\omega<0)
\label{eq:def_spectral}
\end{eqnarray}
It is easily checked that the spectral function is normalized over frequencies for each value of the momentum:
$\int_{-\infty}^{+\infty} A(\vk,\omega) d\omega =1$ and that the quasi-momentum distribution of
particles in the ground-state is given by:
$N(\vk)\equiv\langle\Psi_0|c^\dagger_{\vk}c_{\vk}|\Psi_0\rangle=\int_{-\infty}^{0} A(\vk,\omega) d\omega$.
The spectral function can also be related to the Fourier transform of the {\it retarded Green's function}, defined as:
\begin{equation} \label{eq:greendef}
 G(\vk,t) = - i \step(t) \langle\Psi_0| \acomm{c_{\vk}(t)}{\hc{c}_{\vk}(0)}|\Psi_0 \rangle
\end{equation}
by:
\begin{equation}
A(\vk,\omega) = - \frac{1}{\pi}\,\mathrm{Im} G(\vk,\omega)
\end{equation}

\begin{figure}
\begin{center}
 \includegraphics[width=0.6\linewidth]{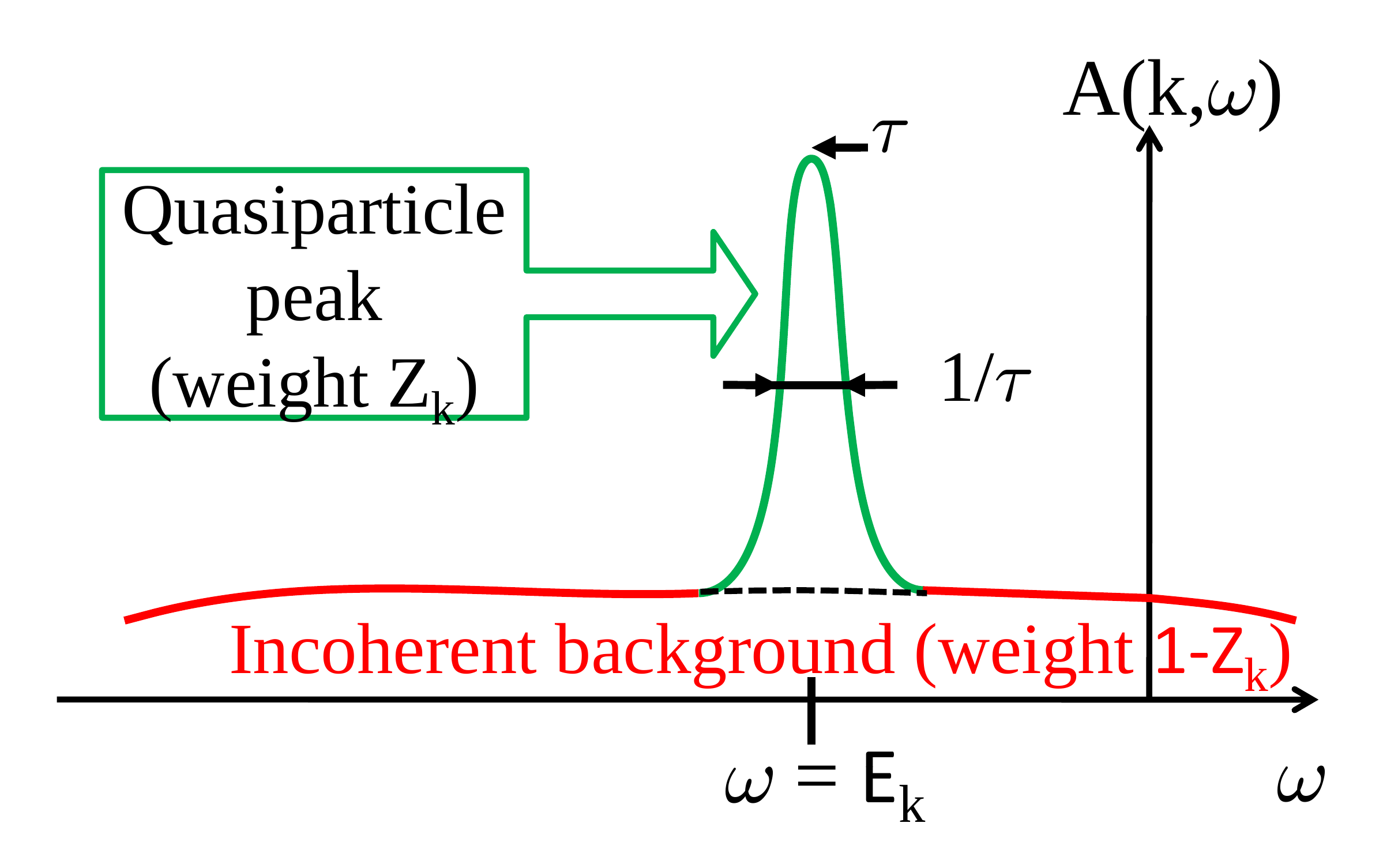}
\end{center}
\caption{\label{fig:spectralqp} A cartoon of the spectral function for interacting particles.
One can recognize several features. There is a continuous background of excitations of total weight
$1-Z_{\vk}$. This part of the spectrum corresponds to incoherent excitations which are not associated with quasiparticles.
In addition to this continuous background, there is a quasiparticle peak. The
total weight of the peak $Z_{\vk}$ is determined by the real part of the self energy.
The center of the peak is at a frequency $\xi_{\vk}$, the renormalized quasiparticle dispersion.
The quasiparticle peak has a lorentzian lineshape that
reflects the finite lifetime of the quasiparticles, inversely proportional to the imaginary part of the self energy.}
\end{figure}

In Fig.~\ref{fig:spectralqp}, we display a cartoon of the spectral function of a Fermi liquid. For momenta not too far
from the Fermi surface, it can be decomposed into two spectral features: a narrow peak corresponding to
quasiparticle excitations, and a broad continuum corresponding to incoherent excitations. The narrow peak is centered at
the excitation frequency $\omega=E_{\vk}-\mu=\xi_{\vk}$ corresponding to the quasiparticle dispersion. It has a spectral weight
$Z_{\vk}\leq 1$ and its width $\gamk=\hbar/\tau_{\vk}$ corresponds to the inverse lifetime of quasiparticle
excitations and can be approximated by a Lorentzian:
\begin{equation}
A_{\mathrm{QP}}(\vk,\omega) \simeq Z_{\vk}\, \frac{\gamk/\pi}{(\omega-\xivk)^2+\gamk^2}
\end{equation}
 Correspondingly, the Green's functions can be separated into two components, involving very different time-scales:
\begin{equation}
G(\vk,\omega) \simeq Z_{\vk}\,e^{-t/\tau_{\vk}}\,e^{-i\xivk\,t/\hbar}\,+\,G_{\mathrm{inc}}(\vk,t)
\end{equation}
The notion of quasiparticle excitations makes sense because their lifetime $\tau_{\vk}$ becomes very large as
$\vk$ approaches the Fermi surface, for phase-space reasons detailed in the next section.
As a result, the first term decays very slowly, while the second ``incoherent'' one
decays fast (corresponding to a broad frequency spectrum).

A very useful quantity is the {\it self-energy}, which is a measure of the difference between the
Green's function of the interacting system and that of the free system. It is defined by
(with $\xivok=\eps_{\vk}-\mu$):
\begin{equation}
G(\vk,\omega) = \frac{1}{\omega-\xivok-\Sigma(\vk,\omega)}
\end{equation}
By expanding this expression close to the FS $\vk\simeq \vk_F$ and at low-frequency
$\omega\simeq 0$, we find that the key quantities characterizing quasiparticles can be read-off
from the self-energy.
The FS of the interacting system are formed by the quasi-momenta which satisfy:
\begin{equation}
\eps_{\vec{k}_F}\,+\,\Sigma(\vkF,0)\,=\, \mu
\end{equation}
in which $\mu$ in the r.h.s should be viewed as a function of the particle density $n$, and of course of the
interaction strength.
The quasiparticle spectral weight, dispersion $\xivk=\vec{v}^*(\vkF)\cdot(\vec{k}-\vkF)$,
and inverse lifetime are given by:
\begin{eqnarray}\nonumber
Z_{\vec{k}}\,&=&\,\left[1-\frac{\partial\Sigma^\prime}{\partial\omega}|_{\omega=0}\right]^{-1}\\ \nonumber
\vec{v}^*(\vkF)\,&=&\, Z_{\vkF}\,
\left[\nabla_{\vec{k}}\xivok + \nabla_{\vec{k}}\Sigma^\prime \right]_{\omega=0,\vec{k}=\vkF}
\\
\gamma_{\vec{k}}\,&=&\,Z_{\vec{k}}\,\Sigma^{\prime\prime}(\vk,\omega=\xivk)
\end{eqnarray}
In these expressions, $\Sigma^\prime$ and $\Sigma^{\prime\prime}$ stand for the real and imaginary parts of the
retarded self-energy, respectively. As expected, the inverse quasiparticle lifetime is related to the latter (but also involves the
weight $Z$, which in contrast would not appear in the scattering rate measured from transport or optical conductivity).
For an isotropic system this leads to the following expression for the effective mass:
\begin{equation}
\frac{m}{m^*}\,=\,Z\,\left[1+\frac{m}{\hbar k_F} \frac{\partial\Sigma^\prime}{\partial k}|_{\omega=0,k=k_F} \right]
\end{equation}
Note that the quasiparticle weight is related only to the frequency-dependence of the self-energy, while
the effective mass involves both the frequency and momentum dependence. Only when the self-energy is momentum-independent
(as e.g. in the limit of large dimensionality, or within the dynamical mean-field theory approximation) do we have $m^*/m=1/Z$.

On general grounds, the following phenomena are clear signatures of strong correlations (and need not necessarily
occur together):
\begin{itemize}
\item A small quasiparticle weight $Z$
\item A large effective mass (low $\vec{v}^*_F$)
\item A short quasiparticle lifetime (large $\gamk$).
\end{itemize}

\subsubsection{Lifetime of quasiparticles:phase-space constraints}

\begin{figure}
\begin{center}
 \includegraphics[width=0.6\linewidth]{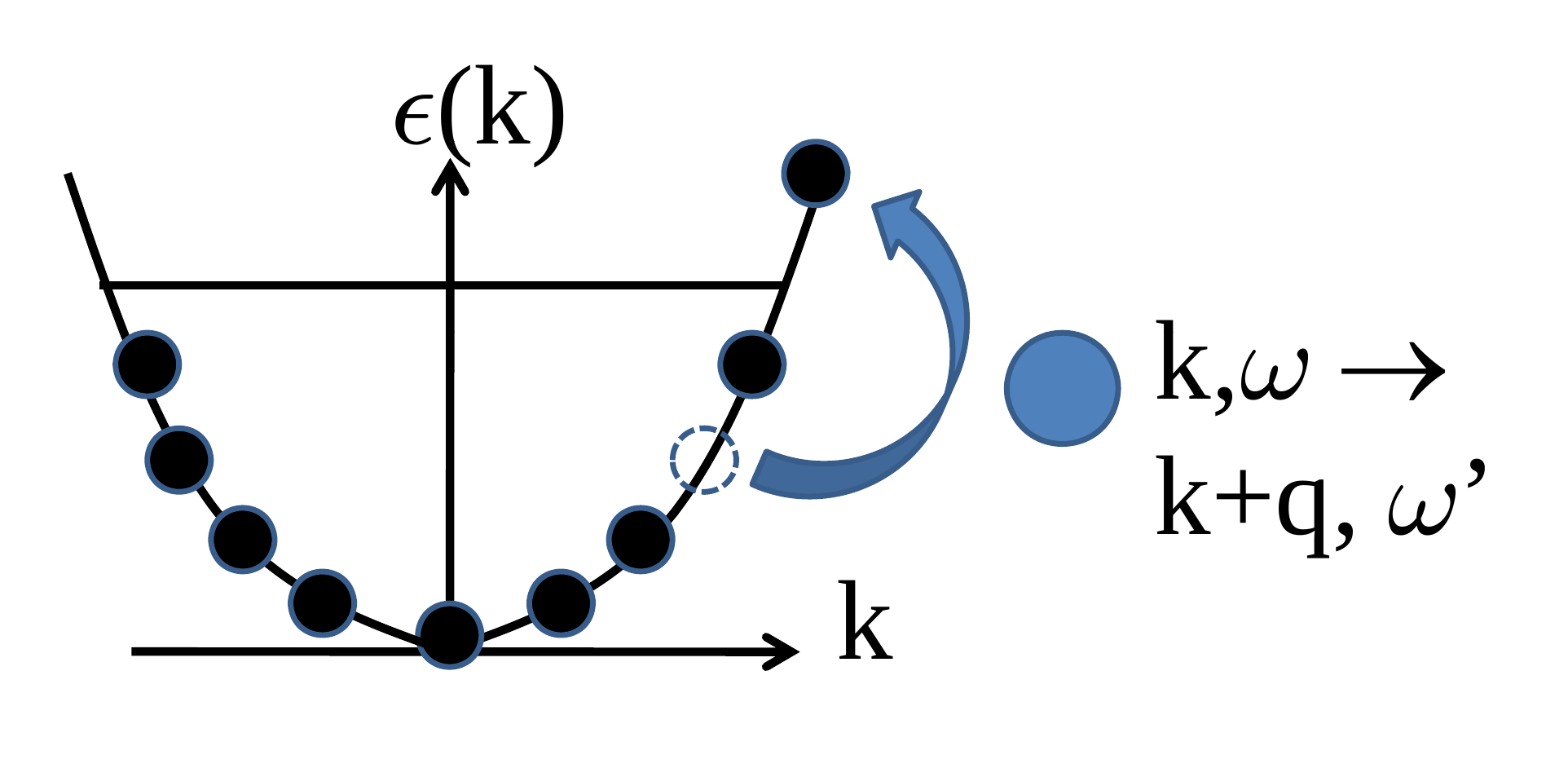}
\end{center}
\caption{\label{fig:parthole} Cartoon of the process giving the lifetime of a particle with energy
$\omega$. The ground-state of the free system has all single-particle states filled below
the Fermi energy $\eps_F$. The excitations are thus particle-hole excitations where a particle is promoted
from below the Fermi level to above the Fermi level. Due to the presence of the sharp Fermi
level, the phase space available for making such a particle-hole excitations is severely restricted.}
\end{figure}

\begin{figure}
\begin{center}
 \includegraphics[width=0.6\columnwidth]{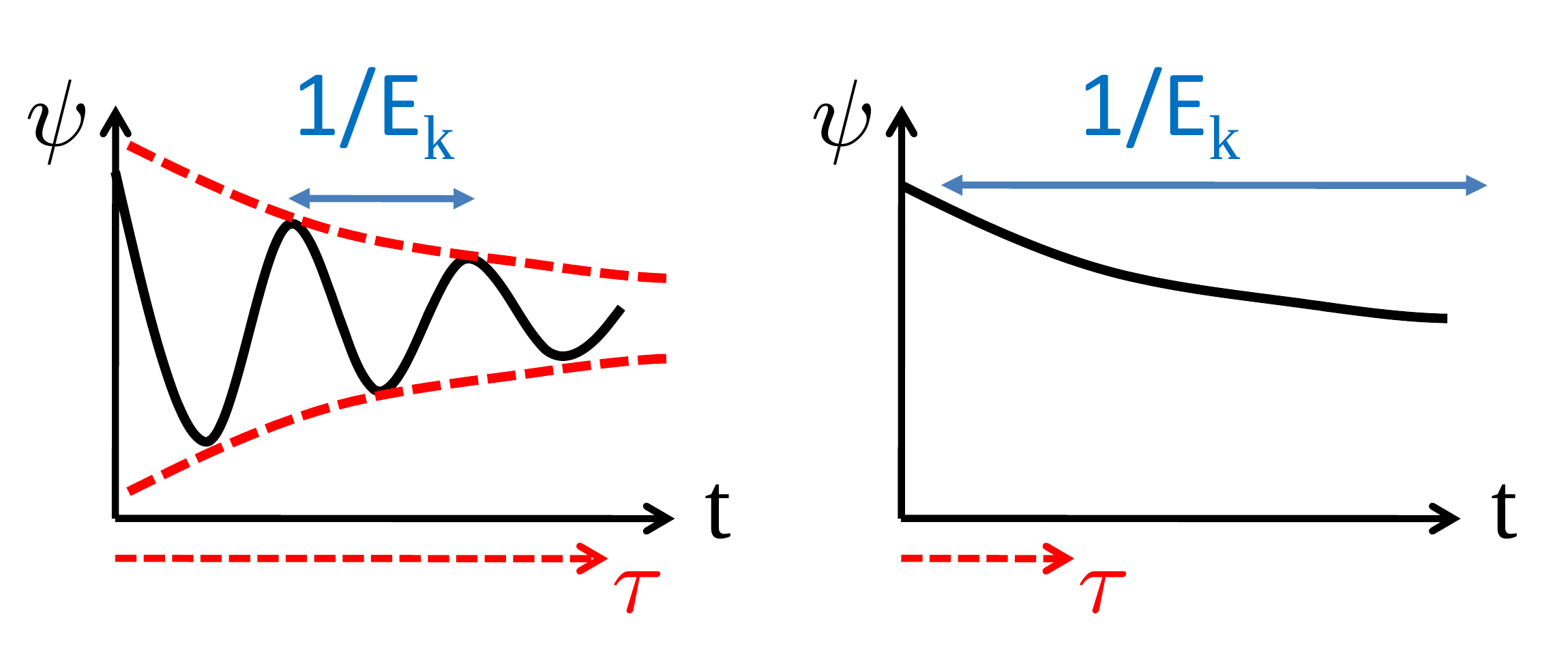}
\end{center}
\caption{\label{fig:amortiss} For particles with an energy $E_k$ and a finite lifetime $\tau$, the energy controls the oscillations in time of the wavefunction.
(Left:) In order to properly identify an excitation as a particle it is mandatory that the wavefunction oscillates
several time before being damped by the lifetime.
(Right:) In contrast, if the damping is too fast, it is impossible to define precisely the
frequency of the oscillations, and thus a precise excitation energy associated with a long-lived quasiparticle.}
\end{figure}

In order to estimate the lifetime of a quasiparticle let us look at  the scattering of a particle from a
state $\vk$ to another state. Let us start from the non
interacting ground state in the spirit of a perturbative
calculation in the interactions. As shown in
\fref{fig:parthole}
a particle coming in the system with an energy $\omega$ and a
momentum $\vk$ can excite a particle-hole excitation, taking a
particle below the Fermi surface with an energy $\omega_1$ and
putting it above the Fermi level with an energy $\omega_2$. The
process is possible if the initial state is occupied and the
final state is empty. One can estimate the probability of
transition using the Fermi golden-rule. The probability of the
transition gives directly the inverse lifetime of the particle,
and thus the imaginary part of the self energy. We will not
care here about the matrix elements of the transition, assuming
that all possible transitions will effectively happen with some
matrix element. The probability of transition is thus the sum
over all possible initial states and final states that respect
the constraints (energy conservation and initial state
occupied, final state empty). Since the external particle has
an energy $\omega$ it can give at most $\omega$ in the
transition. Thus $\omega_2 - \omega_1 \leq \omega$. This
implies also directly that the initial state cannot go deeper
below the Fermi level than $\omega$ otherwise the final state
would also be below the Fermi level and the transition would be
forbidden. The probability of transition is thus
\begin{equation}
 P \propto \int_{-\omega}^{0} d\omega_1 \int_0^{\omega+\omega_1} d\omega_2 = \frac12 \omega^2
\end{equation}
One has thus the remarkable result that because of the
\emph{discontinuity} due to the Fermi surface and the Pauli
principle that only allows the transitions from below to above
the Fermi surface, the inverse lifetime behaves as $\omega^2$.
This has drastic consequences since it means that contrarily to
the naive expectations, when one considers a quasiparticle at
the energy $\omega$, the lifetime grows much \emph{faster} than
the period $\tau_\omega= 2\pi/\omega$ characterizing the oscillations of
the wavefunction (\fref{fig:amortiss}). In fact
\begin{equation}
\frac{\tau_{\mathrm{QP}}}{\tau_\omega} \propto \frac{1}{\omega} \to \infty
\end{equation}
when one approaches the Fermi level. In other words the Landau
quasiparticles become \emph{better and better defined} as one
gets closer to the Fermi level. This is a remarkable result
since it confirms that we can view the system as composed of
single particle excitations that resemble the original
electrons, but with renormalized parameters (effective mass
$m^*$, quasiparticle weight $Z_k$, etc.).

\subsubsection{Probing quasiparticles: photoemission and outcoupling spectroscopies}
Experimental spectroscopic techniques are available, which to a good approximation realize in
practice the {\it gedanken} experiment of Fig.~\ref{fig:gedankensingle}, and hence allow for a direct
imaging of quasiparticle excitations.

\begin{figure}
\begin{center}
 \includegraphics[width=0.6\columnwidth]{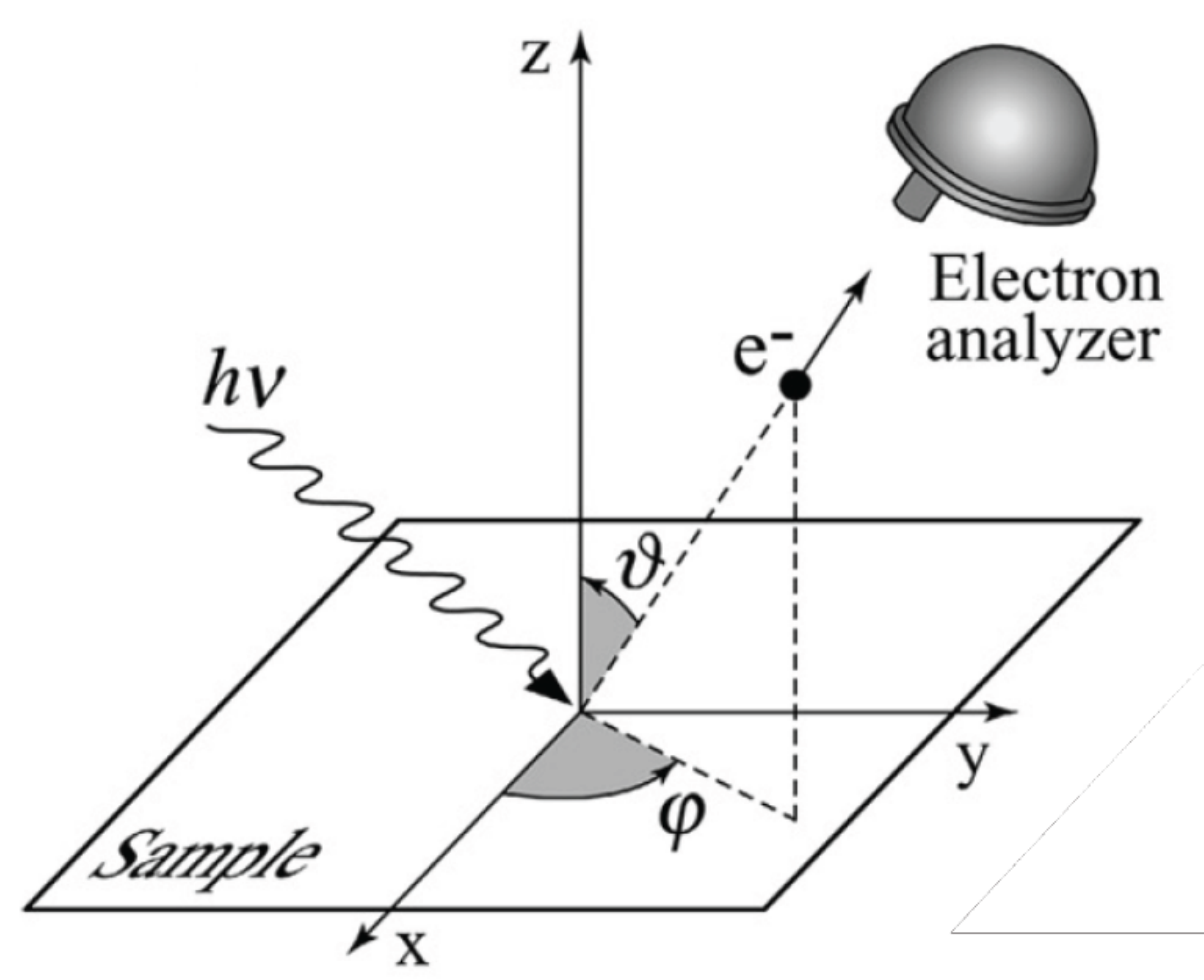}
\end{center}
\caption{\label{fig:photoemission_principle} Basic principle of photoemission spectroscopy.
A photon beam is sent onto the (carefully cleaved) surface of the sample. An electron is extracted (photoelectric effect) and
its energy and momentum is recorded from the electron analyzer. (Adapted from \protect\cite{damascelli_ARPESintro_physscripta_2004}).}
\end{figure}

In the solid-state context, angular-resolved photoemission spectroscopy (ARPES) is a remarkable experimental
method which has undergone considerable development over the past two decades (stimulated to a large extent
by the study of high-$T_c$ superconductors)
see e.g.~\cite{damascelli_rmp_2003,damascelli_ARPESintro_physscripta_2004}.
The basic principle of this method is illustrated on \fref{fig:photoemission_principle}.
Under certain conditions and approximations, the measured photoemission intensity is given by:
\begin{equation}
I(\vk,\omega)\,=\,M(\vk,\omega)\,A(\vk,\omega)\,f(\omega)
\label{eq:photointensity}
\end{equation}
In this expression, $M$ is a matrix element, $A(\vk,\omega)$ is the one-particle spectral function introduced above, and
$f(\omega)$ is the Fermi distribution.
In addition, because of the finite energy resolution, the measured signal is a convolution of (\ref{eq:photointensity}) with a Gaussian
of a certain width. Currently available energy resolutions depend a lot on the incident photon energy: it is typically of
order $50-100$~meV when using X-rays with energies of several hundred eV's at the synchrotron, of order $5-10$~meV for
laboratory sources such as a Helium lamp ($\sim 21$~eV), and as low as a fraction of a meV for the recently developed laser-based
photoemission ($h\nu\sim 6$~eV). These different sources provide complementary information, since there is a trade-off
between bulk vs. surface sensitivity, energy resolution, and the momentum-space constraints limiting the area of the Brillouin
zone that can be probed.

The Fermi function appears in expression (\ref{eq:photointensity})  because this spectroscopy measures the probability
for extracting an electron from the system, and hence {\it mostly probes hole-like excitations}.
Momentum-resolved spectroscopies of particle-like excitations have unfortunately  a much poorer resolution. Scanning tunneling
microscopy (STM), in contrast, does probe both $\omega<0$ and $\omega>0$, but in a momentum-integrated way.

\begin{figure}
\begin{center}
 \includegraphics[width=0.9\columnwidth]{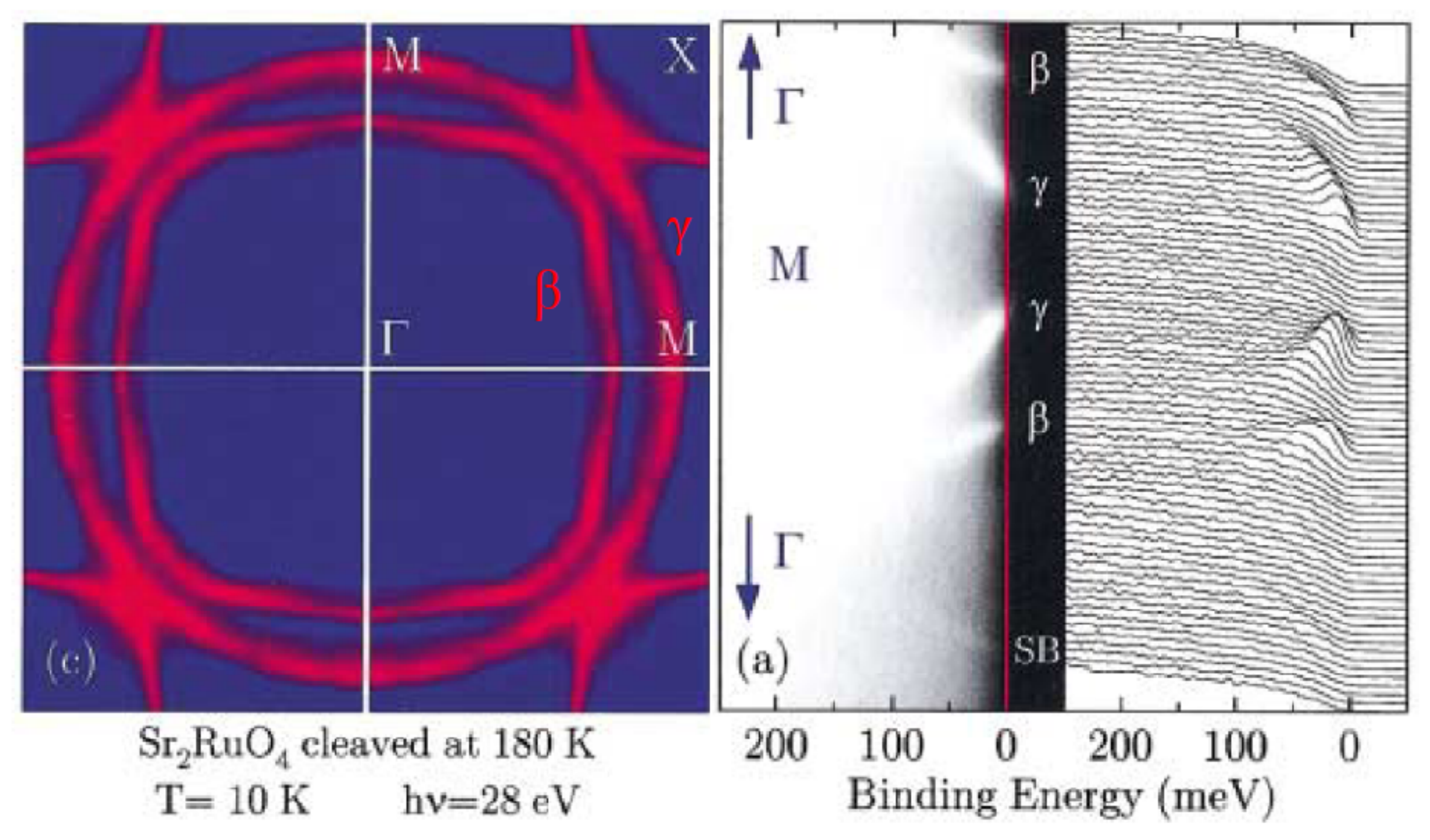}
\end{center}
\caption{\label{fig:sr2ruo4_arpes}ARPES spectroscopy of Sr$_2$RuO$_4$.
Left: ARPES intensity map providing a determination of the Fermi surface, which has three sheets ($\alpha,\beta,\gamma$).
Right: Energy-dependence of the photoemission signal (energy-distribution curves, or EDCs) for several momenta
along the $\Gamma$-$M$-$\Gamma$ direction in the Brillouin zone. Clear quasiparticle peaks are seen when approaching the
FS crossing of the $\beta$- and $\gamma$-sheets. After \protect\cite{damascelli_sr2ruo4_prl_2000}.}
\end{figure}

As an example, \fref{fig:sr2ruo4_arpes} displays ARPES measurements on Sr$_2$RuO$_4$, a two-dimensional
transition-metal oxide with strong electronic correlations. This material has a three-sheeted FS, which can be beautifully
imaged with ARPES (as well as with other techniques, such as quantum oscillations in a magnetic field, with good agreement
between these two determinations of the FS). On \fref{fig:sr2ruo4_arpes} (right side), the photoemission signal is displayed along
a certain cut ($M-\Gamma$) in momentum-space which reveals quasiparticle peaks corresponding to two of these FS sheets.
For momenta $\vk$ far from the FS, only a broad incoherent signal is seen. With $\vk$ approaching $\vk_F$,  a peak develops revealing the
quasiparticles. When $\vk$ crosses the FS into empty states, the signal disappears because of the Fermi factor.
Careful examination of these spectra show that the quasiparticle peak becomes more narrow as the FS is approached, as
expected from the (Landau) phase-space arguments above.

\begin{figure}
\begin{center}
 \includegraphics[width=\columnwidth]{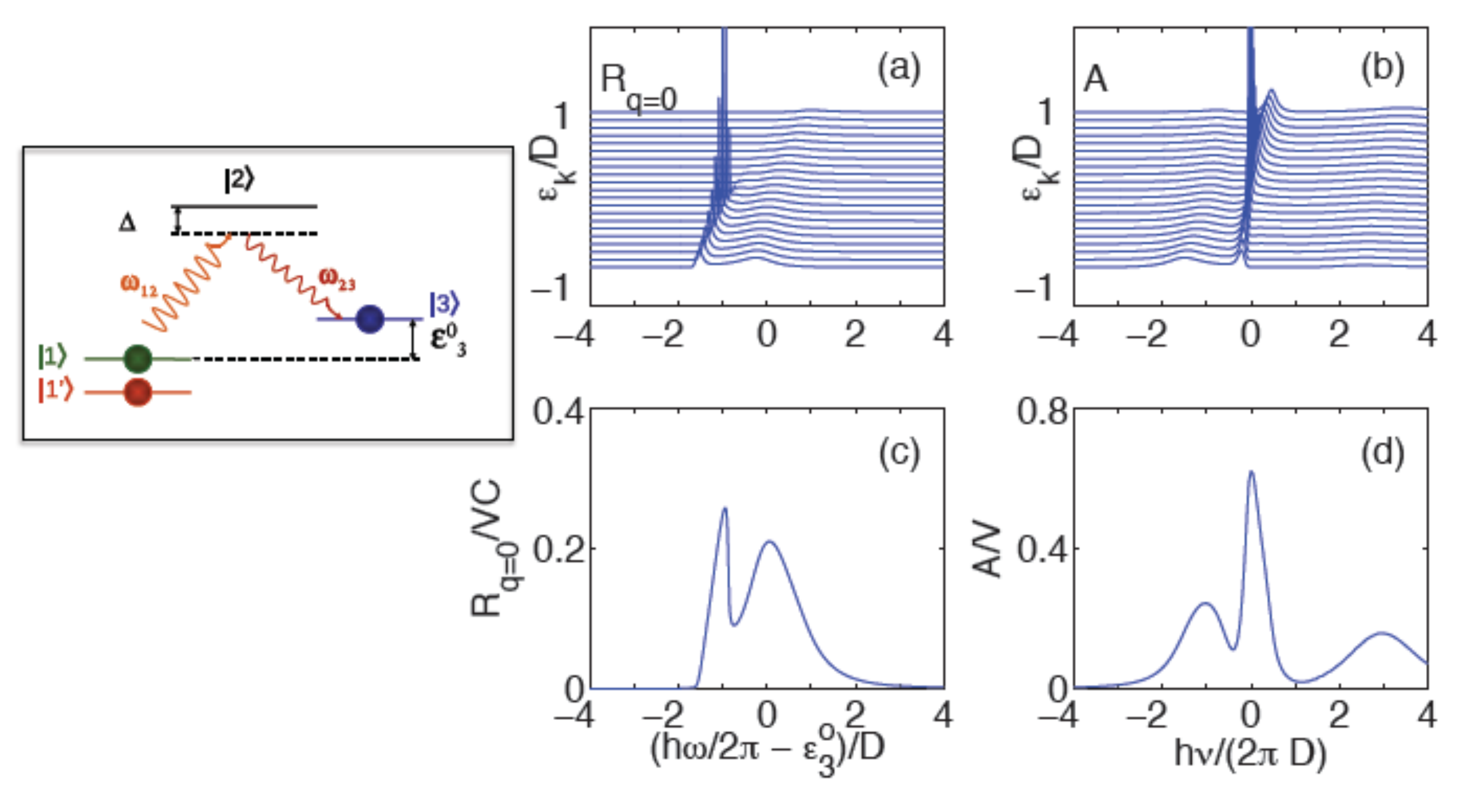}
\end{center}
\caption{\label{fig:outcoupling}
Left: Raman outcoupling process. Atoms in hyperfine state $|1\rangle$ are transferred to state $|3\rangle$ using
two laser beams with frequencies $\omega_{12}$, $\omega_{23}$.
Right: Theoretical spectra for a homogenous Hubbard model in a strongly-correlated regime
$U/W=1.75$ (with $W$ the bandwidth) and total density per site $n=0.85$, as obtained from dynamical mean-field theory.
(a-b): momentum-resolved rf-spectra (a) and spectral function (b).
(c-d): momentum integrated rf-spectrum (c) and spectral function (d).
Three main features are seen on the spectral function: upper and lower Hubbard bands corresponding to incoherent, high-energy,
quasi-local excitations, and a sharp dispersing quasi-particle peak near the Fermi level.
Both the quasiparticle peak and lower Hubbard band are seen in the outcoupled spectra.
From~\protect\cite{bernier_Raman_thermometry_pra_2010}.
}
\end{figure}

In the context of cold atomic gases, an analogue of photoemission spectroscopy
can also be performed~\cite{dao_raman_prl_2007,StewartJin2008},
see also \cite{dao_Raman_long_pra_2009,levin_physrep_2009} and references therein.
The idea there is to trigger the conversion of one of the hyperfine state (say, $|1\rangle$) present in the system of interest
into an out-coupled state $|3\rangle$. This can be achieved either by exciting the system with radio-frequencies (rf spectroscopy)
or by inducing a stimulated Raman transition using two laser beams (\fref{fig:outcoupling}).
A time-of-flight measurement can then be performed which allows for a determination of the initial momentum
$\vk$ of the outcoupled atom.
When studying for example an interacting mixture of two hyperfine species $|1\rangle,|1^\prime\rangle$, one would
ideally like to pick the out-coupled state $|3\rangle$ such that it has only very weak interactions with either
$|1\rangle$ or $|1^\prime\rangle$.
Under such conditions, the production rate of outcoupled atoms is obtained from Fermi's golden-rule as:
\begin{eqnarray}
R_{\vq}(\vk,\omega) &=&\frac{2\pi}{\hbar} \sum_\br W_{\vk}^{\vq}~
|\Omega(\vr)|^2~f(\varepsilon^{\vr}_{3,{\vk}}-\hbar\omega-\mu_0)
\nonumber \\
&& ~~~~~~ \times
A({\vk}-{\vq},\varepsilon^{\vr}_{3,{\vk}} - \mu_0 - \hbar\omega; \mu_{\vr}).
\label{eq:rate}
\end{eqnarray}
In this expression, $\vq=\vk_1-\vk_2$ is the momentum difference between the two laser beams in a Raman setup
(the case of rf-spectroscopy amounts simply to set $\vq=\vec{0}$). $\vec{r}$ denotes a given point in the trap, with
$\mu_{\vr}=\mu_0-V_1(\vr)$ the local chemical potential (using the LDA approximation) and $\mu_0$ the chemical potential
at the center of the trap. $\Omega(\vr)$ is the Rabi frequency of the transition and $\omega=\omega_{12}-\omega_{23}$
(\fref{fig:outcoupling}). $W_{\vk}^{\vq}$ is a matrix element involving Wannier functions in the lattice and
$\varepsilon^{\vr}_{3,{\vk}}=\varepsilon_{3,\vk}+V_3(\vr)$ is the dispersion of the outcoupled atom corrected by its
trapping potential.

The main message of this expression is that, as in photoemission spectroscopy,
measuring the outcoupling rate provides access to the spectral function (provided certain conditions are met).
On \fref{fig:outcoupling}, we display theoretical results for
the fermionic Hubbard model in a strongly correlated regime, which demonstrate
that the key features seen in the spectral function (quasiparticle peak and incoherent lower Hubbard band)
can be detected by rf or Raman outcoupling spectroscopy.

\begin{figure}
\begin{center}
 \includegraphics[width=\columnwidth]{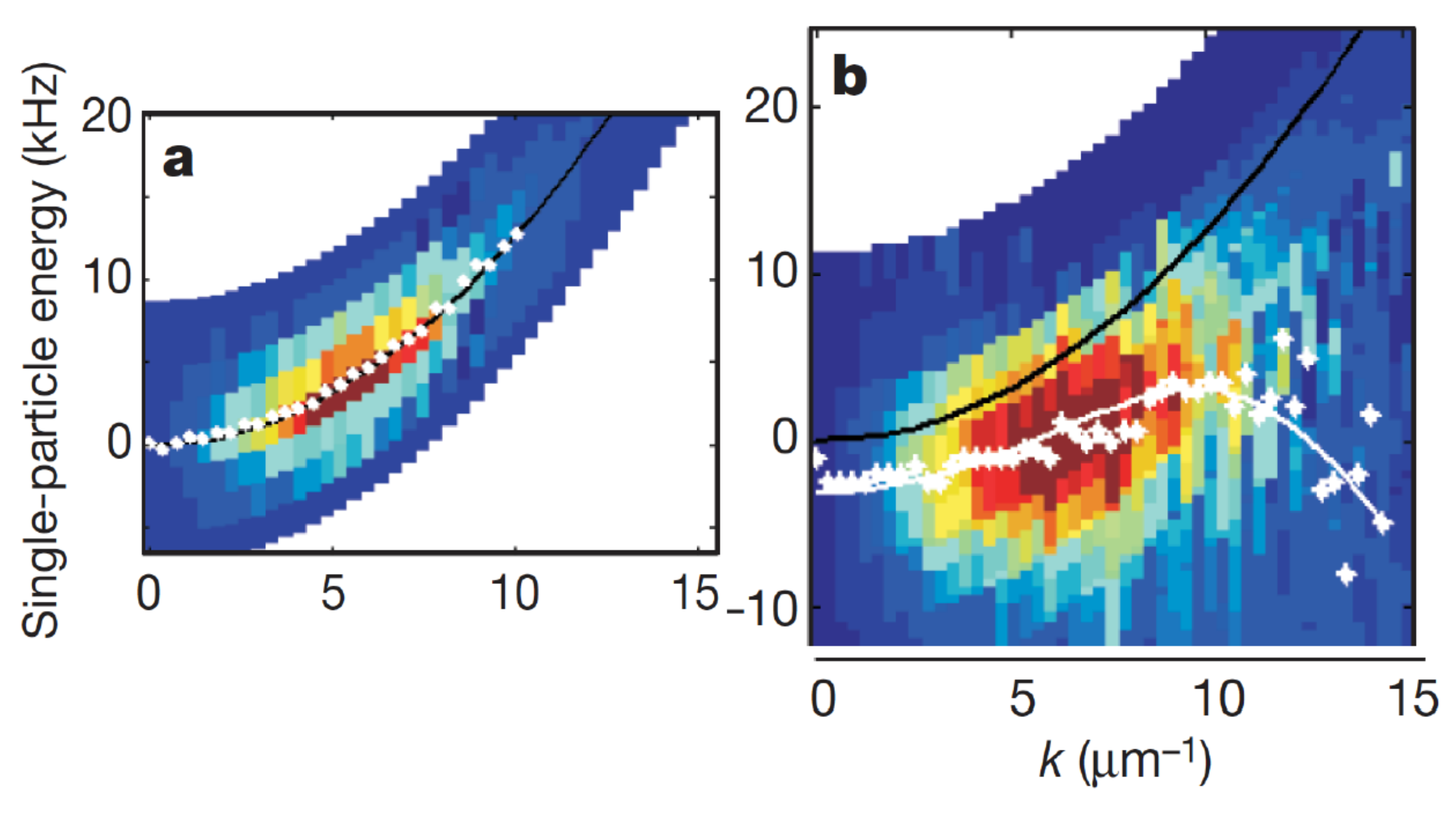}
\end{center}
\caption{\label{fig:arpes_jin}``Photoemission'' spectroscopy of a two-component trapped gas of $^{40}$K fermionic atoms.
Displayed are intensity maps obtained by energy- and momentum- resolved rf-spectroscopy.
(a): Data for a very weakly interacting gas, showing the expected parabolic dispersion of excitations.
(b): Data close to unitarity $1/k_F a\sim 0$. From \protect\cite{StewartJin2008}.
}
\end{figure}

Recently, the JILA group performed a beautiful experiment~\cite{StewartJin2008} in which the single-particle excitations
of a trapped fermionic gas were measured using energy- and momentum- resolved rf-spectroscopy, hence
demonstrating the usefulness of such spectroscopic probes. Some of their results are reproduced on
\fref{fig:arpes_jin}.


\section{Mott transition of fermions: three dimensions}
\label{sec:mottfermion}
%

In this section, we consider again the physics of Mott localization, this time in the context of a two-component gas of
fermions in an optical lattice with a repulsive interaction.
In comparison to the bosonic case considered in Sec.\ref{sec:mottboson}, a major novelty here is the existence
of an internal degree of freedom (the two hyperfine states, or the spin in the case of electrons in a solid).
This leads to the possibility of long-range ``magnetic'' ordering. Even so, it is important to keep in mind that the basic
physics behind the Mott localization of fermions is identical to the bosonic case, at least at strong coupling $U\gg \t$. Namely,
the repulsive interaction makes it unfavorable for particles to hop, resulting in an incompressible state with suppressed density
fluctuations.

\subsection{Homogeneous system: the half-filled Hubbard model}
\begin{figure}
\begin{center}
 \includegraphics[width=\columnwidth]{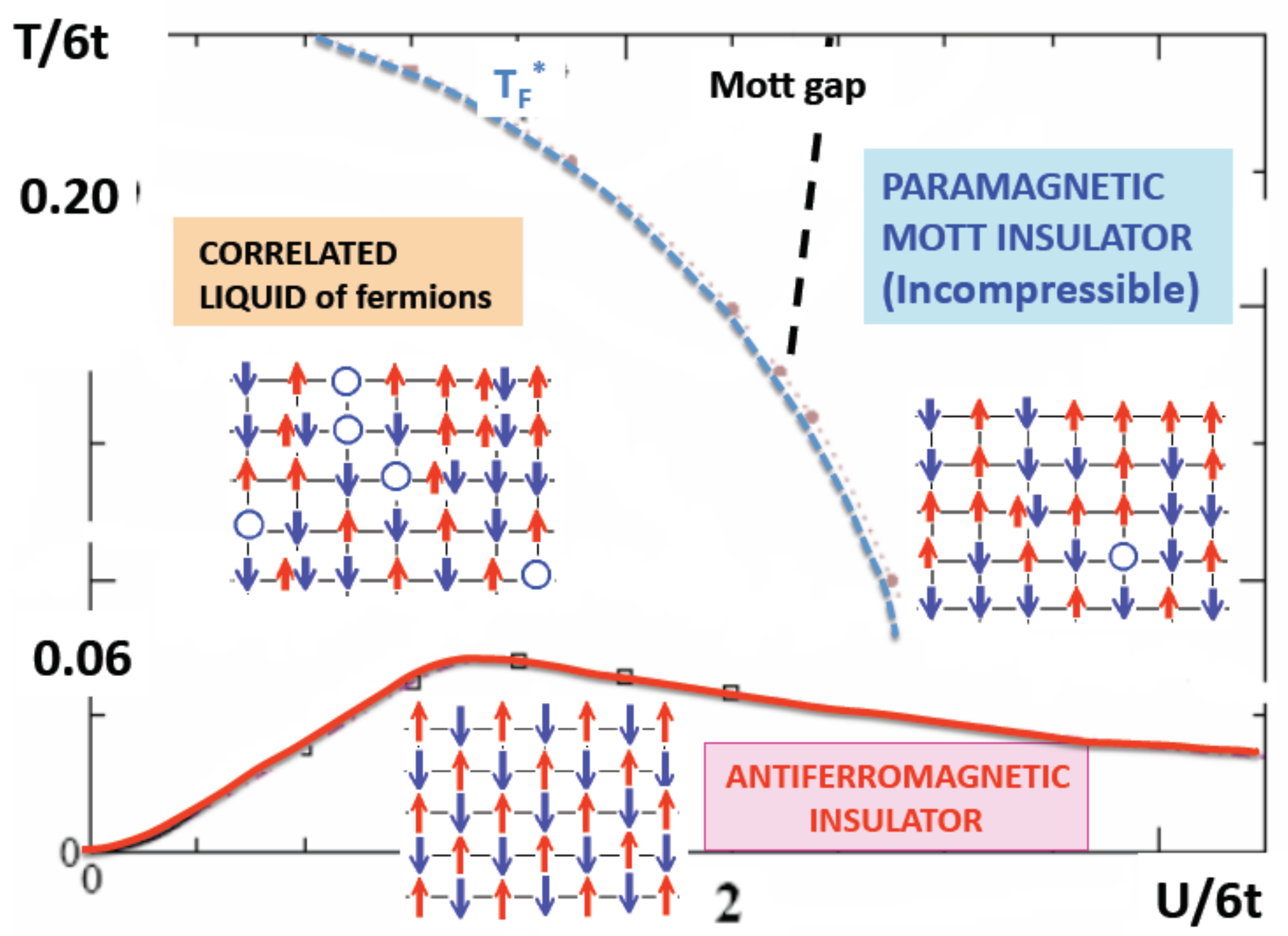}
\end{center}
\caption{Phase diagram of the homogeneous Hubbard model, for a three-dimensional cubic lattice with one-particle
per site on average. The plain line (red) denotes the phase transition into a long-range ordered antiferromagnet
(N\'eel temperature). The long-dashed line (black) denotes the Mott gap: to the right of this line
the paramagnetic phase behaves as an incompressible Mott insulator. The short-dashed line (blue) denotes the
quasiparticle coherence scale. To the right of this line, the paramagnetic phase behaves as an itinerant fermionic liquid
with long-lived quasiparticles.
Typical snapshots of the wave-function in real space are displayed for each regime.
\label{fig:hubbard_phasediag} }
\end{figure}
On \fref{fig:hubbard_phasediag}, we display the phase diagram of the fermionic Hubbard model
for a three-dimensional cubic lattice and a homogeneous density of one particle per site on average
(``half-filled band'').
From the point of view of symmetry breaking and long-range order, there are only two phases.
Below the N\'eel temperature $T_N(U)$ (plain/red line), antiferromagnetic long-range order occurs,
in which spin and translational symmetries are broken. At $T=0$ this phase has a gap and is an
insulator.
The phase $T>T_N$ is a paramagnet (no long-range spin ordering). However, physically important {\it crossovers}
take place within this phase. The short-dashed line in \fref{fig:hubbard_phasediag} denotes the coherence scale of
quasiparticles $T_F^*(U)$. For $T<T_F^*$ (i.e. low-enough temperature and weak-enough coupling)
one has an itinerant fermionic liquid, with long-lived quasiparticles (essentially a Fermi liquid, apart from possible
subtleties associated with perfect nesting).
Another key energy scale is the Mott gap (long-dashed line), which
is of order $\Delta_g\sim U$ at large $U$. For $T<\Delta_g$, one has an essentially incompressible Mott insulator (up to
very rare thermal excitations), with frozen density fluctuations and a high spin entropy, i.e. a localized paramagnet.

Hence, when increasing the strength of the repulsive coupling $U/\t$, one crosses over from a fermionic liquid to an
incompressible localized paramagnet (through an intermediate incoherent state which is a ``bad metal'' or a poor insulator).
In a situation where magnetic long-range order is suppressed (e.g. due to geometrical frustration of the lattice), this crossover
may be replaced by a true phase transition. Because no symmetry breaking distinguishes a metal from an insulator at
finite temperature, this transition is expected to be first-order at $T\neq 0$, similar to a liquid-gas transition.
The precise description of this crossover or transition is not so easy theoretically. Indeed, in contrast to the phase transition between
a superfluid and a Mott insulator of bosons, there is no evident order parameter associated with a static correlation function
which discriminates between a metal and a paramagnetic Mott insulator of fermions.
Possible order parameters are all related to frequency-dependent (dynamical) response functions: for example the
Drude weight associated with the $\omega\rightarrow 0$ component of the ac-conductivity, or the quasiparticle weight $Z$
introduced above and associated with the low-frequency behavior of the one-particle Green's function.
For this reason, a mean-field theory of this crossover or transition must focus on one- or two-particle
response functions. Currently, the most complete approach of this kind is dynamical mean-field theory (DMFT),
which has allowed for many successes in understanding strongly-correlated fermion systems. For brevity, we refer the
reader to review articles for a presentation of this theoretical approach.

\subsubsection{From Mott to Slater} The transition into the antiferromagnetic state deserves some further remarks. At strong coupling $U/\t\gtrsim u^*$,
there is a clear separation of energy scales: the gap $\Delta_g$ ($\sim U$ at large $U$) is much larger than the antiferromagnetic
superexchange $J_{\mathrm{AF}}\sim \t^2/U$ which also controls $T_N\propto J_{\mathrm{AF}}$.  Hence for $T\ll \Delta_g$, density
fluctuations are frozen out, particles are localized into a Mott insulating state and only the spin degrees of freedom are active
which are described by an effective Heisenberg model. In this Mott regime, localization precedes spin ordering which is a
low-energy instability of the insulating paramagnet.
In contrast, at weak coupling $U/\t\ll u^*$, long-range magnetic order
and the blocking of translational degrees of freedom cannot be distinguished: in this regime the opening of a gap
is intimately connected to spin ordering and can be described using a simple spin-density wave mean-field theory (Slater regime).
The characteristic coupling $u^*$ separating these two regimes is also the one at which the crossover from a liquid to
an insulating state takes place in the paramagnetic state. The Slater and Mott-Heisenberg regimes are connected by a smooth crossover.

\subsubsection{From repulsion to attraction}
There  is actually a direct formal analogy between this physics and that of the BCS-BEC crossover in the Hubbard
model with an {\it attractive interaction}. Indeed, on a bipartite lattice (i.e. a lattice made of two sublattice $A$ and $B$) with
nearest-neighbor hopping, one can perform the following symmetry operation:
\begin{equation}
c_{i\uparrow}\,\rightarrow\,\,\widetilde{c}_{i\uparrow}\,\,\,,\,\,\,
c_{i\downarrow}\,\rightarrow\,(-1)^i\,\widetilde{c}^\dagger_{i\downarrow}
\end{equation}
with $(-1)^i=+1$ on the A-sublattice and $=-1$ on the B-sublattice. At half-filling, this symmetry simply changes the sign of
the coupling $U$, hence establishing an exact connection between the two cases. Long-range AF order along the $x$- or $y$-axis
is mapped onto superconducting long-range order, while AF order along the $z$-axis is mapped onto a charge-density wave state
in which pairs reside preferentially on one sublattice. The two types of ordering are degenerate at half-filling in the attractive case.
The BCS regime of the attractive case maps onto the Slater regime of the repulsive one, and the BEC regime onto
the Mott-Heisenberg one.
In fact, the symmetry also maps the attractive model away from half-filling onto the repulsive model at half-filling
in a uniform magnetic field \cite{ho_attractive_hubbard}.

\subsection{Trapped system} In the presence of a trapping potential $V(\vr)=V_t\,(r/a)^2$, the local density
changes as one moves away from the trap center, so that different phases can coexist in the system.
When the trap potential varies slowly, the density profile is accurately predicted within the
local density approximation (LDA), which relates the local state of the system to that of the homogeneous
system with a chemical potential $\mu(\vr)=\mu_0-V(\vr)$, so that the local density reads:
$n(\vr)=n_{\mathrm{hom}}[\mu=\mu_0-V(\vr)]$.
Furthermore, for a large system, one can replace the summation over lattice sites by an integral
over the chemical potential, so that the relation between the total particle number $N$ and the chemical
potential at the center of the trap reads:
\begin{equation}
\rho\,\equiv\,N \left(\frac{V_t}{6\t}\right)^{3/2}
\,=\,2\pi\int_{-\infty}^{\omu_0} d\omu (\omu_0-\omu)^{1/2}\,n_{\mathrm{hom}}[\omu]
\end{equation}
where $\omu$ is the chemical potential normalized to the half-bandwidth of the lattice
$\omu=\mu/(6\t)$.
From this expression, we see that the state diagram of the system can be discussed in terms of the
scaled particle number $\rho=N\,(V_t/6\t)^{3/2}$: increasing the number of particles or compressing the system by
increasing $V_t$ accordingly has the same effect.

\begin{figure}
\begin{center}
 \includegraphics[width=\columnwidth]{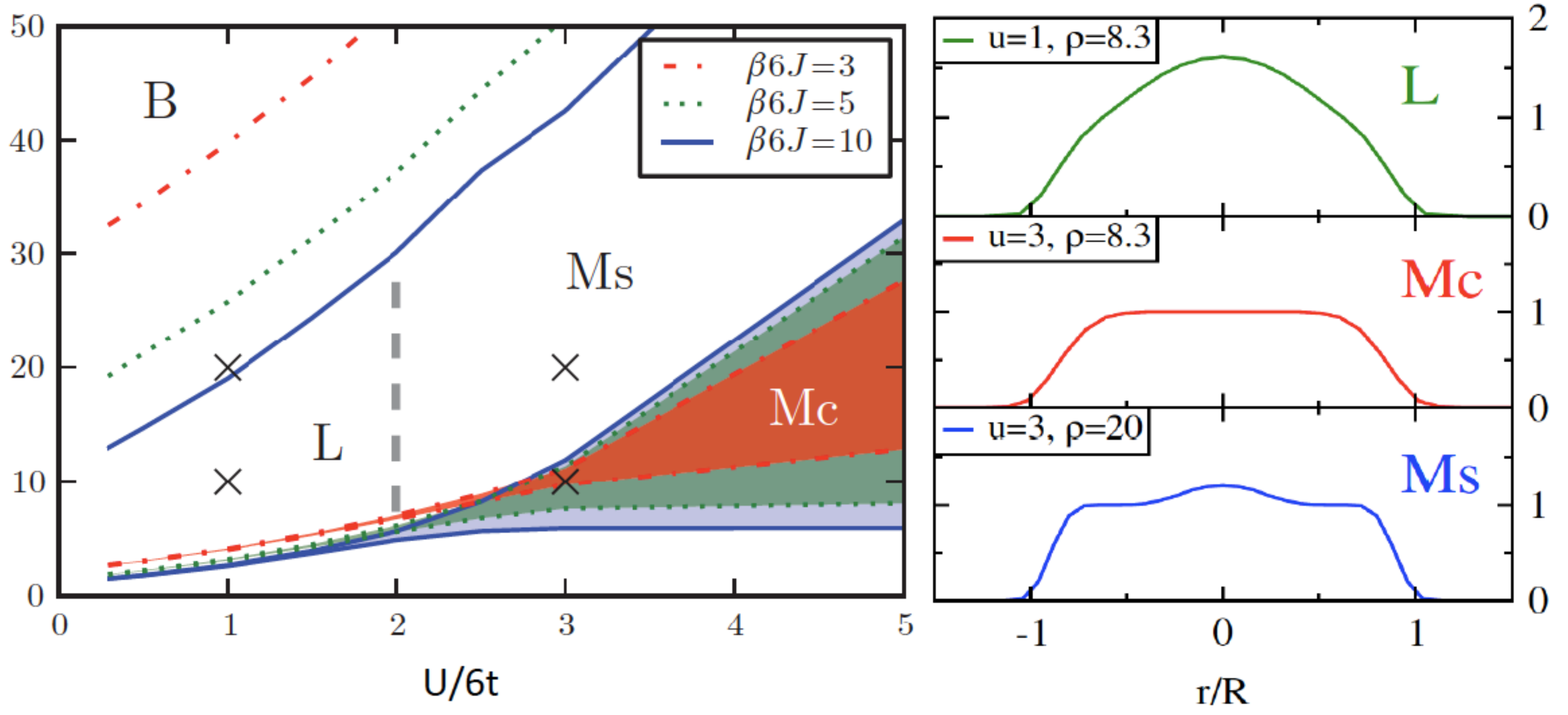}
\end{center}
\caption{\label{fig:mott_statediagram}
State diagram of two-component repulsive fermions in a cubic
optical lattice with parabolic confinement, for different
    temperatures $\beta=1/k_BT$ (DMFT). The four characteristic regimes (see text) are labeled
    by: B (band insulator in the center of the trap), Mc (Mott insulator in the
    center of the trap, shaded areas), Ms (shell of Mott insulator away from the center) and
    L (liquid state).
    For each temperature the (crossover) lines indicate, from bottom to top,
    the $\rho$ values at which the central density takes the values 0.995,
    1.005 and 1.995. The gray dashed line marks
    the crossover from the liquid to the Mott state with increasing
    interaction.
    The crosses indicate points at which the density
    profiles are plotted (right). After ~\protect\cite{DeLeoParcollet2008,deleo_thermodynamics_pra_2011}.
}
\end{figure}

On \fref{fig:mott_statediagram}, we display the state diagram of a two-component fermionic gas confined to
a cubic optical lattice in a harmonic trap, as a function of $\rho$ and interaction strength $u=U/6\t$. This
state diagram was obtained~\cite{DeLeoParcollet2008,deleo_thermodynamics_pra_2011} using DMFT
calculations for the homogeneous Hubbard model.
Different temperatures in the currently accessible range are considered. At still lower temperature
(not displayed on \fref{fig:mott_statediagram}),
antiferromagnetic long-range order will occur in the regimes with a commensurate Mott plateau, as also
discussed below.
The state diagram displays four characteristic regimes (labeled L, B, Mc and Ms).
Three of them are illustrated by the corresponding density profiles $n(r)$ calculated
at representative points.
For low interaction strength (regime `L') the density profile adjusts
to the trapping profile and the system remains a Fermi liquid everywhere in the trap.
For very large values of the scaled particle number $\rho$, a band insulator with $n=2$ forms in the
center of the trap (regime `B').
The pinning to $n=2$, and hence the band insulator, is destroyed by
increasing the temperature.
For larger interaction strength (regime `Mc') a Mott-insulating region appears in the center of the trap,
in which the density is pinned to $n=1$ particle per site.
Close to the boundary of the trap, the Mott insulating region is surrounded by a liquid region.
Increasing the number of atoms in the trap at large interaction strength can increase the
pressure exerted on the atoms, and can cause the occurrence of a liquid region with
filling larger than one in the center, surrounded by a shell of Mott insulator with $n=1$
(regime `Ms').

\begin{figure}
\includegraphics[width=0.55\columnwidth]{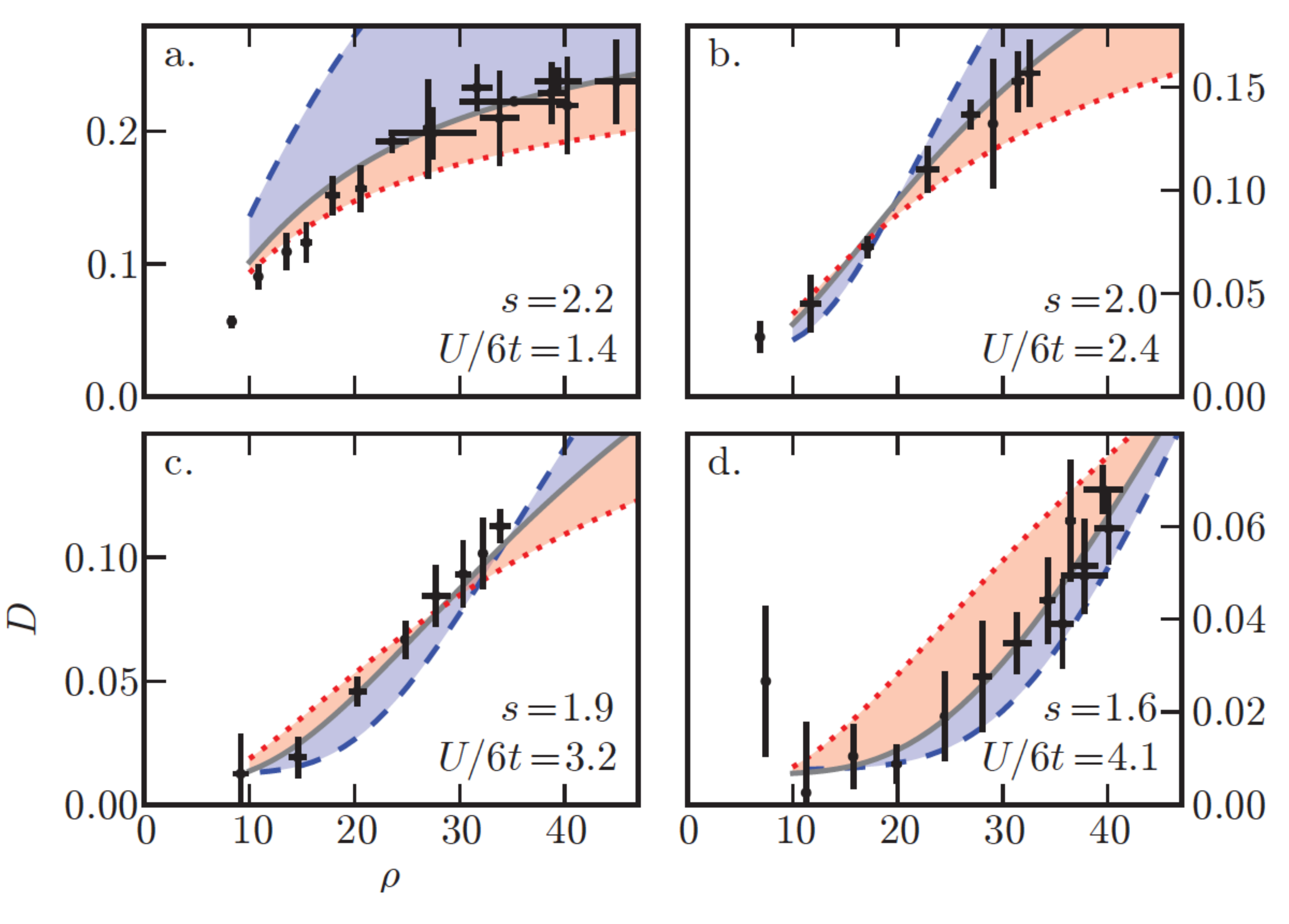}
\hfill
\includegraphics[width=0.55\columnwidth]{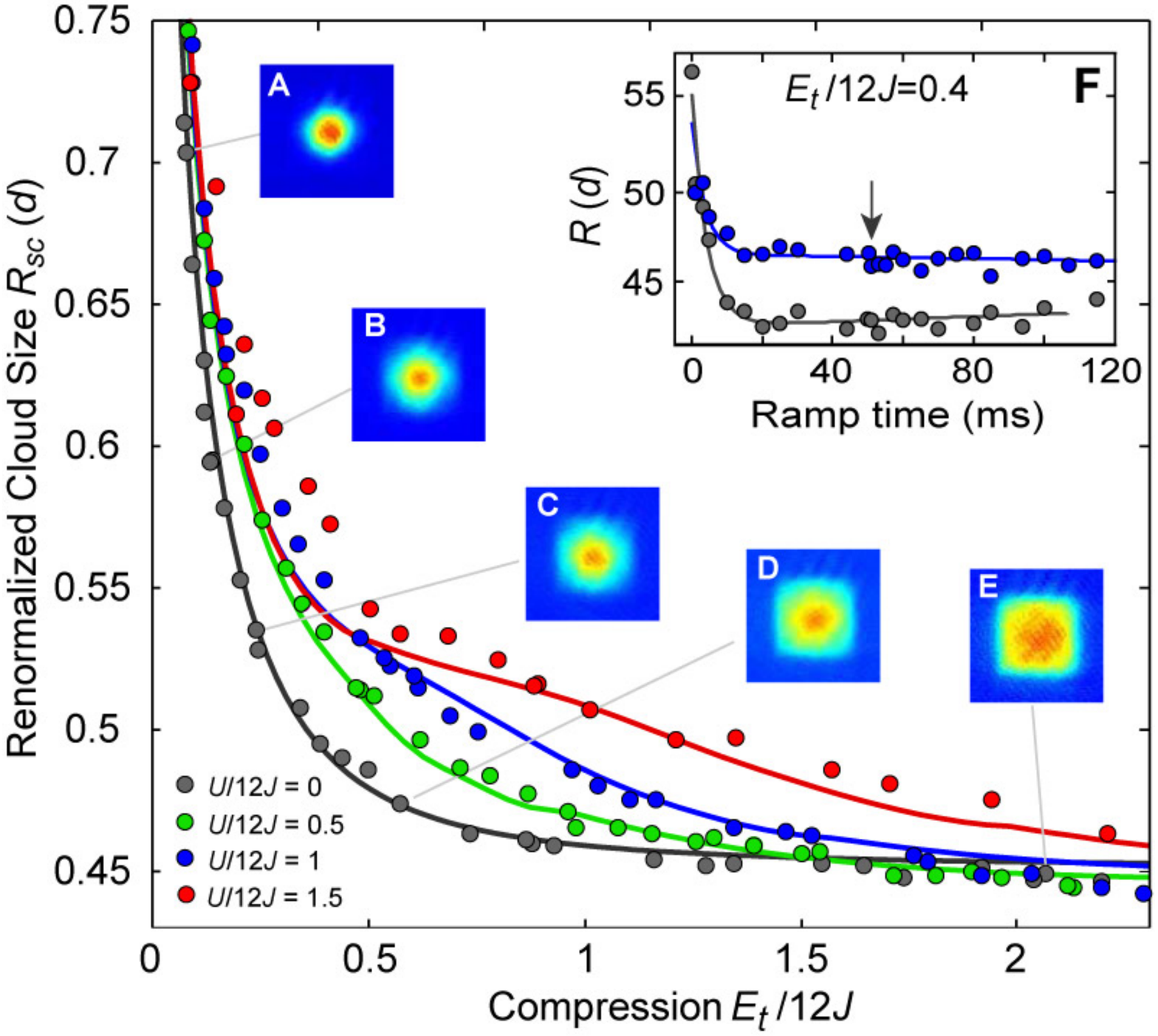}
\caption{Experiments on cold fermionic atoms with repulsive interactions in a three-dimensional optical lattice, revealing the
crossover into a Mott insulating regime.
{\bf Left)}  Double occupancy: experiment versus theory. Points and error
    bars are the mean and standard deviation of at least three experimental
    runs. The solid curve in each panel is the best fit of the second order
    high-temperature series to the experimental data and yields specific
    entropies of $s=2.2(2), 2.0(5), 1.9(4), 1.6(4)$ for the different
    interactions strengths of $U/6t=1.4(2), 2.4(4), 3.2(5), 4.1(7)$.
    Curves for $s=1.3$ (dashed curve) and $2.5$ (dotted curve) represent
    the interval of specific entropy measured before and after the ramping
    of the lattice. Reproduced from ~\protect\cite{jordens_temperature_prl_2010}.
    {\bf Right)} Cloud sizes versus compression.
    Measured cloud size $R_{sc}$ in a $V_{lat} = 8\,E_r$ deep lattice as a function of the external trapping potential for various interactions $U/12t=0$ (black), $U/12t=0.5$ (green), $U/12t=1$ (blue), $U/12t=1.5$ (red) - in this figure the hopping is designated by $J$. Dots denote single experimental shots, lines the theoretical expectation from DMFT for $T/T_F=0.15$ prior to loading. The insets {\bf (A-E)} show the quasi-momentum distribution of the non-interacting clouds (averaged over several shots). {\bf (F)} Resulting cloud size for different lattice ramp times at $E_t/12t=0.4$ for a non-interacting and an interacting Fermi gas. The arrow marks the ramp time of 50\,ms used in the experiment.
    Reproduced from \protect\cite{SchneiderRosch2008}.}

\label{fig:mottfermion_exp}
\end{figure}

Recently, as displayed on \fref{fig:mottfermion_exp}, experiments have reported the observation of the Mott insulating
region for fermionic atoms~\cite{JoerdensEsslinger2008,SchneiderRosch2008}.
%
\fref{fig:mottfermion_exp} displays a comparison between experimental data and theoretical calculations.
In the left panel~\cite{jordens_temperature_prl_2010},
the measured double occupancy as a function of atom number~\cite{JoerdensEsslinger2008} is
compared to theoretical calculations performed at constant entropy (assuming that turning on the optical lattice
corresponds to an adiabatic process). High-temperature series expansions were actually sufficient for this comparison, with DMFT
yielding identical results. Fitting theory to experiment allows for a determination of the actual value
of the entropy, and ultimately of the temperature attained after the lattice is turned on, for a given particle number.
This analysis reveals that the lowest temperature that was reached in this experiment (at small atom number)
is comparable to the hopping amplitude ($T\sim t$). The regime with very small double occupancy at the two largest values of $U/6\t$ actually
corresponds to the formation of a Mott plateau in the center of the trap. This is more clearly revealed in the
measurement of the cloud size as a function of trap compression (right panel)~\cite{SchneiderRosch2008}, see
also \cite{ScarolaTroyer2009})
as a plateau signalling the onset of an incompressible regime for the largest value of $U$ displayed.
Those experiments provide us with an `analog quantum simulator' validation of theoretical methods for strongly correlated fermions
(such as DMFT or high-temperature series), admittedly still in a rather high-temperature regime.
%
%

\subsubsection{Shaking of the lattice}  \label{sec:fershaking}

For the case of bosons (see \sref{sec:shaking}) one can also probe the physics of the fermionic Mott insulator using the shaking of the optical lattice. This probe is complementary with the other spectroscopy probes discussed in this section. For fermions the major difficulty compared to the scheme exposed in \sref{sec:shaking} is to measure the energy absorbed. Indeed for the bosons this could be done by releasing the trap and looking at the width of the central peak. For the fermions the $n(k)$ is a step and looking at how the step is broadened by the absorbed energy is a difficult proposition given the other sources of broadening. Fortunately one can proceed differently and it was shown \cite{kollath_shake_fermions_DMRG} that a measure of the rate of creation of doubly occupied sites (doublon production rate DPR) would give essentially the same information than the measure of the absorbed energy.
Furthermore the total weight of the peak at the Mott gap $U$ was shown to be directly related to the degree of short range antiferromagnetic correlations in the system, making the shaking probe a useful probe for antiferromagnetic correlations as well. This last property can be easily understood by the same arguments than the ones leading to the superexchange (see \fref{fig:superex}). If two neighbors have parallel spins then the kinetic energy term is blocked and thus the perturbation cannot lead to any absorption or DPR. On the contrary if two neighbors have opposite spins, and thus short range antiferromagnetic order, the transition can take place and absorption of energy of DPR occurs.

The proposal of this new way to probe the system by measuring the DPR was very successful since the the counting of the doubly occupied sites can be done with a great accuracy. This allowed to implement this probe and keep the modulation amplitude to small enough rates that the response stayed in the linear response regime \cite{greif_doublon_shaking_fermions}, greatly simplifying the theoretical analysis of this probe and allowing a much simpler and efficient comparison between theory and experiments. Although the position of the peak is clearly at the Mott gap $\Delta_M$ \cite{kollath_shake_fermions_DMRG} computing the shape of the peak is much more complicated than for the bosons. Indeed as shown in \fref{fig:shaking} for the bosons the doublon and holon were moving in a featureless environment of singly occupied sites. On the contrary for the fermions, these two excitations propagate in an antiferromagnetic background, scrambling the spin environment in the process. In order to compute their propagation it was thus necessary to use approximations of such an antiferromagnetic background. Fortunately such approximations existed in the condensed matter context, and made more efficient by the relatively high temperature present in the cold atomic systems. The simplest version is the so-called retraceable approximation where the holon and doublon simply retrace their steps to go back to their point of origin \cite{sensarma_doublon_shaking_fermions}. More recently a more sophisticated approximation using slave boson techniques allowed to treat both the effects of temperature and the trapping and to provide a very good comparison with the experimental data as shown in \fref{fig:shaking}. Since the shaking amplitude depends on the temperature, this allows to use the shaking as a thermometer as well.

\subsubsection{Reaching the antiferromagnetic state}
While these experiments have evidenced the crossover into a paramagnetic Mott insulator, reaching the phase with
antiferromagnetic long-range order (\fref{fig:hubbard_phasediag}) requires further cooling.
In order to estimate how much further effort is needed, and assuming an adiabatic process, an analysis of
the entropy of that phase in the trap is needed.
As pointed out in \cite{werner_cooling_2005}, an important consideration in this respect is the entropy per site of
the homogeneous half-filled Hubbard model on the N\'eel critical line. This quantity is very small at
small $U/\t$, passes through a shallow maximum for $U/\t\simeq u^*$ (due to additional density fluctuations)
and reaches a finite value $\simeq \ln2/2$ in the strong-coupling Heisenberg
limit~\cite{werner_cooling_2005,wessel_entropy_prb_2010,fuchs_thermodynamics_Neel_prl_2011}.
Note that, in contrast, the N\'eel temperature becomes very small at large $U/\t$, illustrating the importance of
thinking rather in terms of entropy.
In the trap, the entropy of liquid wings (in the Mc regime above) need to be taken into account as well.
Theoretical studies~\cite{DeLeoParcollet2008,fuchs_thermodynamics_Neel_prl_2011}
indicate that, in the favorable case of intermediate coupling, the trapped system must be cooled down to an entropy per atom
of order $s=S/N\simeq 0.66$ in order to reach the antiferromagnetic state in the center of the trap, about
three times smaller than the entropy that was reached in the experiments above.

\begin{figure}
\begin{center}
 \includegraphics[width=\columnwidth]{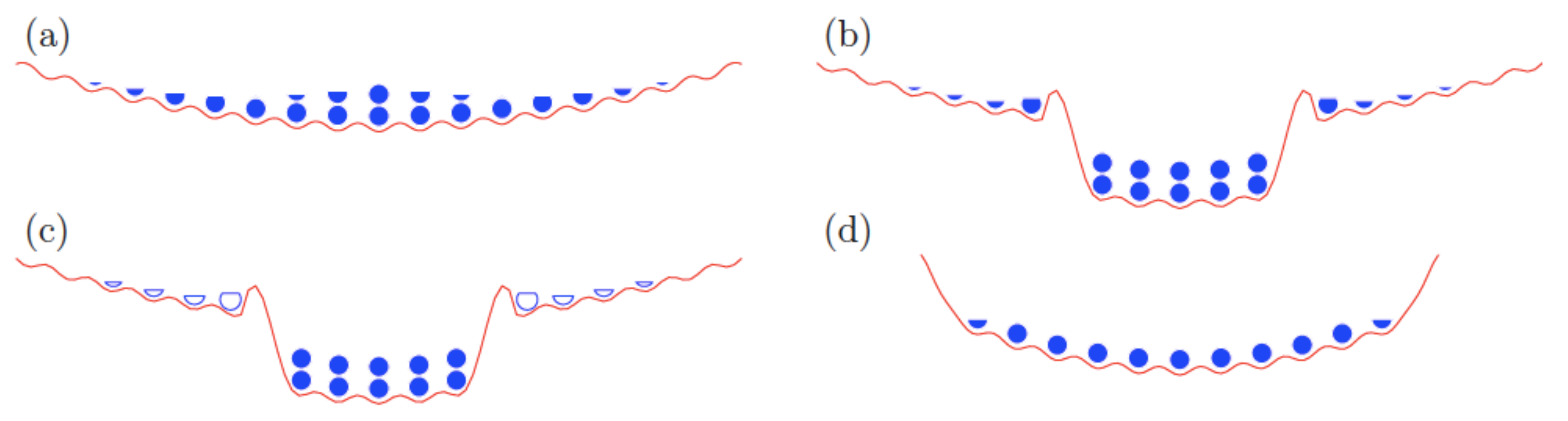}
\end{center}
\caption{\label{fig:cooling}
Cooling scheme by trap shaping, following \protect\cite{bernier_cooling_pra_2009}.
(a) The atoms trapped in a parabolic profile are loaded into an optical lattice.
(b) A band insulator (hence with a very low entropy) is created in a dimple at the center of the trap.
This core region is isolated from the rest of the system, the storage region, by rising potential
barriers. (c) If needed, the storage region is removed from the system. (d) The band insulator
is relaxed adiabatically (hence preserving the low entropy) to the desired quantum phase,
e.g. a Mott insulator by flattening the dimple and turning off or pushing outwards the barriers.
}
\end{figure}

Obviously, cooling further fermionic atoms trapped in an optical lattice is a key current challenge.
Several proposals have been put forward to this effect, e.g. in ~\cite{HoZhou2009} , \cite{bernier_cooling_pra_2009}
A discussion and a number of relevant references on the issue of cooling can be found in those articles, as well as in
~\cite{deleo_thermodynamics_pra_2011}.
In \fref{fig:cooling}, we display the basic idea behind the proposal for cooling by shaping the trap potential made
in \cite{bernier_cooling_pra_2009}.


\section{One dimensional Fermions}
\label{sec:mottfermion1D}


In a similar was than for the bosons, let us examine the case of one-dimensional Fermions.

As one can easily guess, there will be no Fermi liquid in one dimension. Indeed the Fermi liquid
theory rests on the fact that individual excitations very similar to the ones for free fermions exist.
Clearly this cannot be the case in 1D where only collective excitations can live. One again only the general
idea will be given and the reader referred to \cite{giamarchi_book_1d} for more details and references.

\subsection{Luttinger liquid and Mott insulators}

The bosonization formulas of \sref{sec:1dtech} can be easily modified to deal with bosons. The density
is strictly identical and can obviously be expressed in the same way
in terms of the field $\phi$. For the the single-particle operator
one has to satisfy an anticommutation relation instead of
(\ref{eq:comphen}). We thus have to introduce in representation
(\ref{eq:singlephen}) something that introduces the proper
minus sign when the two fermions operators are commuted. This
is known as a Jordan--Wigner transformation. Here, the operator
to add is easy to guess. Since the field $\phi_l$ has been
constructed to be a multiple of $2\pi$ at each particle,
$e^{i\frac12\phi_l(x)}$ oscillates between $\pm 1$ at the
location of consecutive particles. The Fermi field can thus be
easily constructed from the boson field (\ref{eq:singlephen})
by
\begin{equation}
    \psi^\dagger_F(x) = \psi^\dagger_B(x) e^{i\frac12\phi_l(x)}
\end{equation}
This can be rewritten in a form similar to
(\ref{eq:singlephen}) as
\begin{equation} \label{eq:singlephenfer}
 \psi^\dagger_F(x) = [\rho_0 - \frac1\pi \nabla \phi(x)]^{1/2}
 \sum_{p} e^{i (2p+1) (\pi \rho_0 x - \phi(x))}e^{-i \theta(x)}
\end{equation}
For fermions note that the least
oscillating term in (\ref{eq:singlephenfer}) corresponds to
$p=\pm 1$. This leads to two terms oscillating with a period
$\pm\pi\rho_0$ which is nothing but $\pm\kF$. These two terms
thus represent the Fermions leaving around their respective
Fermi points $\pm\kF$, also known as right movers and left
movers.

The action keeps exactly the same form than (\ref{eq:lutacphen}).
The important difference is that since the single particle operator contains already $\phi$
and $\theta$ at the lowest order (see (\ref{eq:singlephenfer}))
the kinetic energy alone leads to $K=1$ and interactions
perturb around this value, while for bosons non-interacting
bosons correspond to $K = \infty$. Attraction corresponds to $K > 1$ while
repulsion leads to $K < 1$. The correlation functions can thus easily be obtained.
For the density-density correlations we have exactly the same form than for the bosons
(\ref{eq:singleboscor}), the only difference being the different potential values for the LL parameter $K$.
In particular for the non-interacting fermions $K=1$ and one
recovers the universal $1/r^2$ decay of the Friedel
oscillations in a free electron gas. For repulsive interactions
$K<1$ and density correlations decay more slowly, while for
attractive interactions $K>1$ they will decay faster, being
smeared by the superconducting pairing.

The situation is different for the single particle correlations.
Contrarily to the case of bosons, for fermions the correlation contains
the terms $p = \pm 1$, corresponding to fermions close to $\pm
\kF$ respectively. If we compute the correlation for the right
movers we get
\begin{equation}
\begin{split}
 G_R(x,\tau) &= -e^{i \kF x} \langle T_\tau e^{i(\theta(x,\tau)-\phi(x,\tau)} e^{-i(\theta(0,0)-\phi(0,0))} \rangle \\
             &= e^{i \kF x} e^{-[\frac{K+K^{-1}}2 \log(r/\alpha) - i {\rm Arg}(y + i x)]}
\end{split}
\end{equation}
The single particle correlation thus decays as a non-universal
power law whose exponent depends on the Luttinger liquid
parameter. For free particles ($K=1$) one recovers
\begin{equation}
 G_R(r) =  - e^{i\kF x}
 e^{-\log[(y_\alpha - i x)/\alpha]} =
 -ie^{i\kF x} \frac1{x+i(v_F \tau
 +\alpha\;{\rm Sign}(\tau))}
\end{equation}
which is the normal function for ballistic particles with
velocity $u$. For interacting systems $K\neq 1$ the decay of
the correlation is always faster, which shows that single
particle excitations do not exist in the one dimensional world, and thus of course that no Fermi liquid
can exist.

\begin{figure}
\begin{center}
 \includegraphics[width=0.7\linewidth]{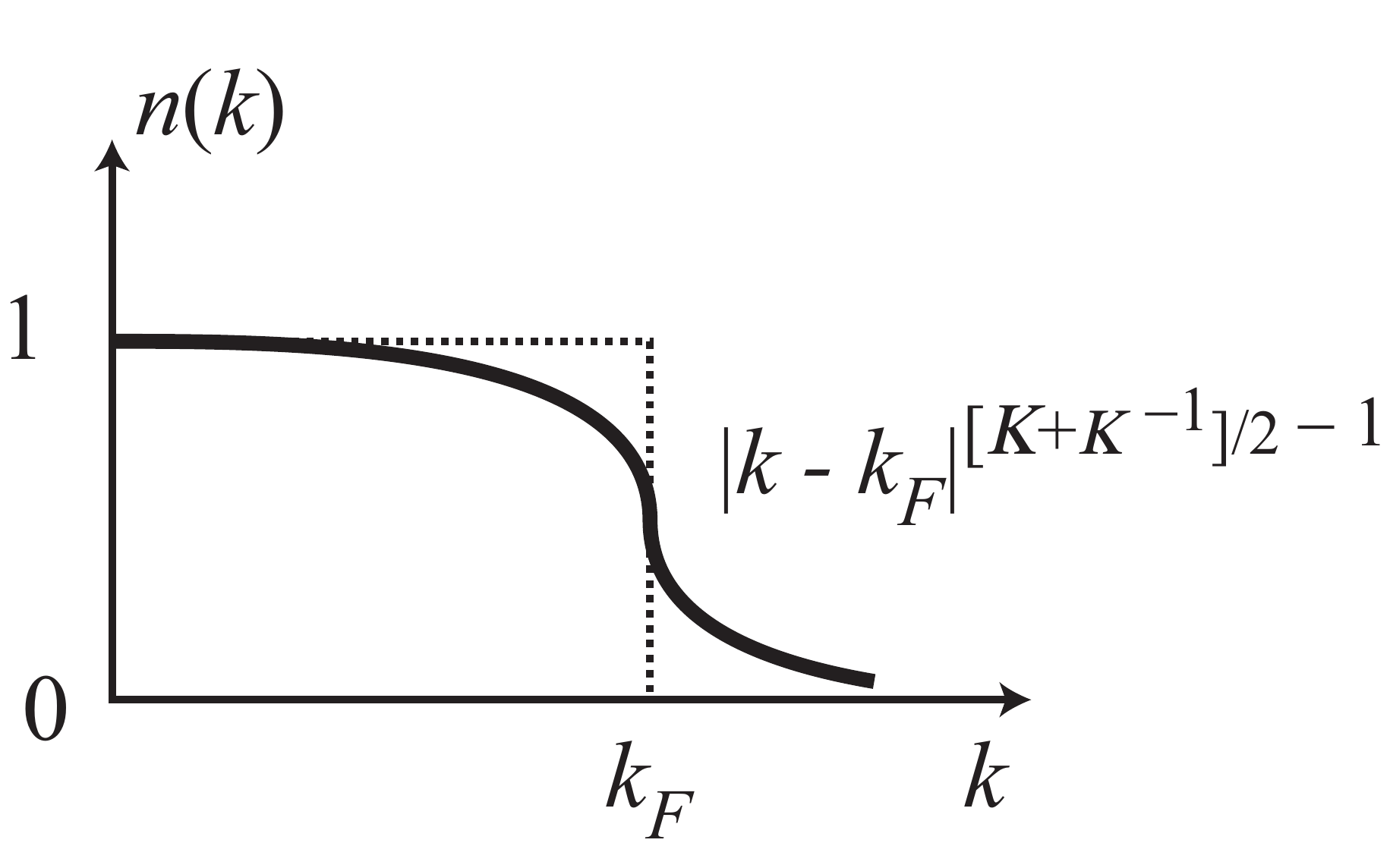}
\end{center}
 \caption{\label{fig:occuplut} The occupation factor $n(k)$. Instead of the usual
 discontinuity at $\kF$ for a Fermi liquid, it has a power law
 essential singularity. This is the signature that fermionic
 quasiparticles do not exist in one dimension. Note that the position of the
 singularity is still at $\kF$. This is a consequence of Luttinger's theorem stating that the volume of the Fermi surface cannot be changed
 by interactions.}
\end{figure}

One important consequence is the occupation factor $n(k)$ which
is given by the Fourier transform of the equal time Green's
function
\begin{equation}
 n(k) = \int dx \; e^{-i k x} G_R(x,0^-) =
  - \int dx \;e^{i (\kF - k) x}
 \left(\frac{\alpha}{\sqrt{x^2+\alpha^2}}\right)^{\frac{K+K^{-1}}2}
 e^{i\;{\rm Arg}(-\alpha + i x)}
\end{equation}
The integral can be easily determined by simple dimensional
analysis. It is the Fourier transform of a power law and thus
\begin{equation}
 n(k) \propto |k-\kF|^{\frac{K+K^{-1}}2-1}
\end{equation}
The occupation factor is shown in \fref{fig:occuplut}.
Instead of the discontinuity at $\kF$ that signals in a Fermi
liquid that fermionic quasiparticles are sharp excitations, one
thus finds in one dimension an essential power law singularity.
Formally, this corresponds to $Z=0$, another signature that all
excitations are converted to collective excitations and that
new physics emerges compared to the Fermi liquid case.

In practice this difference on $n(k)$ is relatively difficult to see unless the interaction is quite large, since the discontinuity of
$n(k)$ is smeared by the temperature. There are thus better ways to check for the LL properties for fermions \cite{giamarchi_book_1d}.

In a similar way that for bosons, one can add to the problem a lattice and check for the presence of a Mott insulator. The problem and properties are essentially the same than for bosons and we will not repeat the analysis here, but refer the reader to \cite{giamarchi_mott_shortrev,giamarchi_book_1d}. The essential difference comes again from the different values of the LL parameter $K$ for the two systems. So for example for the Hubbard model, the Mott insulator can be obtained for any values of $K < 1$, i.e. for any repulsive interactions. This is very similar to what happens in higher dimensions (see \fref{fig:hubbard_phasediag}. The perfect antiferromagnetic order is replaced by a powerlaw decay of the antiferromagnetic correlation functions. As for the case of bosons string order parameters can exist.

\subsection{Two component fermions: spin-charge separation}

\begin{figure}
 \begin{center}
  \includegraphics[width=0.7\linewidth]{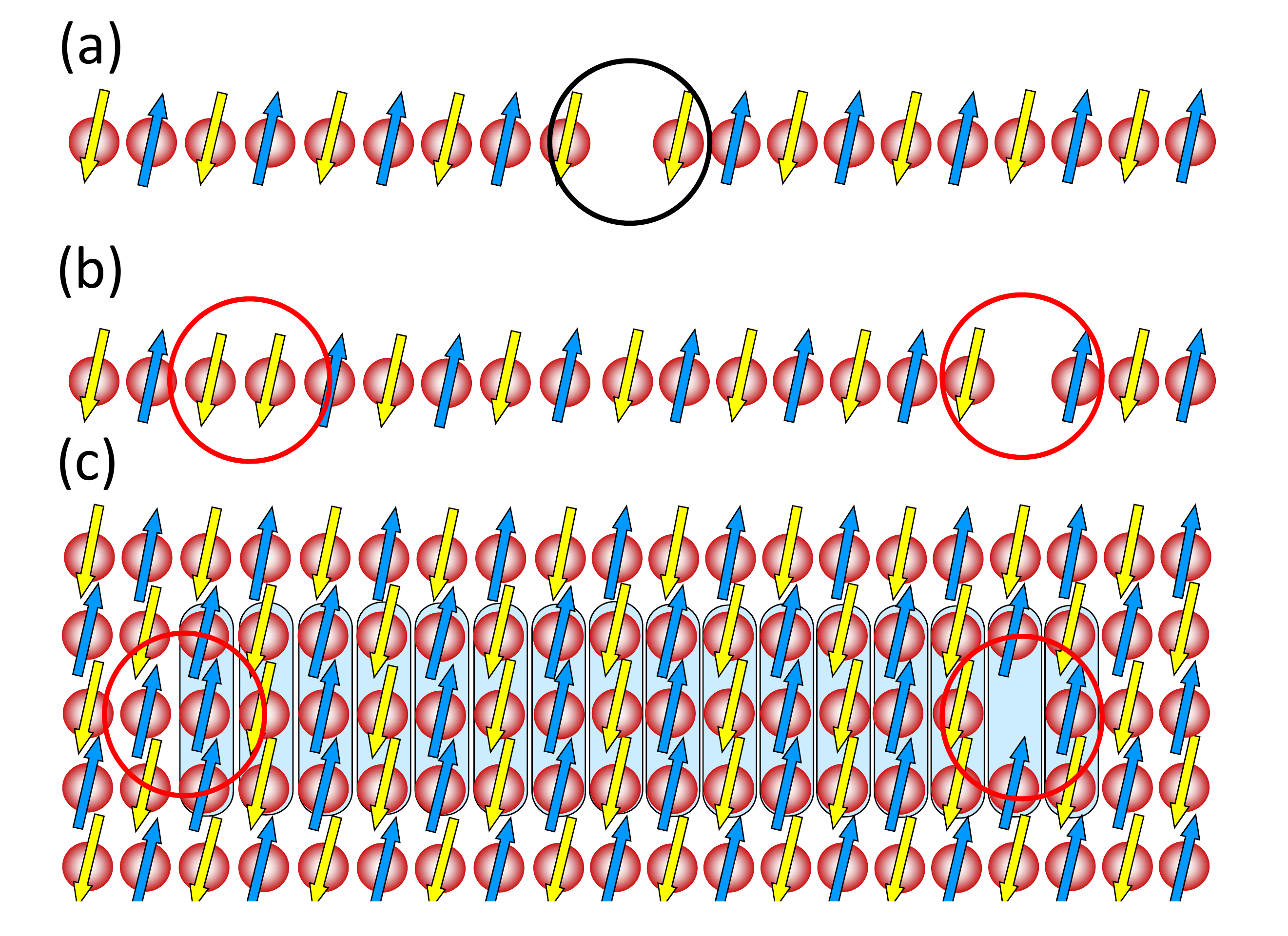}
 \end{center}
 \caption{\label{fig:fractional}
 A cartoon of the spin-charge separation (fractionalization of excitations) that naturally occurs in one dimension; a) one removes a particle
 which carries a spin and a charge; b) after the excitations have propagated we see that there is a place in the system where two parallel spins exist but no charge is missing. This is the spinon which carries spin but no charge. There is also a hole but with no distortion of the surrounding antiferromagnetic environment. This is the holon with a charge but no spin. The particle has thus fractionalized into to more elementary (collective) excitations.
 c) on the contrary to what happens in 1D in higher dimensions the holon and the spinon are held together by a series of frustrated bonds. They are thus bound and form the Fermi liquid quasiparticle.}
\end{figure}

A very interesting properties of one dimensional systems can be seen on two component systems (such as e.g. the Hubbard model). In that case one can represent the excitations by introducing collective variables for each component of the spins. One has thus four collective variables $(\phi_\up,\theta_\up)$ and $(\phi_\down,\theta_\down)$. However one can see that something remarkable happens. If one introduces the variables
\begin{equation}
\begin{split}
 \phi_\rho(x) &= \frac1{\sqrt2} [\phi_\up(x) + \phi_\down(x)] \\
 \phi_\sigma(x) &= \frac1{\sqrt2} [\phi_\up(x) - \phi_\down(x)]
\end{split}
\end{equation}
the first variable represents fluctuations of the total density $\rho_\up(x) + \rho_\down(x)$ while the second represents fluctuations of the spin density $\rho_\up(x) + \rho_\down(x)$. In terms of these variables the interaction in the Hubbard model completely decouples. Indeed
\begin{equation}
\begin{split}
 H &= U \sum_i \hn_{i\up} \hn_{i\down} \to \frac{U}{\pi^2} \int dx (\nabla\phi_\up(x))(\nabla\phi_\down(x)) \\
   &= \frac{U}{2\pi^2} \int dx [(\nabla\phi_\rho(x))^2 - (\nabla\phi_\sigma(x))^2]
\end{split}
\end{equation}
A similar decoupling occurs for the kinetic energy \cite{giamarchi_book_1d}. This means that the full Hilbert space of the problem decouples into two sectors, one sector only involving \emph{charge} excitations, and another involving \emph{spin} excitations. It immediately shows that a single particle excitation such as the Fermi liquid quasiparticle, which carries charge \emph{and} spin cannot exist. It shows that in one dimension what we could naively think as of an elementary excitation, namely an electron which carries both a charge and a spin, is in fact not the most elementary one. The electron fractionalize into two more elementary excitations: a) a holon which carries a charge but no spin; b) a spinon which carries a spin but no charge.
These excitations are directly linked to the fields $\phi_\rho$ and $\phi_\sigma$. Such a fractionalization is thus one of the important hallmark of the one dimensional world. It occurs in a variety of systems and context \cite{giamarchi_book_1d}. For the case of fermions with spin, one can make a cartoon
to visualize it. Such a cartoon is indicated in \fref{fig:fractional}. We also see that such a mechanism does not occur naturally in higher dimensions.
One of the important consequences of the spin-charge separation would be the occurrence in photoemission of a double singularity structure at the energies of holon and the spinon and not the single one that one expects in a Fermi liquid (see \fref{fig:spectralqp}). Probing for such effect is thus an extremely interesting and challenging question. In the condensed matter context only one experiment performing tunneling between two quantum wires could observe such a spin-charge separation \cite{auslaender05_quantumwire_tunneling_new,tserkovnyak_quantumwire_tunneling}. Cold atoms could thus be a very nice system to observe this effect. For fermions the temperature is still an issue, thus proposals to use two components bosonic systems instead have been put forward \cite{kleine_2velocities_bosons,kleine_2velocities_bosons_long} and remain to be tested.

\section{Conclusion}
\label{sec:conclusion}

This concludes our brief tour of interacting quantum fluids.
We have presented the basic concepts that underlay our understanding of quantum
interacting systems, both fermionic and bosonic.
Two major cornerstones are the Fermi liquid and the Luttinger liquid theories, which are effective
theories of the low-energy excitations of the system. They apply in two and higher dimensions,
and in one dimension, respectively.
They constitute references to which any novel properties or novel system must be compared.
Important effects of the interactions, such as the superfluid, Mott insulating and antiferromagnetic phases have been
discussed, and are at the forefront of current research.

A full solution or complete understanding of interacting quantum models beyond these low-energy effective models is still
a tremendously difficult task today, although recent years have witnessed significant progress in the field.
Indeed, the arsenal of tools at our disposal to tackle such questions,
both on the analytic and on the numerical side, has increased considerably and those tools have undergone
considerable development.
Notwithstanding,
the physics of such a simple model as the Hubbard model is still a formidable challenge, especially in two dimensions.
Cold atomic systems in optical lattices have provided a remarkable realization of such models and it is
certain that the ``quantum simulators'' realised in this novel experimental setup will help driving the field forward.

Of course many more challenges remain and these notes cannot even list all the exciting new
subjects that are connected to this physics. It is clear that questions such as cooling,  thermometry and new experimental
probes or spectroscopies are of central interest in order to make progress.
Cold atoms, by the control one can exert on the dimensionality of the lattice, have also opened the way to the study of
dimensional crossovers between low and higher dimensional situations.
For example the passage from a one dimensional situation to a two- or three- dimensional one remains a challenge which is of course of direct interest
to many systems in condensed matter physics.
In a similar way, cold atomic systems have opened the possibility to tackle much richer situations involving
several internal degrees of freedom, e.g. bosons with two ``spin'' components, Bose-Fermi mixtures, multi-component pairing states, etc...
All these systems potentially display a very rich and novel physics.
Cold atoms have also provided remarkable isolated quantum systems allowing to tackle in a different
way than in condensed matter the question of the out of equilibrium behavior of interacting quantum systems.
They also open the possibility of  dealing in a controlled manner with the influence of an external dissipative bath.
Last but not least, and because of the extreme control on the properties of the system they have allowed to study
in a controlled way the influence of disorder and the combined effects of disorder and interactions.

All these subjects go far beyond, but build upon the material exposed in these notes and constitute
the heart of the research on strongly correlated quantum systems.
Cold atoms have opened all these avenues and new frontiers for us, we are only at the beginning of the trip,
and we can surely expect beautiful surprises and discoveries in the years to come.

\section{Acknowledgements}

We are especially grateful to: J.-S.~Bernier, I.~Bloch, I.~Carusotto, M.~Cazalilla, V.~Cheianov, J.~Dalibard, T.L.~Dao, L. De Leo, E.~Demler, T.~Esslinger, M.~Ferrero, F.~Gerbier, A.F.~Ho, A.~Iucci, C.~Kollath,
M.~K\"{o}hl, O.~Parcollet, C.~Salomon, U.~Schollwoeck, A.~Tokuno, M.~Zvonarev for discussions and collaborations.
We acknowledge the support of the Agence Nationale de la Recherche, France (under programs
GASCOR, FABIOLA and FAMOUS), the Swiss National Science Foundation under MaNEP and Division II,
and the Army Research Office (DARPA-OLE program).

\bibliographystyle{OUPnamed_notitle}

\end{document}